\documentclass[11pt]{article}

\RequirePackage[numbers]{natbib}
\RequirePackage[colorlinks,citecolor=blue,urlcolor=blue]{hyperref}

\usepackage[bottom=1.5in,top=0.5in,left=1in, right=1in]{geometry}

\usepackage{pdfpages}
\usepackage{graphicx}
\usepackage{times}
\usepackage{algorithm}
\usepackage[algo2e]{algorithm2e} 
\RequirePackage{amsmath, amsthm}
\usepackage{enumitem}
\usepackage{xcolor}
\usepackage{amssymb}
\usepackage{times}
\usepackage{bm}
\usepackage{natbib}
\usepackage{subcaption}
\usepackage{color}
\usepackage{booktabs}
\usepackage{siunitx}
\usepackage{float}
\usepackage{tabularx}

\usepackage{comment}

\def\reviseone{\color{black}}
\def\pico{\color{black}}
\def\revisetwo{\color{black}}
\def\picotwo{\color{black}}

\newcommand{\footremember}[2]{%
	\footnote{#2}
	\newcounter{#1}
	\setcounter{#1}{\value{footnote}}%
}

\counterwithin{equation}{section}

\definecolor{green}{rgb}{0, 0.4, 0}

\newcommand{\mathbbm}[1]{\text{\usefont{U}{bbm}{m}{n}#1}}

\makeatletter
\renewcommand{\algocf@captiontext}[2]{#1\algocf@typo. \AlCapFnt{}#2} 
\def\@algocf@capt@plain{top}
\renewcommand{\algocf@makecaption}[2]{%
	\addtolength{\hsize}{\algomargin}%
	\sbox\@tempboxa{\algocf@captiontext{#1}{#2}}%
	\ifdim\wd\@tempboxa >\hsize
	\hskip .5\algomargin%
	\parbox[t]{\hsize}{\algocf@captiontext{#1}{#2}}
	\else%
	\global\@minipagefalse%
	\hbox to\hsize{\box\@tempboxa}
	\fi%
	\addtolength{\hsize}{-\algomargin}%
}
\makeatother

\DeclareMathOperator*{\argmax}{arg\,max}
\DeclareMathOperator*{\argmin}{arg\,min}

\newtheorem{theorem}{Theorem}
\newtheorem{lemma}{Lemma}
\newtheorem{corollary}{Corollary}
\newtheorem{proposition}{Proposition}
\newtheorem{remark}{Remark}

\newcommand{\numbereqn}{\addtocounter{equation}{1}\tag{\theequation}} 

\allowdisplaybreaks

\begin{document}

\hypersetup{linkcolor=blue}

\date{\today}

\author{Ray Bai\footremember{GMUStat}{Department of Statistics, George Mason University, Virginia, USA. Email: \href{mailto:rbai2@gmu.edu}{\tt rbai2@gmu.edu}}}

\title{Bayesian Group Regularization in Generalized Linear Models with a Continuous Spike-and-Slab Prior \thanks{Keywords and phrases:
		{generalized linear model},
		{group sparsity},
		{spike-and-slab},
		{spike-and-slab group lasso},
		{posterior contraction},
		{rate of convergence},
		{variable selection}
	}
}

\maketitle

\begin{abstract}
\noindent We study Bayesian group-regularized estimation in high-dimensional generalized linear models (GLMs) under a continuous spike-and-slab prior. Our framework covers both canonical and non-canonical link functions and subsumes logistic, Poisson, negative binomial, and Gaussian regression with group sparsity. We obtain the {\reviseone minimax $\ell_2$} convergence rate for both {\reviseone a} maximum \emph{a posteriori} (MAP) estimator \emph{and} the full posterior distribution {\reviseone under our prior}. Our theoretical results thus justify the use of the posterior mode as a point estimator. The posterior distribution {\reviseone also} contracts at the same rate as the MAP estimator, an attractive feature of our approach which is not the case for the group lasso. For computation, we propose {\reviseone expectation-maximization (EM) and Markov chain Monte Carlo (MCMC) algorithms}. We illustrate our method through simulations and a real data application on predicting human immunodeficiency virus (HIV) drug resistance from protein sequences. 
\end{abstract}

\section{Introduction} \label{intro}

\subsection{Motivation}

Generalized linear models (GLMs) \citep{McCullaghNelder1989} are widely used in practice and provide a unified way to model both continuous and discrete responses given a set of covariates. Suppose that we observe $n$ independent observations $\{ (\mathbf{x}_i, y_i) \}_{i=1}^{n}$, where $\mathbf{x}_i  = (x_{i1}, \ldots, x_{ip})^\top \in \mathbb{R}^{p}$ denotes a vector of $p$ covariates for the $i$th observation. GLMs assume that the response variable $y_i$ belongs to the exponential family with {\reviseone conditional} density,
\begin{equation} \label{exponentialfamily}
	f_i(y_i \mid {\reviseone \mathbf{x}_i}) = \exp \left\{ y_i \theta_i - b(\theta_i) + c(y_i) \right\}.
\end{equation}
where $\theta_i {\reviseone = \theta_i(\mathbf{x}_i)} \in \Theta \subset \mathbb{R}$ is the natural parameter {\reviseone depending on $\mathbf{x}_i$}, $b(\cdot)$ and $c(\cdot)$ are known functions, and the cumulant function $b$ is assumed to be twice differentiable with $b''(\theta) > 0$ for all $\theta \in \Theta$.  The family \eqref{exponentialfamily} includes the Gaussian, Bernoulli, binomial, Poisson, negative binomial, and gamma distributions \citep{McCullaghNelder1989}. {\reviseone The conditional mean $\mathbb{E}(y_i \mid \mathbf{x}_i) = b'(\theta_i)$} is related to a linear combination of the covariates $\mathbf{x}_i$ through a {\reviseone differentiable} link function $h$ so that $(h \circ b'): \Theta \mapsto \mathbb{R}$ is a strictly increasing function, i.e.
\begin{equation} \label{GLMnogroups}
	(h \circ b')(\theta_i) = {\reviseone \sum_{j=1}^{p} x_{ij} \beta_j },~~ i = 1, \ldots, n,
\end{equation}
where $\bm{\beta} = {\reviseone (\beta_1, \ldots, \beta_p)^\top} \in \mathbb{R}^{p}$ is the vector of regression coefficients to be estimated. In practice, $h$ is often chosen to be the canonical link function $h = (b')^{-1}$. This is the case for Gaussian linear regression, logistic regression, and Poisson regression. However, $h$ can also be chosen to be a \emph{non}-canonical link function. Probit regression and negative binomial regression with a log link are examples of GLMs with non-canonical link functions. 

Nowadays, it is common to collect datasets where the total number of covariates $p$ is large. In this ``large $p$'' setting, the covariates also often exhibit a grouping structure. For example, in genomic data, genes within the same biological pathway function together as a group to affect a clinical phenotype such as disease status or survival time \citep{HuangBrehenyMa2012}. At the individual gene level, mutations in an amino acid sequence can also be represented with group structure \citep{Rhee2006}. In our motivating data application in Section \ref{application}, each position of a protease gene sequence is represented by a group of binary variables, where each binary variable indicates the presence or absence of a specific amino acid. In MRI imaging, voxels within the same brain region naturally form a group \citep{LeeCao2021, WenYuYuLi2019}. In semiparametric GLMs, continuous functions of the covariates are often estimated by groups of nonlinear basis functions \citep{Lian2012}. Finally, categorical covariates with multiple levels can be represented as groups of dummy variables for each non-baseline category \citep{BrehenyHuang2015, MeierVanDeGeerBuhlmann2008}.

In these scenarios, it is desirable to take advantage of the {\reviseone known} grouping structure in order to improve estimation and prediction. Suppose that we have $G$ groups. Then we can express each $i$th covariate vector $\mathbf{x}_i \in \mathbb{R}^{p}$ as $\mathbf{x}_i = (\mathbf{x}_{i1}^\top, \ldots, \mathbf{x}_{iG}^\top)^\top$, where $\mathbf{x}_{ig} \in \mathbb{R}^{m_g}$ is a group of covariates of size $m_g$ and $\sum_{g=1}^{G} m_g = p$. In this case, we relate the covariates to the {\reviseone conditional mean $\mathbb{E}(y_i \mid \mathbf{x}_i) = b'(\theta_i)$} through the linear relationship,
\begin{equation} \label{GLMgroups}
	(h \circ b')(\theta_i) = \sum_{g=1}^{G} \mathbf{x}_{ig}^\top \bm{\beta}_g,~~ i = 1, \ldots, n,
\end{equation}
where $\bm{\beta}_g \in \mathbb{R}^{m_g}$ is the $g$th vector of regression coefficients corresponding to the $g$th group. It is clear that \eqref{GLMnogroups} is a special case of the grouped regression model \eqref{GLMgroups} where $G=p$, and $m_g = 1$ for all $g \in \{ 1, \ldots, G \}$ (i.e. each regression coefficient in \eqref{GLMnogroups} is its own ``group'' of size one). Therefore, \eqref{GLMgroups} provides a very natural generalization of the traditional GLM structure \eqref{GLMnogroups}. To distinguish these two structures, we henceforth refer to the model \eqref{GLMgroups} as a \emph{grouped GLM} and the model \eqref{GLMnogroups} as an \emph{unstructured GLM}.

When the number of groups $G$ is moderate or large in \eqref{GLMgroups}, some form of regularization is often desired. In the frequentist literature, penalized group estimators have been extended to GLMs with group structure \eqref{GLMgroups} \citep{MeierVanDeGeerBuhlmann2008, Lian2012, BlazereLoubesGamboa2014, BrehenyHuang2015}. In the Bayesian literature, spike-and-slab priors \citep{TangBioinformatics2017, LeeCao2021} and automatic relevance determination priors \citep{WenYuYuLi2019} have been used as sparsity-inducing priors in grouped GLMs. These methods all shrink a large number of the groups in \eqref{GLMgroups} towards zero, so that only a few of the groups of covariates are significantly associated with the mean response. 

In this paper, we adopt the Bayesian approach and employ a group spike-and-slab prior (to be introduced in Section \ref{method}) for estimating the $\bm{\beta}_g$'s in \eqref{GLMgroups}. We study our method theoretically and introduce {\reviseone computational algorithms for implementing it}. Our theory and algorithms apply to any member of the exponential family \eqref{exponentialfamily} and encompass both canonical and non-canonical link functions. Further, even when there is no known group structure, our results can still be applied to the traditional unstructured GLM \eqref{GLMnogroups}. Thus, our theoretical and computational framework is quite broad.

\subsection{Related work and our contributions} \label{litreview}

The literature on theory for \emph{Bayesian} high-dimensional GLMs is quite sparse. \citet{Jiang2007}, \citet{JeongGhosal2021}, and \citet{TangMartin2023} have also studied {\reviseone posterior} contraction rates for Bayesian GLMs in high dimensions. However, our work departs from these other papers in several important ways. {\reviseone First, we study GLMs under \emph{group} sparsity \eqref{GLMgroups}.} This setting is more general than the unstructured setting \eqref{GLMnogroups} considered by \citet{Jiang2007}, \citet{JeongGhosal2021}, and \citet{TangMartin2023}. However, since our model subsumes the traditional GLM \eqref{GLMnogroups} (where all $G$ groups in \eqref{GLMgroups} have size one), we obtain results for \emph{both} models \eqref{GLMnogroups} and \eqref{GLMgroups}.

{\reviseone Secondly, our study is conducted under a \emph{continuous} spike-and-slab prior}, to be introduced in Section \ref{method}. In contrast, \citet{Jiang2007}, \citet{JeongGhosal2021}, and \citet{TangMartin2023} all study {\reviseone \emph{point-mass} (discontinuous) spike-and-slab priors}. In these other papers, a model complexity prior is used to first select a random subset of $s < n$ predictors. Conditionally on the chosen set, the $s$ coordinates are then endowed with a multivariate prior, while the other regression coefficients are modeled with a Dirac delta density at zero. {\reviseone In contrast}, we {\reviseone employ} an absolutely continuous prior {\revisetwo which needs to be handled differently from these other papers}.

{\reviseone Finally, we characterize the convergence rate of \emph{both} a maximum \emph{a posteriori} (MAP) estimator \emph{and} the full posterior distribution.} In contrast, \citet{Jiang2007}, \citet{JeongGhosal2021}, and \citet{TangMartin2023} studied \emph{only} the full posterior distribution but \emph{not} any specific Bayesian point estimators.  One may wonder why it is necessary to study {\reviseone MAP estimators} and the posterior distribution separately. First, practitioners typically report a point estimate (e.g. the posterior mean, median, or mode) when they employ Bayesian methods. Thus, it is important to study the properties of these point estimators. Secondly, many researchers have shown that different Bayesian point estimates may have different asymptotic properties or behave very differently from the full posterior. For example, in the sparse normal means model, \citet{JohnstoneSilverman2004} showed that the posterior median under a point mass spike-and-slab prior attains the minimax risk, whereas the posterior mean converges at a slower, suboptimal rate. Under a different empirical Bayes spike-and-slab prior, \citet{CastilloMismer2018} showed that the posterior mean and median both obtain the optimal rate, but the full posterior converges at a suboptimal rate. In high-dimensional linear regression under a Laplace prior, \citet{CastilloSchmidtHieberVanDerVaart2015} showed that the posterior mode converges at the near minimax rate but the full posterior distribution converges much more slowly than the mode.  These examples reinforce the argument that Bayesian point estimators need to be analyzed separately from the full posterior. {\revisetwo In this work, we focus on a local MAP estimator rather than the posterior mean or median, because of its practical appeal. Unlike the posterior mean or median, local MAP estimators under our spike-and-slab prior result in \textit{exact} sparsity and are computationally faster to calculate. Given these appealing features, we aim to elucidate the theoretical optimality of MAP estimation for our model.} 

{\reviseone In short, our main contributions can be summarized as follows:
	\begin{enumerate}
		\item We study a heavy-tailed, \emph{continuous} spike-and-slab prior for \textit{grouped} GLMs \eqref{GLMgroups} with both canonical \emph{and} non-canonical link functions. This class of models is much broader than the grouped linear regression model studied by other authors \citep{YuanLin2006, BaiMoranAntonelliChenBoland2022, NardiRinaldo2008}, since it encompasses any member of the exponential family \eqref{exponentialfamily} and allows for many possible link functions besides the identity link.  
		\item We show that both a local MAP estimator \emph{and} the full posterior distribution under our spike-and-slab prior converge at the minimax-optimal $\ell_2$ error rate in GLMs with group sparsity. Our results thus justify the use of the posterior mode as a point estimator in high-dimensional Bayesian GLMs. We are not aware of any previous theoretical studies of Bayesian \emph{point estimates} under spike-and-slab priors in high-dimensional GLMs. At the same time, the posterior contraction rate also implies that the full posterior distribution provides valid inference in the sense that posterior credible sets have radius of an optimal size.
		\item For efficient point estimation under our model, we introduce an expectation-maximization (EM) algorithm. For fully Bayesian posterior inference, we provide Markov chain Monte Carlo (MCMC) algorithms. 
\end{enumerate}}
\noindent The rest of this paper is structured as follows. In Section \ref{method}, we describe our prior specification and discuss Bayesian estimation of $\bm{\beta}$. In Section \ref{MAPcharacterization}, we study {\reviseone the convergence rate of a local MAP estimator} under our approach. In Section \ref{fullposteriorcharacterization}, we characterize the convergence rate for the \emph{entire} posterior distribution and show that it can overcome the slow posterior contraction rate of the group lasso. Section \ref{computation} discusses how to implement our method. We conduct simulation studies in Section \ref{illustrations} and an analysis of an HIV drug resistance dataset in Section \ref{application}. Most of the proofs are deferred to the Appendix.

\subsection{Notation and preliminaries}

For two sequences of positive real numbers $a_n$ and $b_n$, we write $a_n = o(b_n)$ or $a_n \prec b_n$ if $\lim_{n \rightarrow \infty} a_n / b_n = 0$, $a_n = O(b_n)$ or $a_n \lesssim b_n$ if $|a_n / b_ n| \leq M$ for some positive real number $M$ independent of $n$, and $a_n \asymp b_n$ if $b_n \lesssim a_n \lesssim b_n$. For a set $\mathcal{S}$, we denote its cardinality by $|\mathcal{S}|$, and for a subset $\mathcal{T} \subset \mathcal{S}$, $\mathcal{T}^{c}$ means $\mathcal{T}^{c} = \mathcal{S} \setminus \mathcal{T}$. 

The $\ell_{\infty}$, $\ell_2$ and $\ell_1$ norms of a vector $\mathbf{v}$ are denoted by $\lVert \mathbf{v} \rVert_{\infty}$, $\lVert \mathbf{v} \rVert_2$ and $\lVert \mathbf{v} \rVert_1$ respectively. For an $m \times n$ matrix $\mathbf{A}$ with entries $a_{ij}$, we denote $\lVert \mathbf{A} \rVert_{2, \infty} = \max_{1 \leq i \leq n} (\sum_{j=1}^{m} a_{ij}^2)^{1/2}$ as the maximum row length of $\mathbf{A}$ and $\lVert \mathbf{A} \rVert_{\max} = \max_{i,j} | a_{ij} |$ as the maximum entry in absolute value. For a symmetric matrix $\mathbf{C}$, we denote its minimum and maximum eigenvalues by $\lambda_{\min}(\mathbf{C})$ and $\lambda_{\max}(\mathbf{C})$ respectively. For a vector $\mathbf{x}$, {\reviseone $\textrm{diag} \{ \mathbf{x} \}$} denotes the diagonal matrix determined by entries of $\mathbf{x}$. If $f$ is a univariate function, then $f(\mathbf{x})$ means that $f$ is applied elementwise to the entries in the vector $\mathbf{x}$. The notation {\reviseone $\mathbf{1}_n$} means an {\reviseone $n$}-dimensional vector of all ones, while {\reviseone $\mathbf{0}_n$} denotes an {\reviseone $n$}-dimensional zero vector. 

To succinctly express a GLM with group sparsity \eqref{GLMgroups} under the exponential family \eqref{exponentialfamily}, we denote $\mathbf{Y} = (y_1, \ldots, y_n)^\top \in \mathbb{R}^{n}$, $\mathbf{X} = ( \mathbf{X}_1, \ldots, \mathbf{X}_G ) \in \mathbb{R}^{n \times p}$, and $\bm{\beta} = (\bm{\beta}_1^\top, \ldots, \bm{\beta}_G^\top)^\top \in \mathbb{R}^{p}$, where $p = \sum_{g=1}^{G} m_g$. We define the function $\xi$ as $\xi = (h \circ b')^{-1}$. Then the log-likelihood for \eqref{GLMgroups} can be written (up to normalizing constant) as
\begin{equation} \label{ll}
	\ell_n(\bm{\beta}) = \mathbf{Y}^\top \xi(\mathbf{X}\bm{\beta}) - \mathbf{1}_n^\top b(\xi(\mathbf{X}\bm{\beta})).
\end{equation}
The gradient of $\ell_n(\bm{\beta})$ is then
\begin{equation} \label{llgradient}
	\nabla \ell_n(\bm{\beta}) = \mathbf{X}^\top \textrm{diag} \{ \xi'(\mathbf{X}\bm{\beta}) \} (\mathbf{Y} - {\reviseone b'(\xi(\mathbf{X}\bm{\beta}))}).
\end{equation}
while the {\reviseone observed information matrix} is
\begin{equation} \label{llhessian}
	- \nabla^2 \ell_n(\bm{\beta}) = \mathbf{X}^\top \bm{\Sigma}(\bm{\beta}) \mathbf{X},
\end{equation}
where 
\begin{equation} \label{Sigmamatrix}
	\bm{\Sigma}(\bm{\beta}) = {\reviseone \bm{\Omega}(\bm{\beta}) - \text{diag} \{ \xi '' (\mathbf{X}\bm{\beta}) \} \text{diag} \{ \mathbf{Y} - b'(\xi(\mathbf{X}\bm{\beta})) \},}
\end{equation} 
{\reviseone and} 
\begin{equation} \label{Omegamatrix}
	{\reviseone \bm{\Omega}(\bm{\beta}) = \text{diag} \{ (h^{-1})' (\mathbf{X}\bm{\beta}) \} \text{diag} \{ \xi'(\mathbf{X}\mathbf{\beta}) \} }.
\end{equation}
{\reviseone Note that $\mathbf{X}^\top \bm{\Omega}(\bm{\beta}) \mathbf{X}$ is also the \textit{Fisher} information matrix, or the expected value of the Fisher information matrix \eqref{llhessian}. Furthermore, if the canonical link function is used, then {\revisetwo $\bm{\Sigma}(\bm{\beta}) = \bm{\Omega}(\bm{\beta})$}, and \eqref{Sigmamatrix} can be greatly simplified to  $\bm{\Sigma}(\bm{\beta}) = \textrm{diag} \{ b''(\mathbf{X}\bm{\beta}) \}$.}

\section{Prior specification and Bayesian estimation}\label{method}

\subsection{Spike-and-slab group lasso} \label{spikeandslab}

Given a high-dimensional GLM with group structure \eqref{GLMgroups}, a Bayesian approach to estimation and variable selection is to put a prior on the parameter $\bm{\beta}$. For an $m_g \times 1$ random vector $\bm{\beta}_g$, we first define the multivariate density function,
\begin{equation} \label{multivariateLaplace}
	\bm{\Psi}(\bm{\beta}_g \mid \lambda) = \frac{\lambda^{m_g} e^{-\lambda \lVert \bm{\beta}_g \rVert_2}}{2^{m_g} \pi^{m_g-1} \Gamma((m_g+1)/2) }.
\end{equation} 
It is important to note that \eqref{multivariateLaplace} is a multivariate \emph{Laplace} distribution {\reviseone \citep{FangSymmetric1990}}, \emph{not} a multivariate Gaussian. The exponent term of \eqref{multivariateLaplace} contains the $\ell_2$ norm $\lVert \bm{\beta}_g \rVert_2$ rather than the squared $\ell_2$ norm $\lVert \bm{\beta}_g \rVert_2^2$. As a result, the density \eqref{multivariateLaplace} has tails that are \emph{heavier} than normal. It is easy to see that if $m_g=1$, then \eqref{multivariateLaplace} reduces to a univariate \emph{Laplace} density with scale parameter $\lambda^{-1}$. The hyperparameter $\lambda$ controls how concentrated $\bm{\beta}_g$ is around the zero vector $\bm{0}_{m_g}$, with larger values of $\lambda$ leading to a density that is more peaked around $\bm{0}_{m_g}$.

To induce group sparsity in $\bm{\beta}$ under \eqref{GLMgroups}, we endow $\bm{\beta}$ with the spike-and-slab group lasso (SSGL) prior of \citet{BaiMoranAntonelliChenBoland2022},
\begin{equation} \label{SSGL}
	\pi(\bm{\beta}) = \prod_{g=1}^{G} \left[ (1-\theta) \bm{\Psi}(\bm{\beta}_g \mid \lambda_0) + \theta \bm{\Psi}(\bm{\beta}_g \mid \lambda_1 ) \right],
\end{equation}
where $\theta \in (0,1)$ is a mixing proportion. In \eqref{SSGL}, $\lambda_0$ is set to be a large value so that $\bm{\Psi}(\bm{\beta}_g \mid \lambda_0)$, i.e. the ``spike,'' is highly concentrated around the zero vector $\bm{0}_{m_g}$. Meanwhile, $\lambda_1 \ll \lambda_0$ is set to be small so that $\bm{\Psi}(\bm{\beta}_g \mid \lambda_1)$, i.e. the ``slab,'' is a diffuse and relatively flat density. In \eqref{SSGL}, the slab density models the nonzero groups, while the spike density models the zero groups. The SSGL prior was originally introduced by \citet{BaiMoranAntonelliChenBoland2022} in the Gaussian linear regression model, $\mathbf{Y}= \mathbf{X}\bm{\beta} + \bm{\varepsilon}, \bm{\varepsilon} \sim \mathcal{N}_n(\bm{0}, \sigma^2 \mathbf{I}_n)$. {\reviseone This paper extends the work of \citet{BaiMoranAntonelliChenBoland2022} to the much more general GLM setting where the response variables $\mathbf{Y}$ can also be discrete or non-Gaussian (e.g. binary, binomial, Poisson, negative binomial, gamma, etc.). Using the prior \eqref{SSGL}, we also develop asymptotic theory for \emph{point estimation} in high-dimensional Bayesian GLMs, which to our knowledge, has not been studied before.}

When $m_1 = \ldots = m_G = 1$, the SSGL prior \eqref{SSGL} reduces to a two-component mixture of \emph{univariate} Laplace densities,
\begin{equation} \label{SSL}
	\pi(\bm{\beta}) = \prod_{j=1}^{p} \left[ (1-\theta) \psi(\beta_j \mid \lambda_0) + \theta \psi(\beta_j \mid \lambda_1 ) \right].
\end{equation}
where $\psi(\beta_j \mid \lambda) = (\lambda / 2) \exp ( - \lambda | \beta_j |)$ denotes the density of a univariate Laplace distribution. The prior \eqref{SSL} is the spike-and-slab lasso (SSL) originally introduced by \citet{RockovaGeorge2018} in non-grouped linear regression. In order to conduct Bayesian inference for unstructured GLMs \eqref{GLMnogroups}, we can place the SSL prior \eqref{SSL} on the individual regression coefficients in \eqref{GLMnogroups}. \citet{TangShenZhangYi2017} extended the SSL \eqref{SSL} to high-dimensional GLMs. However, the theoretical properties for the SSL in GLMs have thus far not been investigated. As a byproduct of our theoretical analysis of the SSGL \eqref{SSGL}, we \emph{also} obtain the rates of convergence for the SSL \eqref{SSL} in GLMs in Sections \ref{MAPcharacterization} and \ref{fullposteriorcharacterization}.

\subsection{Bayesian estimation} \label{Bayesianestimation}

After endowing the groups of regression coefficients $\bm{\beta}$ in  \eqref{GLMgroups} with an appropriate prior distribution $\pi(\bm{\beta})$, we obtain the posterior distribution for $\bm{\beta}$,
\begin{equation} \label{posteriorSSGL}
	\pi(\bm{\beta} \mid \mathbf{Y}) = \frac{\exp(\ell_n(\bm{\beta})) \pi(\bm{\beta})}{\int \exp(\ell_n(\bm{\beta})) \pi(\bm{\beta}) d \bm{\beta} },
\end{equation}
where $\ell_n(\bm{\beta})$ is the log-likelihood \eqref{ll}. The posterior \eqref{posteriorSSGL} is typically intractable, but Markov chain Monte Carlo (MCMC) can be used to draw samples from the approximate posterior. From \eqref{posteriorSSGL}, we also see that the log-posterior (up to normalizing constant) is 
\begin{equation} \label{logposterior-main} 
	\log \pi(\bm{\beta} \mid \mathbf{Y}) = \ell_n(\bm{\beta}) + \log \pi(\bm{\beta}).
\end{equation}
Hence, a very natural point estimator for $\bm{\beta}$ is a local MAP estimator $\widehat{\bm{\beta}}$, i.e. 
\begin{equation} \label{MAPestimatorSSGL}
	\widehat{\bm{\beta}} \text{ such that } \nabla \log \pi(\widehat{\bm{\beta}} \mid \mathbf{Y}) = \bm{0}_p.
\end{equation} 
Local MAP estimators (or local posterior modes) are useful point estimates to consider because standard optimization algorithms can be used to rapidly obtain an estimate $\widehat{\bm{\beta}}$. These optimization algorithms are often faster and much more scalable than MCMC algorithms. In general, local MAP estimators may not be unique, {\revisetwo and the log-likelihood function $\ell_n(\boldsymbol{\beta})$ may be unbounded}. However, if the prior $\pi(\bm{\beta})$ in \eqref{logposterior-main} is proper, does not depend on the observed data, {\revisetwo and sufficiently restricts the parameter space for $\boldsymbol{\beta}$ (e.g. by shrinking most of the coefficients in $\boldsymbol{\beta}$ towards zero)}, then the posterior density $\pi( \bm{\beta} \mid \mathbf{Y})$ will be proper and bounded, i.e. ${\revisetwo \sup_{\bm{\beta}}} \pi( \bm{\beta} \mid \mathbf{Y} ) < \infty$. This enables us to find a local MAP estimator \eqref{MAPestimatorSSGL}. 

In particular, if $\pi(\bm{\beta})=\prod_{g=1}^{G} \bm{\Psi}(\bm{\beta}_g \mid \lambda)$, where $\bm{\Psi}(\cdot \mid \lambda)$ is the multivariate Laplace prior \eqref{multivariateLaplace}, then {\reviseone \eqref{logposterior-main}} becomes
\begin{equation} \label{grouplassoobj}
	{\reviseone \log \pi(\bm{\beta} \mid \mathbf{Y}) = \ell_n(\bm{\beta}) - \lambda \sum_{g=1}^{G} \lVert \bm{\beta}_g \rVert_2}, 
\end{equation}
which is the objective function for the group lasso {\reviseone estimator} of \citet{YuanLin2006}. Thus, the group lasso {\reviseone estimator} corresponds to the MAP estimator under independent multivariate Laplace priors \eqref{multivariateLaplace} on the $\bm{\beta}_g$'s, and the MAP estimator for each {\reviseone group} $\bm{\beta}_g$ is either exactly $\bm{0}_{m_g}$ or nonzero. 

If instead, we use the SSGL prior \eqref{SSGL} for $\pi(\bm{\beta})$ in {\reviseone \eqref{logposterior-main}}, then {\reviseone local SSGL MAP estimators \eqref{MAPestimatorSSGL}} will \emph{also} be exactly sparse, since the mixture components in \eqref{SSGL} are both multivariate Laplace. However, whereas the group lasso \eqref{grouplassoobj} applies the same amount of shrinkage $\lambda$ to every group, the SSGL \eqref{SSGL} allows for \emph{adaptive} shrinkage. This is because the slab density $\bm{\Psi}(\cdot \mid \lambda_1)$ of the SSGL \eqref{SSGL} prevents groups with larger coefficients from being downward biased. The combination of \emph{exact} group sparsity and adaptive shrinkage of {\reviseone local MAP estimators \eqref{MAPestimatorSSGL}} under the SSGL prior \eqref{SSGL} makes the SSGL very appealing for both group selection \emph{and} estimation. {\reviseone In contrast, neither the posterior mean nor the posterior median under the SSGL prior is exactly sparse. Estimating the posterior mean or median also often requires the use of MCMC.} In Sections \ref{SSGLvsGL} and \ref{GLsuboptimality}, we {\reviseone further} demonstrate the theoretical advantages of the SSGL prior \eqref{SSGL} over the group lasso prior \eqref{multivariateLaplace}. SSGL is also empirically shown to significantly outperform the group lasso in Sections \ref{illustrations} and \ref{application}. 

\section{Characterization of the MAP estimator} \label{MAPcharacterization}

\subsection{Convergence rate of the MAP estimator} \label{MAPconvergence}

We first {\reviseone theoretically study MAP estimation under the SSGL prior \eqref{SSGL}. Our goal is to establish the existence of a local MAP estimator \eqref{MAPestimatorSSGL} with the minimax estimation rate in terms of $\ell_2$ risk.} To the best of our knowledge, our work is the first one to investigate {\reviseone \emph{point estimation} under a spike-and-slab prior} in high-dimensional \emph{Bayesian} GLMs. Other authors \citep{Jiang2007, JeongGhosal2021, TangMartin2023} have only studied the convergence rate for the \emph{full} posterior distribution in Bayesian GLMs, which we will also {\reviseone consider} in Section \ref{fullposteriorcharacterization}. 

Suppose the true regression coefficients vector is $\bm{\beta}_0 = (\bm{\beta}_{01}^\top, \ldots, \bm{\beta}_{0G}^\top)^\top \in \mathbb{R}^{p}$, where $\bm{\beta}_{0g} \in \mathbb{R}^{m_g}$ {\reviseone is the subvector corresponding to the $g$th group of size $m_g$}. Then the true model is
\begin{equation} \label{truemodel}
	(h \circ b')(\theta_{0i}) = \sum_{g=1}^{G} \mathbf{x}_{ig}^\top \bm{\beta}_{0g},~~i=1, \ldots, n,
\end{equation}
where $h$ and $b$ are the known link function and cumulant function respectively, while $\theta_{0i}$ is the true natural parameter in \eqref{exponentialfamily}. Further, let $S_0 \subset \{1, \ldots, G\}$ be the set of indices of the true nonzero groups in $\bm{\beta}_0$, with cardinality $s_0 = |S_0|$. {\reviseone Then $S_0^c = \{1, \ldots, G\} \setminus S_0$.}  {\reviseone For a $p$-dimensional vector $\bm{\beta}$, let $\bm{\beta}_{S_0}$ denote the subvector of $\bm{\beta}$ with the groups in $S_0$, and let $\bm{\beta}_{S_0^c}$ be the subvector with groups in $S_0^c$. Let $\mathbf{X}_{S_0}$ denote the submatrix of the design matrix $\mathbf{X}$ with the $\sum_{g \in S_0} m_g$ columns of $\mathbf{X}$. Recall that $\bm{\Sigma}(\bm{\beta})$ and $\bm{\Omega}(\bm{\beta})$ are the matrices defined respectively in \eqref{Sigmamatrix} and \eqref{Omegamatrix}. } 

{\reviseone Without loss of generality, assume that the first $s_0$ groups in $\bm{\beta}_0$ are nonzero, so that $\bm{\beta}_0 = (\bm{\beta}_{0S_0}^\top, \bm{0}^\top)^\top$.} We {\reviseone additionally} make the following set of assumptions:

\begin{enumerate}[label=(A\arabic*)]
	\item $G \gg n$ and {\reviseone $\log G = o(n^{1/2})$.}
	\item {\reviseone $s_0  = o(n^{1/2} / \log G)$} and $m_{\max} = O(\log n \wedge (\log G / \log n))$, {\reviseone where $m_{\max} = \max_{1 \leq g \leq G} m_g$.}
	\item The design matrix $\mathbf{X}$ satisfies the following conditions:
	\begin{enumerate}[label=(\roman*)]
		\item All the entries $x_{ij}$ of $\mathbf{X}$ satisfy {\reviseone $|x_{ij}| = O(\log G)$}.
		\item Define the neighborhood $\mathcal{N}_0 = \{ \bm{\delta} \in \mathbb{R}^{p} : \lVert \bm{\delta} - \bm{\beta}_0 \rVert_2 \leq (s_0 \log G /n)^{1/2} \}$. For any $\bm{\delta} \in \mathcal{N}_0$, {\reviseone $\lambda_{\min}(n^{-1}  \mathbf{X}_{S_0}^\top \bm{\Sigma}(\bm{\delta}) \mathbf{X}_{S_0}) \gtrsim 1$, $\lambda_{\min} (n^{-1} \mathbf{X}_{S_0}^\top \bm{\Omega}(\bm{\delta}) \mathbf{X}_{S_0}) \gtrsim 1$, and $\lambda_{\max} (n^{-1} \mathbf{X}_{S_0}^\top \bm{\Omega}(\bm{\delta}) \mathbf{X}_{S_0}) \prec \log G / \log n$.}
		\item For any group $g \in S_0^c$ and $\bm{\delta} \in \mathcal{N}_0$, $\lVert \mathbf{X}_{S_0}^\top \bm{\Sigma}(\bm{\delta}) \mathbf{X}_g \rVert_{2, \infty} = O(n)$.
	\end{enumerate}
	\item The observations $\{ (\mathbf{x}_i, y_i) \}_{i=1}^{n}$ satisfy the following conditions:
	\begin{enumerate}[label=(\roman*)]
		\item The responses $\{ y_i \}_{i=1}^{n}$ satisfy $\mathbb{E}(|y_i - b'(\xi(\mathbf{x}_i^\top \bm{\beta}_0))|^k) \leq \frac{k!}{2} \mathbb{E}[y_i^2] L^{k-2}$ for some $L >0$ and every integer $k \geq 2$. In addition, $\textrm{Var}(y_i \mid \mathbf{x}_i) \lesssim G$ for all $i = 1, \ldots, n$.
		\item For the function $\xi = (h \circ b')^{-1}$ in \eqref{ll}, $\xi'(\mathbf{x}_i^\top \bm{\beta}) < \infty$ for all {\revisetwo $\boldsymbol{\beta} \in \mathbb{R}^{p}$ and} $i = 1, \ldots, n$.
	\end{enumerate}
\end{enumerate}
Assumption (A1) allows the number of groups $G$ to diverge at a nearly exponential rate with $n$ {\reviseone and is analogous to the growth rate for univariate GLMs in \citet{FanLv2011}}. Assumption (A2) allows the number of nonzero groups $s_0$ and the group sizes $m_g$'s to diverge but at rates slower than $n$. {\reviseone Note that Assumption (A2) implies a sparsity condition on $\bm{\beta}_0$, where the true $\bm{\beta}_0$ only contains a small number of nonzero groups. Assumption (A3)(i) holds automatically if the entries of $\mathbf{X}$ are uniformly bounded. If the entries of $\mathbf{X}$ are sub-Gaussian with scale factor $\sigma$, then this assumption also holds with probability greater than $1 - 2 \exp (-D n / 2 \sigma^2)$ for some constant $D>0$. 
	
	Assumption (A3)(ii) gives restricted eigenvalue conditions for the submatrix $\mathbf{X}_{S_0}$ of $\mathbf{X}$. Note that the maximum eigenvalue for $n^{-1} \mathbf{X}_{S_0}^\top \bm{\Omega}(\bm{\delta})\mathbf{X}_{S_0}$ can diverge as $n \rightarrow \infty$.} Restricted eigenvalue conditions are routinely employed in the high-dimensional GLM literature \citep{FanLv2011, TangMartin2023}. These conditions ensure the identifiability of $\bm{\beta}_0$. {\reviseone It should also be noted that since our theory covers both canonical \textit{and} non-canonical link functions $h(\cdot)$, we require eigenvalue conditions for both $n^{-1} \mathbf{X}_{S_0}^\top \bm{\Sigma}(\bm{\delta}) \mathbf{X}_{S_0}$ and $n^{-1} \mathbf{X}_{S_0}^\top \bm{\Omega}(\bm{\delta})\mathbf{X}_{S_0}$. If a non-canonical link function is used, then in general, $\bm{\Sigma}(\bm{\beta}) \neq \bm{\Omega}(\bm{\beta})$. However, if the canonical link function is used, then $\bm{\Sigma}(\bm{\beta}) = \bm{\Omega}(\bm{\beta})$, and the first two eigenvalue conditions in (A3)(ii) are the same.}

Assumption (A3)(iii) is an {\reviseone irrepresentability-type condition} \citep{ZhaoYu2006, FanLv2011}, which limits the correlations between the active covariates $\mathbf{X}_{S_0}$ and the inactive ones $\mathbf{X}_{S_0^c}$. Since $\mathbf{X}_{S_0}^\top \bm{\Sigma}(\bm{\delta}) \mathbf{X}_g$ has $m_g$ columns and $m_g \ll n$ for all $g \in \{ 1, \ldots, G \}$, the condition that $\lVert \mathbf{X}_{S_0}^\top \bm{\Sigma}(\bm{\delta}) \mathbf{X}_g \rVert_{2, \infty} = O(n)$ is also fairly mild. {\reviseone In fact, Assumption (A3)(iii) is much weaker than the strong irrepresentability condition of \cite{ZhaoYu2006} for the LASSO, which requires that $\lVert \mathbf{X}_{S_0^c}^\top \mathbf{X}_{S_0} (\mathbf{X}_{S_0}^\top \mathbf{X}_{S_0})^{-1} \rVert_{\infty} \leq C <\infty$. In contrast, we can allow $\lVert \mathbf{X}_{S_0}^\top \bm{\Sigma}(\bm{\delta}) \mathbf{X}_g \rVert_{2, \infty}$ to diverge on the order of $n$. A nearly identical assumption as Assumption (A3)(iii) was made by \citet{FanLv2011}. When $G \gg n$, it does not seem as though this type of irrepresentability condition can be removed. In low-dimensional (i.e. fixed $G$) settings, Assumption (A3)(iii) is not needed \citep{NardiRinaldo2008}. However, if $G$ grows much faster than $n$, then conditions like (A3)(iii) may be required \citep{HuangMaZhang2008, FanLv2011}. If we consider weaker notions of convergence such as convergence in $\ell_{\infty}$ norm or if we focus on the problem of support recovery (i.e. the ability to asymptotically identify the correct subset of nonzero groups), then we may also be able to remove the irrepresentability condition \citep{LohWainwright2017}. However, we conjecture that this type of condition is necessary when considering convergence in $\ell_2$ norm in the high-dimensional $G \gg n$ regime.}

Assumption (A4)(i) is an assumption on the central moments of the response variables and implies that the tails decay exponentially. This assumption is satisfied for the Gaussian, Poisson, Bernoulli, gamma, and Laplace distributions, among others \citep{Baraud2010}. In addition, the assumption that $\textrm{Var}(y_i \mid \mathbf{x}_i) \lesssim G$ is a very weak assumption, in light of Assumption (A1) that allows $G = O(e^{n^{\xi}})$, for some $\xi \in (0,1/2)$. Assumption (A4)(ii) is also satisfied for many GLMs, even if $\mathbf{x}_i^\top \bm{\beta}$ is unbounded. For example, the canonical link function $h = (b')^{-1}$ is usually used in practice, e.g. in logistic, Poisson, and Gaussian regression. In this case, $\xi(u) = (h \circ b')^{-1}(u)=u$ and $\xi'(u) = 1$ for all $u \in \mathbb{R}$. {\reviseone In negative binomial regression with the log link $h(u) = \log u$ and a given number of failures $r$, we have $b(u) = -r \log(1-e^{u})$, $\xi(u) = -\log(re^{-u}+1)$. Thus, $\xi'(u) = r/(r+e^{u}) \leq 1$ for all $u \in \mathbb{R}$.} However, even if $\xi'(u)$ is unbounded in $\mathbb{R}$, we can still satisfy $\xi' (\mathbf{x}_i^\top \bm{\beta}_0) < \infty$ if we make a stronger assumption that $\lVert \mathbf{X}\bm{\beta}_0 \rVert_{\infty} < \infty$.

{\reviseone The minimax-optimal $\ell_2$ risk for estimating $\bm{\beta}_0$ in the grouped regression model \eqref{truemodel} is $\{ [s_0 \log (G / s_0) + \sum_{g \in S_0} m_g]/n\}^{1/2}$ \citep{HuangZhangAoS2010}. Under Assumptions (A1)-(A2), the minimax $\ell_2$ convergence rate is of the same order as $(s_0 \log G / n)^{1/2}$. {\reviseone Our first theorem below} certifies that there exists a local MAP estimator \eqref{MAPestimatorSSGL} under the SSGL prior \eqref{SSGL} which achieves this minimax-optimal rate with respect to the $\ell_2$ risk. This justifies the use of MAP estimation for Bayesian grouped regression with the SSGL prior.}

\begin{theorem}[convergence rate of {\reviseone a local MAP estimator} under SSGL] \label{Th:1}
	Suppose that we have a grouped GLM \eqref{truemodel}, and we endow $\bm{\beta}_0$ with the SSGL prior \eqref{SSGL} where the hyperparameters $(\lambda_0, \lambda_1, \theta)$ satisfy $\lambda_0 = (1-\theta)/\theta \asymp G^{c}$, where $c>2$, and $\lambda_1 \asymp 1/n$.  Further, assume that Assumptions (A1)-(A4) hold. Then {\reviseone there exists a strict local MAP estimator $\widehat{\bm{\beta}} = (\widehat{\bm{\beta}}_{S_0}^\top, \widehat{\bm{\beta}}_{S_0^c}^\top)^\top$ of the log-posterior \eqref{logposterior-main} such that, as $n \rightarrow \infty$, $\widehat{\bm{\beta}}_{S_0^c} = \bm{0}$ and}
	\begin{equation}
		\lVert \widehat{\bm{\beta}} - \bm{\beta}_0 \rVert_2 = O_p \left(  \sqrt{\frac{s_0 \log G}{n}} \right).
	\end{equation} 
\end{theorem}
\begin{proof}
	Appendix \ref{App:A}.
\end{proof}

\begin{remark} \label{Remark:1}
	We have treated the mixing weight $\theta$ as a deterministic quantity which depends on $G$. However, Theorem \ref{Th:1} still holds if we instead put a prior $\pi(\theta)$ on $\theta$, as long as $\pi(\theta)$ satisfies $P((1-\theta)/\theta \geq G^c) \geq 1 - e^{- D s_0 \log G}$ where $D > 0$ and $c > 2$. Then, since with probability tending to one, $\lambda_0 = (1-\theta)/\theta \geq G^{c}$, we can condition our analysis on the high probability event $\mathcal{A} = \{ (1-\theta) / \theta \geq G^{c} \}$ and our theory still holds. This will be satisfied, for example, when $\theta \sim \textrm{Beta}(1, G^c)$ \citep{BaiMoranAntonelliChenBoland2022}.
\end{remark}

\begin{remark} \label{Remark:new-1}
	{\revisetwo  By the continuous mapping theorem, $f(\widehat{\boldsymbol{\beta}})$ is also a consistent estimator of $f(\boldsymbol{\beta}_0)$ for any function $f$ that is continuous at $\boldsymbol{\beta}_0$. For example, let $S \subset \{1, \ldots, p\}$, and let $\boldsymbol{\beta}_S$ be the subvector of $\boldsymbol{\beta}$ with indices in $S$. 
		Taking $f(\boldsymbol{\beta}) = \mathbf{A}_S \boldsymbol{\beta}$, where $\mathbf{A}_S \in \mathbb{R}^{|S| \times p}$ is a matrix with unit vector rows such that $\mathbf{A}_S \boldsymbol{\beta} = \boldsymbol{\beta}_S$, it follows that $\widehat{\boldsymbol{\beta}}_S$ is a consistent estimator of $\boldsymbol{\beta}_S$.}
\end{remark}


\begin{remark}
	{\reviseone Theorem \ref{Th:1} implies that as $n \rightarrow \infty$, $\theta \asymp 1/(G^c+1) \rightarrow 0$. Since $\theta$ can be interpreted as the prior probability that the group $\bm{\beta}_g$ belongs to the slab density $\bm{\Psi}(\cdot \mid \lambda_1)$ rather than the spike density $\bm{\Psi}(\cdot \mid \lambda_0)$, this implies that the SSGL prior \eqref{SSGL} will classify a smaller proportion of the groups as nonzero as $n \rightarrow \infty$. This matches our assumption that the true proportion of nonzero groups, i.e. $s_0 / G$, also decays to zero as $n \rightarrow \infty$. However, the diffuse slab $\bm{\Psi}(\cdot \mid \lambda_1)$ still enables us to identify the true signals. To see this, notice that the prior probability that $\bm{\beta}_g$ belongs to the slab can be written as
		\begin{align*} \label{pstar}
			p_{\theta}^{\star} (\bm{\beta}_g) & = \frac{\theta \bm{\Psi} (\bm{\beta}_g \mid \lambda_1)}{\theta \bm{\Psi} (\bm{\beta}_g \mid \lambda_1) + (1-\theta) \bm{\Psi} (\bm{\beta}_g \mid \lambda_0)} \\
			& = \frac{1}{1 + \left( \frac{1-\theta}{\theta} \right) \left( \frac{\lambda_0}{\lambda_1} \right)^{m_g} \exp \left[ - (\lambda_0-\lambda_1) \lVert \bm{\beta}_g \rVert_2 \right] }. \numbereqn
		\end{align*}
		We can see from \eqref{pstar} that if $\bm{\beta}_g \neq \bm{0}_{m_g}$, then $p_{\theta}^{\star}(\bm{\beta}_g) \approx 1$ as $n \rightarrow \infty$. On the other hand, if $\bm{\beta}_g = \bm{0}_{m_g}$, then $p_{\theta}^{\star}(\bm{\beta}_g) \approx 0$ as $n \rightarrow \infty$.}
\end{remark}

We also have the following corollary which gives the convergence rate of {\reviseone a local MAP estimator} for $\bm{\beta}$ under the SSL prior \eqref{SSL} of \citet{RockovaGeorge2018} on the regression coefficients in unstructured GLMs \eqref{GLMnogroups}.  

\begin{corollary}[convergence rate of {\reviseone a local MAP} estimator under SSL] \label{Cor:1}
	Suppose that we have an unstructured GLM \eqref{GLMnogroups}, and we endow $\bm{\beta}_0$ with the SSL prior \eqref{SSL} where the hyperparameters $(\lambda_0, \lambda_1, \theta)$ satisfy $\lambda_0 = (1-\theta)/\theta \asymp p^{c}$, where $p > 2$, and $\lambda_1 \asymp 1/n$. Further, assume that $p \gg n$, $\log p = {\reviseone o(n^{1/2})}$, $s_0 = {\reviseone o(n^{1/2} / \log p)}$, and Assumptions (A3)-(A4) hold with $G$ replaced by $p$. Then there exists a {\reviseone strict local MAP estimator $\widehat{\bm{\beta}} = (\widehat{\bm{\beta}}_{S_0}^\top, \widehat{\bm{\beta}}_{S_0^c}^\top)^\top$ of the log-posterior \eqref{logposterior-main} such that, as $n \rightarrow \infty$, $\widehat{\bm{\beta}}_{S_0^c} = \bm{0}$ and}
	\begin{equation}
		\lVert \widehat{\bm{\beta}} - \bm{\beta}_0 \rVert_2 = O_p \left( \sqrt{\frac{s_0 \log p}{n}} \right).
	\end{equation}
\end{corollary}
\begin{proof}
	The model \eqref{GLMnogroups} is a special case of the grouped model \eqref{GLMgroups} where $m_1 = \ldots = m_G = 1$. Since we can also treat the SSL prior \eqref{SSL} as a special case of the SSGL prior \eqref{SSGL} with $m_1 = \cdots = m_G = 1$, the result follows from Theorem \ref{Th:1}. \hfill $\qed$
\end{proof}

Theorems \ref{Th:1} and Corollary \ref{Cor:1} justify {\reviseone using MAP estimation for point estimation} under the SSGL \eqref{SSGL} and SSL \eqref{SSL} priors in high-dimensional GLMs. In Section \ref{fullposteriorcharacterization}, we will turn our attention to the asymptotic behavior of the \emph{full} posterior distribution. 

\subsection{Comparison of our work to the group lasso} \label{SSGLvsGL}

The group lasso estimator \eqref{grouplassoobj} of \citet{YuanLin2006} has been studied theoretically in GLMs by \citet{BlazereLoubesGamboa2014}. Similar to the SSGL, \citet{BlazereLoubesGamboa2014} obtained the convergence rate of $O((s_0 \log G /n)^{1/2})$ for the group lasso. However, \citet{BlazereLoubesGamboa2014} require the assumption that $\sum_{g=1}^{G} \sqrt{m_g} \lVert \bm{\beta}_{0g} \rVert_2 < \infty$ (condition (H.3) in \citet{BlazereLoubesGamboa2014}) in order to achieve this rate. The condition that $\sum_{g=1}^{G} \sqrt{m_g} \lVert \bm{\beta}_{0g} \rVert_2 < \infty$ seems highly restrictive, especially if the number of groups $G$ diverges to infinity as in Assumption (A1). The only way for this assumption to be satisfied in practice is if \emph{all} of the following conditions hold: (i) the group sizes $m_g$ do \emph{not} diverge with $n$, (ii) the number of nonzero groups $s_0$ is fixed and does \emph{not} diverge with $n$, and (iii) $\lVert \bm{\beta}_0 \rVert_{\infty} < \infty$. 

In contrast, we do not make such a strong assumption. Theorem \ref{Th:1} allows both the group sizes $m_g$ \emph{and} the number of nonzero groups $s_0$ to diverge (Assumption (A2)), while the maximum signal strength $\lVert \bm{\beta}_0 \rVert_{\infty}$ can also grow to infinity. All of these assumptions clearly {\reviseone violate} Condition (H.3) of \citet{BlazereLoubesGamboa2014}. In short, we have derived the convergence rate for the SSGL MAP estimator in GLMs under much weaker conditions than those previously used for the group lasso estimator \eqref{grouplassoobj}. In Section \ref{GLsuboptimality}, we further demonstrate the advantage of SSGL \eqref{SSGL} over the group lasso from a fully Bayesian perspective.

\section{Characterization of the full posterior} \label{fullposteriorcharacterization}

\subsection{Posterior contraction rate} \label{posteriorcontraction}

{\reviseone For fully Bayesian inference, the minimax rate is a useful benchmark for studying contraction rates, because the posterior cannot contract faster than the minimax rate \citep{GhosalAoS2000}.} In this section, we analyze the \emph{full} SSGL posterior \eqref{posteriorSSGL} and show that the full posterior $\pi(\bm{\beta} \mid \mathbf{Y})$ inherits the nice theoretical properties of the SSGL MAP estimator \eqref{MAPestimatorSSGL}. {\reviseone As discussed in Section \ref{litreview}, it is \emph{not} automatically the case that the posterior contracts around the true parameter $\bm{\beta}_0$ in \eqref{truemodel} at the same rate as posterior point estimates \citep{CastilloSchmidtHieberVanDerVaart2015, CastilloMismer2018}. Thus, we require a separate analysis of the full posterior.}

In order to derive theory for the SSGL posterior in high-dimensional GLMs, we require a different set of conditions on the design matrix and on the maximum signal strength of $\bm{\beta}_0$. In particular, we replace conditions (A3)-(A4) with the following conditions. Recall that $S_0 \subset \{ 1, \ldots, G \}$ is the set of true nonzero groups which has cardinality $|S_0| = s_0$, and {\reviseone $\bm{\Omega}(\bm{\beta})$ is defined as in \eqref{Omegamatrix}.}

\begin{enumerate}[label=(B\arabic*)]
	\setcounter{enumi}{+2}
	\item The design matrix $\mathbf{X}$ satisfies the following conditions:
	\begin{enumerate}[label=(\roman*)]
		\item All the entries $x_{ij}$ of $\mathbf{X}$ satisfy {\reviseone $|x_{ij}| = O(\log G)$}.
		\item For a set of indices $S \subset \{1, \ldots, G\}$, let $\mathbf{X}_{S}$ denote the submatrix of $\mathbf{X}$  whose columns contain the groups $g \in S$. For any $S$ where {\reviseone $|S| \leq s_0$}, we have $\lambda_{\min} \left( n^{-1} \mathbf{X}_S^\top {\reviseone \bm{\Omega}(\bm{\beta}_0)} \mathbf{X}_S \right) {\reviseone \gtrsim 1}$. Meanwhile, for any $g \in \{1, \ldots, G \}$ {\reviseone and any $\bm{\beta}$ such that $\lVert \bm{\beta} - \bm{\beta}_0 \rVert_2 \leq G^{M}$ for some $M \geq 1$}, $\lambda_{\max} (n^{-1} \mathbf{X}_g^\top {\reviseone \bm{\Omega}(\bm{\beta})} \mathbf{X}_g ) \lesssim \log G$.
	\end{enumerate}
	\item The maximum signal strength satisfies $\lVert \bm{\beta}_0 \rVert_{\infty} = O(\log G)$.
\end{enumerate}
Assumption (B3)(ii) imposes restricted eigenvalue conditions {\reviseone which are stronger than those in Assumption (A3)(ii).} The assumption $\lambda_{\min}(n^{-1} \mathbf{X}_S^\top {\reviseone \bm{\Omega}(\bm{\beta}_0)} \mathbf{X}_S)~ {\reviseone \gtrsim 1}$ is required to hold for \emph{all} submatrices $\mathbf{X}_S$ where {\reviseone $|S| \leq s_0$}. The condition $\lambda_{\max}(n^{-1} \mathbf{X}_g^\top {\reviseone \bm{\Omega}(\bm{\beta})} \mathbf{X}_g) \lesssim \log G$ also needs to hold for {\reviseone \emph{individual}} submatrices $\mathbf{X}_g$ (not necessarily the entire $n \times p$ design matrix $\mathbf{X}$). In contrast, (A3)(ii) only imposes eigenvalue conditions for a \emph{single} submatrix $\mathbf{X}_{S_0}$. Intuitively, this is because in Theorem \ref{Th:1}, we only need to be able to find one local mode in a small neighborhood around $\bm{\beta}_0$. Contrastingly, Theorems \ref{Th:2} and \ref{Th:3} require the \emph{entire} posterior $\pi(\bm{\beta} \mid \mathbf{Y})$ to concentrate all its mass on configurations where no more than a constant multiple of $s_0$ {\reviseone groups} in $\bm{\beta}$ have magnitude much larger than zero. As a trade-off, however, the irrepresentability condition in Assumption {\reviseone (A3)(iii)} is \emph{not} needed for the full posterior to contract at the minimax-optimal rate. 

Condition (B4) also replaces the moment conditions on the responses in (A4) with a more direct condition on the maximum signal strength for $\bm{\beta}_0$. We need this condition because the SSGL prior \eqref{SSGL} must be able to put sufficient prior mass in a neighborhood of the true $\bm{\beta}_0$ for the posterior to contract around $\bm{\beta}_0$. {\reviseone Taken together, Assumptions (A2) and (B4) imply restrictions on the true parameter space for $\bm{\beta}_0$, both in terms of the sparsity of $\bm{\beta}_0$ and the magnitude of its entries.} Overall, our theoretical results underscore the importance of studying posterior point estimates separately from the full posterior, because different conditions may be required for convergence of these two objects. 

A crucial difference between our theory and that of \citet{Jiang2007}, \citet{JeongGhosal2021}, and \citet{TangMartin2023} is that the SSGL prior \eqref{SSGL} is an absolutely \emph{continuous} spike-and-slab prior. Although the posterior \emph{mode} under \eqref{SSGL} is exactly sparse, the SSGL {\reviseone posterior is continuous and thus puts zero probability on exactly sparse vectors}. Therefore, in order to analyze the \emph{full} posterior \eqref{posteriorSSGL}, we must resort to a notion of ``approximate'' sparsity known as the \emph{generalized dimensionality} \citep{BhattacharyaPatiPillaiDunson2015, RockovaGeorge2018, BaiMoranAntonelliChenBoland2022}. Following \citet{BaiMoranAntonelliChenBoland2022}, we use a small quantity $\omega_g > 0$ to define the generalized inclusion indicator $\nu_{\omega_g}(\bm{\beta}_g)$ and generalized dimensionality $\lvert \bm{\nu} (\bm{\beta}) \rvert$ respectively as
\begin{equation} \label{generalizeddimensionality}
	\nu_{\omega_g}(\bm{\beta}_g) = I(\lVert \bm{\beta}_g \rVert_2 > \omega_g)~~\textrm{and}~~\lvert \bm{\nu}(\bm{\beta}) \rvert = \sum_{g=1}^{G} \nu_{\omega_g}(\bm{\beta}_g).
\end{equation}
For the threshold $\omega_g$ in \eqref{generalizeddimensionality}, we use
\begin{equation} \label{omegag}
	\omega_g = \frac{1}{\lambda_0-\lambda_1} \log \left[ \frac{1-\theta}{\theta}  \frac{\lambda_0^{m_g}}{\lambda_1^{m_g}} \right],
\end{equation}
where $(\lambda_0, \lambda_1, \theta)$ are the hyperparameters in the SSGL prior \eqref{SSGL}. As described in \citet{BaiMoranAntonelliChenBoland2022}, any vectors $\bm{\beta}_g$ that satisfy $\lVert \bm{\beta}_g \rVert_2 = \omega_g$ are the intersection points between the spike density $\bm{\Psi}(\cdot \mid \lambda_0)$ and the slab density $\bm{\Psi}(\cdot \mid \lambda_1)$ in {\reviseone the SSGL prior \eqref{SSGL}}. Therefore, if $\lVert \bm{\beta}_g \rVert_2 > \omega_g$, then $\bm{\beta}_g$ is much more likely to belong to the slab rather than the spike. {\reviseone This also implies that if the threshold $\omega_g$ in \eqref{omegag} tends to zero as $n \rightarrow \infty$ for all $g \in \{ 1, \ldots, G \}$, then $| \bm{\nu}(\bm{\beta}) |$ can estimate the number of nonzero groups in $\bm{\beta}$ as $n \rightarrow \infty$.}

We first {\reviseone establish} in Theorem \ref{Th:2} that the posterior $\pi(\bm{\beta} \mid \mathbf{Y})$ under the SSGL prior \eqref{SSGL} asymptotically puts all of its mass on vectors where the generalized dimensionality \eqref{generalizeddimensionality} is no larger than a constant multiple of the true model size $s_0$. That is, {\reviseone the SSGL posterior concentrates on approximately sparse sets in high-dimensional GLMs}. {\reviseone Theorem \ref{Th:3} then verifies that the SSGL posterior contracts at the minimax-optimal rate around the true parameter $\bm{\beta}_0$.} {\revisetwo In the subsequent theorems, the notation $\mathbb{E}_0$ denotes the expectation operator with the true parameter $\boldsymbol{\beta}_0$ in \eqref{truemodel}.} 

\begin{theorem}[posterior concentration on {\reviseone approximately} sparse sets] \label{Th:2} 
	Assume the same setup as Theorem \ref{Th:1}, and suppose that Assumptions (A1)-(A2) and (B3)-(B4) hold. Then for some {\reviseone $K_1 \geq 1$},
	\begin{equation}
		\mathbb{E}_0 \Pi \left( \bm{\beta} : \lvert \bm{\nu}(\bm{\beta}) \rvert > K_1 s_0 \mid \mathbf{Y} \right) \rightarrow 0
	\end{equation}
\end{theorem}
\begin{proof}
	Appendix \ref{App:B}.
\end{proof}

\begin{theorem}[posterior contraction rate under SSGL]  \label{Th:3}
	Assume the same setup as Theorem \ref{Th:1}, and suppose that Assumptions (A1)-(A2) and (B3)-(B4) hold. Then for some $K_2 > 0$, as $n \rightarrow \infty$,
	\begin{equation}
		\mathbb{E}_0 \Pi \left( \bm{\beta} : \lVert \bm{\beta} - \bm{\beta}_0 \rVert_2 > K_2 \sqrt{\frac{s_0 \log G}{n}}~~\bigg\lvert~~\mathbf{Y} \right) \rightarrow 0.
	\end{equation}
\end{theorem}
\begin{proof}
	Appendix \ref{App:B}.
\end{proof}

\begin{remark} \label{Remark:2}
	Similarly as with {\reviseone the local MAP estimator} in Theorem \ref{Th:1}, we can also obtain the same posterior contraction rate for the full posterior when we endow $\theta$  in \eqref{SSGL} with a prior $\pi(\theta)$. As long as $\pi(\theta)$ satisfies $P((1-\theta)/\theta > G^{c}) > 1 - e^{-D s_0 \log G / n}$, where $D>0$ and $c>2$, then $\lambda_0 = (1-\theta)/\theta \geq G^{c}$ with probability tending to one, and the theory still holds. This will be the case if $\theta \sim \textrm{Beta}(1, G^{c})$. See \citet{BaiMoranAntonelliChenBoland2022} for the proof in the case of Gaussian grouped linear regression, which also holds in the GLM setting.
\end{remark}

\begin{remark} \label{Remark:new-2}
	{\revisetwo If $f(\boldsymbol{\beta})$ is a continuous function, then the posterior distribution for $f(\boldsymbol{\beta})$ is also consistent at $f(\boldsymbol{\beta}_0)$. In particular, this implies that the marginal posteriors for the individual coefficients or subsets of the coefficients in $\boldsymbol{\beta}$ also concentrate around their respective true values (see Remark \ref{Remark:new-1}).}
\end{remark}

\begin{remark} \label{Remark:3}
	In the cases of Gaussian regression and logistic regression, one may be able to remove the restriction (B4) on the maximum signal strength (see, e.g., \citet{CastilloSchmidtHieberVanDerVaart2015}, \citet{RockovaGeorge2018}, and \citet{Atchade2017}). However, since we consider GLMs under the general exponential family \eqref{exponentialfamily}, it seems unlikely that a condition such as (B4) can be totally removed for GLMs in general. To see why, consider Poisson regression, where the cumulant function is $b(\theta_{0i}) = e^{\mathbf{x}_i^\top \bm{\beta}_0}$ and the diagonal entries of $\bm{\Sigma}(\bm{\beta}_0)$ are $\{ e^{\mathbf{x}_i^\top \bm{\beta}_0} \}_{i=1}^{n}$. In this scenario, it seems difficult to control the approximation error without any restrictions on $\lVert \bm{\beta}_0 \rVert_{\infty}$.
\end{remark}

The following corollary is immediate from Theorem \ref{Th:2}. If we have an unstructured GLM \eqref{GLMnogroups} and we endow the regression coefficients in $\bm{\beta}_0$ with the SSL prior \eqref{SSL} of \citet{RockovaGeorge2018}, we obtain the following posterior contraction rate.

\begin{corollary}[posterior contraction rate under SSL]  \label{Cor:2}
	Assume the same setup as Corollary \ref{Cor:1}. Suppose that $p \gg n$, $\log p = {\reviseone o(n^{1/2})}$, $s_0 = {\reviseone o(n^{1/2} / \log p)}$, and Assumptions (B3)-(B4) hold with $G$ replaced by $p$. Then for some $K_3 > 0$, as $n \rightarrow \infty$,
	\begin{equation}
		\mathbb{E}_0 \Pi \left( \bm{\beta} : \lVert {\reviseone \bm{\beta}} - \bm{\beta}_0 \rVert_2 > K_3 \sqrt{\frac{s_0 \log p}{n}}~~\bigg\lvert~~\mathbf{Y} \right) \rightarrow 0.
	\end{equation}
\end{corollary}

Theorem \ref{Th:3} and Corollary \ref{Cor:2} show that for high-dimensional GLMs, the posterior distributions under the SSGL \eqref{SSGL} and SSL \eqref{SSL} priors \emph{also} converge at the {\reviseone fastest possible (i.e. the minimax) rates.} {\reviseone It is not necessarily the case that the full posterior has the same asymptotic behavior as posterior point estimates. For instance, \citet{CastilloMismer2018} and \citet{CastilloSchmidtHieberVanDerVaart2015} provide examples where the full posterior actually contracts much slower than the posterior mean, median, or mode.} 

Our results also suggest that the SSGL and the SSL posteriors $\pi(\bm{\beta} \mid \mathbf{Y})$ provide valid inference in high-dimensional GLMs. Determining the posterior contraction rate of the full posterior is often the first step towards obtaining frequentist guarantees about uncertainty quantification for Bayesian procedures \citep{Rousseau2016}. In particular, the posterior contraction rate gives an indication as to how large we can expect the posterior credible sets to be \citep{Rousseau2016}. However, beyond just their size, more detailed study is typically required to guarantee that these credible sets are also honest, or have asymptotic coverage probability greater than or equal to the prescribed confidence level $1-\alpha, \alpha \in (0,1)$ \citep{Rousseau2016}. The issue of honest coverage of posterior credible sets is beyond the scope of this paper.

\subsection{Suboptimality of the group lasso for fully Bayesian inference} \label{GLsuboptimality}

In Section \ref{SSGLvsGL}, we demonstrated that {\reviseone there exists a local MAP estimator \eqref{MAPestimatorSSGL} for SSGL which converges at the minimax-optimal rate under weaker assumptions than those previously assumed for the group lasso \eqref{grouplassoobj}.} It turns out that from the fully Bayesian perspective, the SSGL also has an advantage over the group lasso. As discussed in Section \ref{Bayesianestimation}, the group lasso estimator \eqref{grouplassoobj} is the MAP estimator under the prior $\pi(\bm{\beta}) = \prod_{g=1}^{G} \bm{\Psi}(\bm{\beta}_g \mid \lambda)$, where $\bm{\Psi}(\cdot \mid \lambda)$ is a \emph{single} multivariate Laplace density \eqref{multivariateLaplace}. In contrast to the two-group SSGL \eqref{SSGL}, {\reviseone however,} the group lasso posterior might contract much \emph{slower} than its posterior mode. This renders the group lasso less useful for uncertainty quantification. {\revisetwo An example of this phenomenon is provided in the following proposition. Recall that $\boldsymbol{\beta}_0 = (\boldsymbol{\beta}_{01}^\top, \ldots, \boldsymbol{\beta}_{0G}^\top)^\top$, where $\boldsymbol{\beta}_{0g}$ is the $g$th group of size $m_g$.}

{\revisetwo 
	\begin{proposition} \label{Prop:1}
		Suppose that $\mathbf{Y} \sim \mathcal{N}_n (\boldsymbol{\beta}_0, \mathbf{I}_n)$ and all the group sizes are constant and equal, i.e. $m_1 = \cdots = m_G = m$ where $1 \leq m < \infty$. Further, assume that $\boldsymbol{\beta}_{01} \neq \bm{0}_{m}$ with $\lVert \boldsymbol{\beta}_{01} \rVert_2 = L < \infty$, while $\boldsymbol{\beta}_{0g} = \bm{0}_{m}$ for all $g \in \{ 2, \ldots, G \}$. Suppose that we endow $\bm{\beta}_{0}$ with the prior $\pi(\bm{\beta}) = \prod_{g=1}^{G} \bm{\Psi}(\bm{\beta}_g \mid \lambda)$, where $\bm{\Psi}(\cdot \mid \lambda)$ is the group lasso prior in \eqref{multivariateLaplace}. Then if $\lambda \asymp \sqrt{2 \log n}$, the MAP estimator $\widehat{\boldsymbol{\beta}}$ satisfies
		\begin{align} \label{group-lasso-MAP-risk}
			\mathbb{E}_0 \lVert \widehat{\boldsymbol{\beta}} - \boldsymbol{\beta}_0 \rVert_2 \lesssim \sqrt {2 \log n} \textrm{ as } n \rightarrow \infty,
		\end{align}
		but for some $K_4 > 0$,
		\begin{align} \label{group-lasso-posterior-contraction}
			\mathbb{E}_{0} \Pi \left( \boldsymbol{\beta} : \lVert \boldsymbol{\beta} - \boldsymbol{\beta}_0 \rVert_2 \leq K_4 \sqrt{\frac{n}{\log n}}~~\bigg|~~ \mathbf{Y} \right) \rightarrow 0 \textrm{ as } n \rightarrow \infty. 
		\end{align}
\end{proposition}}
\begin{proof}
	Appendix \ref{App:B}.
\end{proof}

Proposition \ref{Prop:1} implies that the full posterior distribution under the group lasso prior may asymptotically put all of its mass in an $\ell_2$ ball with radius that is \emph{substantially larger} than the convergence rate of the posterior mode. {\revisetwo Namely, under the grouped GLM \eqref{GLMgroups} where $\mathbf{X}= \mathbf{I}_n$ and the exponential family \eqref{exponentialfamily} distribution is the normal distribution, the group lasso MAP estimator $\widehat{\boldsymbol{\beta}}$ has an expected $\ell_2$ risk of the order $\sqrt{2 \log n}$ when the shrinkage parameter $\lambda$ is chosen of the order $\sqrt{2 \log n}$. This is the minimax rate for estimating $\boldsymbol{\beta}_0$ \citep{DonohoNearlyBlack1992}. However, with this same choice of $\lambda$, the posterior $\pi(\boldsymbol{\beta} \mid \mathbf{Y})$ places zero probability on the ball $\{ \boldsymbol{\beta} : \lVert \boldsymbol{\beta} - \boldsymbol{\beta}_0 \rVert_2 \leq \sqrt{n / \log n} \}$ as $n \rightarrow \infty$, and $\sqrt{2 \log n } \ll \sqrt{n / \log n}$ when $n$ is large.} Intuitively, {\revisetwo the discrepancy in the convergence rate between the MAP estimator and the full posterior} is because there is a conflict between shrinking the coefficients to zero and optimally estimating the nonzero signals. In order to estimate very sparse parameters well, the group lasso needs to set $\lambda$ to be large; but if $\lambda$ is \emph{too} large, then there is too much bias in the resulting estimator. The addition of the slab density $\bm{\Psi} (\cdot \mid \lambda_1)$ in the SSGL prior \eqref{SSGL} alleviates this tension by \emph{preventing} overshrinkage of true signals. Thus, {\revisetwo unlike the group lasso}, the full SSGL posterior \eqref{posteriorSSGL} {\revisetwo also} contracts at the minimax {\revisetwo $\ell_2$ convergence rate}.

\section{Implementation of SSGL for GLMs} \label{computation}

\subsection{EM algorithm for MAP estimation} \label{EMalgorithm}

{\reviseone We first} adopt the penalized likelihood perspective and perform MAP estimation {\reviseone for GLMs under the SSGL model}. The MAP estimator \eqref{MAPestimatorSSGL} is appealing, not just because of its nice theoretical properties, but also because it is \emph{exactly} sparse. Thus, the MAP estimator can be used for both estimation and variable selection in GLMs. To obtain the MAP estimator, we extend the EM variable selection (EMVS) approach of \citet{RockovaGeorge2014} to the GLM setting with grouped variables. 

For the theoretical results in Sections \ref{MAPcharacterization} and \ref{fullposteriorcharacterization}, we treated the mixing proportion $\theta$ in \eqref{SSGL} as a deterministic quantity (that depends on $n$ and $G$). Our theoretical results still hold with a prior on $\theta$, as long as the prior ensures that $\mathcal{A}= \{ (1-\theta)/\theta \geq G^{c}, c > 2 \}$ is a very high probability event (see Remarks \ref{Remark:1} and \ref{Remark:2}). 

For practical implementation, we also recommend endowing $\theta$ with a prior $\pi(\theta)$ in order to model the inherent uncertainty in $\theta$ and \emph{adaptively} learn the true sparsity level from the data. To this end, we endow $\theta$ in \eqref{SSGL} with a beta prior with shape parameters $a>0, b>0$,
\begin{align} \label{thetaprior}
	\theta \sim \mathcal{B}(a,b).
\end{align} 
Our prior specification for $(\bm{\beta}, \theta)$ is then given by $\pi(\bm{\beta}, \theta) = \pi(\bm{\beta} \mid \theta) \pi(\theta)$, where $\pi(\bm{\beta} \mid \theta)$ is as in \eqref{SSGL} and $\pi(\theta)$ is as in \eqref{thetaprior}. The complete log-posterior is then
\begin{equation} \label{logposteriorII}
	\log \pi(\bm{\beta}, \theta \mid \mathbf{Y} ) = \ell_n(\bm{\beta}) + \log \pi(\bm{\beta} \mid \theta) + \log \pi(\theta).
\end{equation}
We use a variant of the EMVS algorithm \citep{RockovaGeorge2014} to iteratively solve for the MAP estimator $(\widehat{\bm{\beta}}, \widehat{\theta})$ in the optimization problem,
\begin{equation} \label{argmaxposterior}
	(\widehat{\bm{\beta}}, \widehat{\theta}) = \argmax_{\bm{\beta} \in \mathbb{R}^{p}, \theta \in(0,1)} \log \pi (\bm{\beta}, \theta \mid \mathbf{Y}).
\end{equation}
The EMVS approach of \citet{RockovaGeorge2014} introduces latent variables $\bm{\gamma} = (\gamma_1, \ldots, \gamma_G)$, $\gamma_G \in \{ 0, 1 \}$, where $\gamma_g = 1$ indicates that the $g$th group of coefficients $\bm{\beta}_g$ should be included in the model. These indicator variables are treated as missing data in the E-step of our algorithm. To be precise, we reparameterize the SSGL prior \eqref{SSGL} as a beta-Bernoulli prior,
\begin{equation} \label{SSGLrepar}
	\begin{array}{rl} 
		\pi ( \bm{\beta} \mid \bm{\gamma} ) = & \displaystyle \prod_{g=1}^{G} \left[ (1-\gamma_g) \bm{\Psi}( \bm{\beta}_g \mid \lambda_0) + {\reviseone \gamma_g} \bm{\Psi} ( \bm{\beta}_g \mid \lambda_1) \right], \\
		\pi ( \bm{\gamma} \mid \theta ) = & \displaystyle \prod_{g=1}^{G} \theta^{\gamma_g} (1-\theta)^{1-\gamma_g},
	\end{array}
\end{equation}
where $\bm{\gamma}$ is a binary vector. As shown in Appendix \ref{App:C}, $\mathbb{E} [ \gamma_g \mid \mathbf{Y}, \bm{\beta}, \theta ] = p_g^{\star} (\bm{\beta}_g, \theta)$, where
\begin{equation} \label{Estep}
	p_g^{\star} ( \bm{\beta}_g, \theta) = \frac{ \theta \bm{\Psi} ( \bm{\beta}_g \mid \lambda_1)}{\theta \bm{\Psi} (\bm{\beta}_g \mid \lambda_1) + (1-\theta) \bm{\Psi} (\bm{\beta}_g \mid \lambda_0)}
\end{equation}
is the conditional posterior probability that $\bm{\beta}_g$ is drawn from the slab distribution rather than from the spike. In the E-step, we compute $p_g^{\star (t-1)} := p_g^{\star} ( \bm{\beta}_g^{(t-1)}, \theta^{(t-1)} ) = \mathbb{E} [ \gamma_g \mid \mathbf{Y}, \bm{\beta}^{(t-1)}, \theta^{(t-1)} ], g = 1, \ldots, G$. {\reviseone With the hyperprior \eqref{thetaprior} on $\theta$, we then update $\theta$  in the M-step as}
\begin{equation} \label{thetaupdate}
	\theta^{(t)} = \frac{a-1 + \sum_{g=1}^{G} p_g^{\star (t-1)}}{a+b+G-2}.
\end{equation}
{\reviseone In the M-step, we update} $\bm{\beta}$ as
\begin{equation} \label{betaupdate}
	\bm{\beta}^{(t)} = \displaystyle \argmax_{\bm{\beta}} \left\{ \ell ( \bm{\beta} ) - \sum_{g=1}^{G} \lambda_g^{\star (t-1)} \lVert \bm{\beta}_g \rVert_2  \right\},
\end{equation}
where each $\lambda_g^{\star (t-1)} = \lambda_1 p_g^{\star (t-1)} + \lambda_0 (1- p_g^{\star (t-1)})$ is an \textit{adaptive} weight ensuring that insignificant groups are shrunk aggressively to zero, while significant groups incur minimal shrinkage. The objective \eqref{betaupdate} is simply a group lasso optimization with known group-specific weights $(\lambda_1^{\star (t-1)}, \ldots, \lambda_G^{\star (t-1)})$. In Appendix \ref{App:C}, we describe how to efficiently solve \eqref{betaupdate}. In summary, our EMVS algorithm proceeds as follows.
\begin{enumerate}
	\item Initialize $(\bm{\beta}^{(0)}, \theta^{(0)})$. For example, we can initialize $\bm{\beta}^{(0)} = \bm{0}_p$ and $\theta^{(0)} = 0.5$.
	\item For $t = 1, 2, \ldots$, repeat until convergence:
	\begin{enumerate}[label=\roman*.]
		\item \textbf{E-step}: For $g = 1, \ldots, G$, compute $p_g^{\star (t-1)} = p_g^{\star}(\bm{\beta}_g^{(t-1)}, \theta^{(t-1)})$ as in \eqref{Estep}.
		\item \textbf{M-step}: Update $\theta^{(t)}$ as in \eqref{thetaupdate} and $\bm{\beta}^{(t)}$ as in \eqref{betaupdate}.
	\end{enumerate}
\end{enumerate}
To determine convergence of the algorithm, we recommend using the criterion $\lVert \bm{\beta}^{(t)} - \bm{\beta}^{(t-1)} \rVert_2^2 / \lVert \bm{\beta}^{(t-1)} \rVert_2^2 < \varepsilon$, where $\varepsilon$ is a small value (e.g. $\varepsilon = 10^{-6}$). {\reviseone Since the EM algorithm has the ascent property, our algorithm is guaranteed to converge to a local mode.}

\subsection{Choice of hyperparameters} \label{hyperparameters}

The performance of the SSGL is mainly governed by the three parameters $(\lambda_0, \lambda_1, \theta)$ in the prior \eqref{SSGLrepar} on $\bm{\beta}$. We recommend fixing the slab hyperparameter $\lambda_1 = 1$, so that the $\bm{\beta}_g$'s with large entries incur very minimal shrinkage. To induce sparsity, the mixing proportion $\theta$ should also be small with high probability, so that most of the $\bm{\beta}_g$'s belong to the spike density. To this end, we recommend setting $a=1, b=G$ for the $\mathcal{B}(a,b)$ prior \eqref{thetaprior} on $\theta$. This ensures that most of the $\bm{\beta}_g$'s will be shrunk to zero. 

The spike hyperparameter $\lambda_0$ in \eqref{SSGLrepar} controls how sparse our final model is, with larger values of $\lambda_0$ leading to more sparsity. For unstructured GLMs \eqref{GLMnogroups} with the SSL prior \eqref{SSL}, \citet{TangShenZhangYi2017} recommended tuning $\lambda_0$ from cross-validation (CV). {\reviseone For SSGL, we follow \citet{TangShenZhangYi2017} and similarly tune $\lambda_0$ using $K$-fold CV, with a default of $K=10$. To accelerate the computational efficiency, we fit the model on each of the $K$ training sets in parallel, allowing for speed-ups on roughly the order of $K$. In practice, we tune $\lambda_0$ from an equispaced grid  $\{ \lambda_{0,1}, \lambda_{0,2}, \ldots, \lambda_{0,\max} \}$, where $0 < \lambda_{0,1} < \lambda_{0,2} < \cdots < \lambda_{0,\max}$ and ${\revisetwo \lambda_{0,\max}}$ is the smallest value of $\lambda_0$ {\revisetwo that results in a MAP estimator of $\widehat{\bm{\beta}} = \bm{0}_p$}. It is not hard to see that ${\revisetwo \lambda_{0, \max}} = \max_{1 \leq g \leq G} \lVert \nabla_g~\ell_n (\bm{0}_p) \rVert_2$, where $\nabla_g$ denotes the subvector of the gradient \eqref{llgradient} corresponding to the $g$th group. Typically, ${\revisetwo \lambda_{0, \max}}$ has an analytical form. For example, in logistic regression, ${\revisetwo \lambda_{0,\max}} = \max_{1 \leq g \leq G} \lVert 0.25 \mathbf{X}_g^\top ( \mathbf{Y} - 0.5 \bm{1}_n) \rVert_2$.}

To account for potentially different group sizes $m_g$, we further rescale $\lambda_0$ for each $g$th group, so that the spike parameter for each $\bm{\beta}_g$ is $\lambda_{0g} = \lambda_0 \sqrt{m_g}$. As discussed in \citet{HuangBrehenyMa2012}, scaling of the regularization penalty by group size is needed to ensure that groups are not unfairly penalized simply for being smaller or erroneously included simply for being larger.

\begin{remark}
	{\reviseone An inspection of the proofs of Theorems 1-3 reveals that the theoretical results in Theorem 1-3 still hold when we replace the single spike parameter $\lambda_0$ with $\lambda_{0g} = \lambda_0 \sqrt{m_g}$ for each $g$th group, as long as we still assume that $\lambda_0 \asymp G^{c}, c>2$, as in Theorems 1-3, and $m_{\max} = O(\log n \wedge (\log G/ \log n))$, as in Assumption (A2).}
\end{remark}

\subsection{Gibbs sampling for fully Bayesian inference} \label{GibbsSampler}

{\reviseone The EM algorithm in Section \ref{EMalgorithm} returns  only a single local MAP estimator \eqref{MAPestimatorSSGL}. However, \emph{fully} Bayesian inference of the grouped GLM model \eqref{GLMgroups} is often desirable, since this allows us to quantify uncertainty for the regression coefficients. For a number of Bayesian GLMs with discrete responses, we can use data augmentation with latent variables \citep{PolsonScottWindle2013, AlbertChib1993} to facilitate Gibbs sampling. In this section, we focus on grouped logistic regression and grouped Poisson regression using P\'{o}lya-gamma data augmentation \citep{PolsonScottWindle2013}.
	
	A random variable $\omega$ with density function $p(\omega)$ is said to follow a P\'{o}lya-gamma (PG) distribution $\mathcal{PG}(a,b)$ with parameters $a > 0$ and $b \in \mathbb{R}$ if 
	\begin{align*}
		\omega = \frac{1}{2\pi^2} \sum_{k=1}^{\infty} \frac{g_k}{ \left( k - \frac{1}{2} \right)^2 + ( \frac{b}{2\pi} )^2}, ~~ \text{and} ~~ g_k \sim \text{Gamma}(a,1).
	\end{align*}
	\citet{PolsonScottWindle2013} established the relation,
	\begin{align} \label{PG-relation}
		\frac{(e^{\psi})^{a}}{(1+e^{\psi})^b} = 2^{-b} e^{\kappa \psi} \int_{0}^{\infty} e^{-\omega \psi^2/2} p(\omega) d \omega,
	\end{align}
	where $\kappa = a - 0.5b$ and $\omega \sim \mathcal{PG}(b, 0)$. In logistic regression, the likelihood contribution from the $i$th observation is
	\begin{align} \label{Bernoulli-likelihood}
		\mathcal{L}_i (\bm{\beta}) = \frac{ (\exp(\mathbf{x}_i^\top \bm{\beta}))^{y_i}}{1+ \exp(\mathbf{x}_i^\top \bm{\beta})}, ~~~ y_i \in \{0, 1 \}.
	\end{align}
	Let $\bm{\omega} = (\omega_1, \ldots, \omega_n)^\top$. Based on \eqref{PG-relation}-\eqref{Bernoulli-likelihood}, the conditional distribution for $\bm{\beta}$ with prior $\pi(\bm{\beta})$ in logistic regression is
	\begin{align} \label{logistic-conditional}
		\pi(\bm{\beta} \mid \bm{\omega}, \mathbf{Y}) & = \pi (\bm{\beta}) \prod_{i=1}^{n} \mathcal{L}_i (\bm{\beta}) \propto \pi(\bm{\beta}) \exp \left\{ -\frac{1}{2} ({\revisetwo \mathbf{Z}} - \mathbf{X} \bm{\beta})^\top \bm{\Omega} ({\revisetwo \mathbf{Z}} - \mathbf{X}\bm{\beta})  \right\}, \numbereqn 
	\end{align}
	where ${\revisetwo \mathbf{Z}} = (\kappa_1 / \omega_1, \ldots, \kappa_n / \omega_n)^\top$, $\bm{\Omega} = \text{diag}( \omega_1, \ldots, \omega_n)$, and for $i = 1, \ldots, n$, $\kappa_i = y_i - 0.5$ and $\omega_i \sim \mathcal{PG}(1,0)$.
	
	Meanwhile, for Poisson regression, \citet{LiBiostatistics2018} showed that the likelihood contribution from the $i$th observation can be approximated as
	\begin{align} \label{poisson-likelihood}
		\widetilde{\mathcal{L}}_i (\bm{\beta}) \approx \frac{\left\{ \exp(\mathbf{x}_i^\top \bm{\beta} - \log M) \right\}^{0.5 y_i}}{ \left\{ 1 + \exp(\mathbf{x}_i^\top \bm{\beta} - \log M) \right\}^{M}}, \numbereqn
	\end{align}
	for a sufficiently large $M > 0$. In practice, we can choose $M = 1+\max_i y_i$.  Based on \eqref{PG-relation} and \eqref{poisson-likelihood}, the approximate conditional distribution for $\bm{\beta}$ with prior $\pi(\bm{\beta})$ in Poisson regression is
	\begin{align} \label{Poisson-conditional}
		\pi(\bm{\beta} \mid \bm{\omega}, \mathbf{Y}) \approx \pi (\bm{\beta}) \prod_{i=1}^{n} \widetilde{\mathcal{L}}_i (\bm{\beta}) \overset{\cdot}{\propto} \pi(\bm{\beta}) \exp \left\{ -\frac{1}{2} ( {\revisetwo \mathbf{Z}} - \mathbf{X} \bm{\beta})^\top \bm{\Omega} ( {\revisetwo \mathbf{Z}} - \mathbf{X}\bm{\beta})  \right\}, \numbereqn
	\end{align}
	where ${\revisetwo \mathbf{Z}} = ( {\revisetwo \kappa_1} / \omega_1 + \log M, \ldots, {\revisetwo \kappa_n} / \omega_n + \log M)^\top$, $\bm{\Omega} = \text{diag}( \omega_1, \ldots, \omega_n)$, and for $i = 1, \ldots, n$, ${\revisetwo \kappa_i} = 0.5(y_i - M)$ and $\omega_i \sim \mathcal{PG}(M,0)$.  
	
	The exponential terms in \eqref{logistic-conditional} and \eqref{Poisson-conditional} are Gaussian kernels. Therefore, if the prior $\pi(\bm{\beta})$ is a Gaussian or a Gaussian scale-mixture, then we can easily draw samples from $\pi(\bm{\beta} \mid \bm{\omega}, \mathbf{Y})$ in our Gibbs sampling algorithm. 
	
	\begin{remark}
		The conditional distribution \eqref{Poisson-conditional} is not exact. However, \eqref{Poisson-conditional} greatly simplifies posterior sampling for $\bm{\beta}$ in Poisson regression, since the updates for $\bm{\beta}$ can be obtained in closed form. Sampling from the \emph{exact} conditional distribution of $\bm{\beta}$ in Poisson regression typically requires Metropolis-Hastings, which is not practical when $\bm{\beta}$ is high-dimensional. {\revisetwo This is because we cannot decouple the individual components of $\boldsymbol{\beta}$ in the likelihood function, and therefore, we need to update the entire vector $\boldsymbol{\beta}$ in each MCMC iteration. When $\bm{\beta}$ is a large vector, there is a very low probability of accepting a proposal draw for $\bm{\beta}$, regardless of the proposal density used. Therefore, we sample $\boldsymbol{\beta}$ directly from the approximate conditional distribution \eqref{Poisson-conditional}}. In simulation studies, we found that this approximation was very adequate.
	\end{remark}
	
	Suppose that $\bm{\beta}_g \sim \bm{\Psi} ( \bm{\beta}_g \mid \lambda)$, where $\bm{\Psi}(\bm{\beta}_g \mid \lambda)$ is the multivariate Laplace density \eqref{multivariateLaplace}. Then $\boldsymbol{\beta}_g$ is the marginal prior of the scale mixture,
	\begin{equation} \label{SSGLscalemixture}
		\bm{\beta}_g \mid \tau_g \sim \mathcal{N}( \bm{0}, \tau_g \bm{I}_{m_g}), ~~ \tau_g \sim \text{Gamma} \left( \frac{m_g+1}{2}, \frac{\lambda^2}{2} \right).
	\end{equation}
	Let $\bm{\tau} = (\tau_1, \ldots, \tau_G)^\top$ and $\bm{\gamma} = (\gamma_1, \ldots, \gamma_G)^\top$, and let $\text{Bdiag}$ denote a block-diagonal matrix. Based on \eqref{SSGLscalemixture} and the reparameterization \eqref{SSGLrepar} of the SSGL prior, we can rewrite the SSGL prior \eqref{SSGL} for $\bm{\beta}$ as a Gaussian scale-mixture hierarchical model,
	\begin{equation} \label{SSGLrepar2}
		\begin{array}{ll}
			\bm{\beta} \mid \bm{\tau} \sim \mathcal{N}_p( \bm{0}_p, \bm{D}_{\bm{\tau}}),& \text{where } \bm{D}_{\bm{\tau}} = \text{Bdiag} (\tau_1 \bm{I}_{m_1}, \ldots, \tau_G \bm{I}_{m_G}), \\
			\tau_g \mid \gamma_g \sim \text{Gamma} \left( \frac{m_g+1}{2}, \frac{(\lambda_g^{\star})^2}{2} \right), & \text{where } {\revisetwo \lambda_g^{\star} = \gamma_g \lambda_1 + (1-\gamma_g) \lambda_0}, \\
			\gamma_g \mid \theta \sim \text{Bernoulli}(\theta), & g = 1, \ldots, G.
		\end{array}
	\end{equation}
	Due to the hierarchical representation \eqref{SSGLrepar2} of the SSGL prior, we can exploit conditional conjugacy to obtain the conditional distributions for $\bm{\beta}$ in \eqref{logistic-conditional} and \eqref{Poisson-conditional} in closed form. With a Beta hyperprior \eqref{thetaprior} for $\theta$ in \eqref{SSGLrepar2}, the conditional distributions for $\theta$, $\bm{\tau}$, $\bm{\gamma}$, and $\bm{\omega}$ are also all available in closed form. 
	
	Our Gibbs sampling algorithm proceeds by sequentially drawing samples from the conditional distributions of $\bm{\beta}$, $\bm{\tau}$, $\bm{\gamma}$, $\theta$, and $\bm{\omega}$, holding the other variables fixed at their current values. If $p = \sum_{g=1}^{G} m_g$ is greater than $n$, then we can also use the fast sampling algorithm of \citet{bhattacharya2016fast} to sample $\bm{\beta}$ in $\mathcal{O}(n^2p)$ time rather than $\mathcal{O}(p^3)$ time. The complete MCMC algorithms for Bayesian grouped logistic regression and grouped Poisson regression with the SSGL prior \eqref{SSGL} are provided in Appendix \ref{App:D}.
	
	\begin{remark}
		Although we focused on logistic and Poisson regression in this section, data augmentation with P\'{o}lya-gamma latent variables also works for other Bayesian GLMs such as negative binomial regression and multinomial regression \citep{PolsonScottWindle2013}. Data augmentation with other types of latent variables can also be used for other GLMs, e.g. Bayesian probit regression models can be implemented using the approach of \cite{AlbertChib1993}.
	\end{remark}
}

\section{Simulation studies} \label{illustrations}

\subsection{Setup and performance metrics} \label{setup}

We investigated the performance of the SSGL prior \eqref{SSGL} in numerical experiments with $G<n$ and $G>n$. We considered grouped logistic regression for binary data and grouped Poisson regression for count data. In particular, Experiments 3 and 4 in Sections \ref{logisticregression} and \ref{Poissonregression} are meant to mimic two real applications where our methodology is especially useful: a) semiparametric additive models with continuous covariates, and b) genetic association studies involving single nucleotide polymorphisms (SNPs) \citep{BrehenyHuang2015}. In Appendix \ref{App:E}, we also present some simulation results for grouped negative binomial regression with a log link, which is an example of our method with a \emph{non}-canonical link function.

In semiparametric additive models (Experiment 3 in Sections \ref{logisticregression} and \ref{Poissonregression}), we flexibly model the effects of continuous covariates $x_j$ on the mean response as univariate functions $f_j(x_j)$. The $f_j$'s are approximated using linear combinations of $K$ basis functions $g_{jk}(x_j)$, i.e. $f_j(x_j) \approx \sum_{k=1}^{K} \beta_{jk} g_{jk}(x_j)$. The $j$th main effect $f_j$ is then estimated as $\widehat{f}_j(x_j) = 0$ if $\bm{\beta}_j = (\beta_{j1}, \ldots, \beta_{jK})^\top = \bm{0}_K$ or as $\widehat{f}_j(x_j) \neq 0$ if $\bm{\beta}_j \neq \bm{0}_K$. 

In the simulated genetic association studies (Experiment 4 in Sections \ref{logisticregression} and \ref{Poissonregression}), the responses are either binary or count phenotypes, and the covariates are simulated SNPs. SNPs are categorical variables coded as one of three values $\{$``0'', ``1'', or ``2''$\}$, depending on the number of minor alleles present \citep{BrehenyHuang2015}.  We thus represent each SNP as a factor with two levels, i.e. a group of two indicator variables. Assuming that ``2'' is the baseline, we can represent each $j$th SNP $x_j$ with two dummy variables $\mathbb{I}(x_j=0)$ and $\mathbb{I}(x_j=1)$. If $x_j = 2$, then $\mathbb{I}(x_j=0) = \mathbb{I}(x_j=1) = 0$. 

We have implemented SSGL for GLMs in the \textsf{R} package \texttt{SSGL}. In all of our experiments, we chose the hyperparameters in the SSGL prior as described in Section \ref{hyperparameters}. We compared the performance of SSGL to {\reviseone other group-regularized estimators of the form,
	\begin{equation} \label{group-regularized}
		\argmax_{\bm{\beta} \in \mathbb{R}^{p}} \ell_n(\bm{\beta}) + \sum_{g=1}^{G} \text{pen}_{\lambda}( \lVert \bm{\beta}_g \rVert_2),
	\end{equation}
	where $\ell_n(\bm{\beta})$ is the grouped GLM log-likelihood function defined in \eqref{ll} and $\text{pen}_{\lambda}( \lVert \bm{\beta}_g \rVert_2)$ is a penalty function on $\lVert \bm{\beta}_g \rVert_2$ which depends on tuning parameter $\lambda > 0$. Our competitors were} the group lasso {\reviseone (GL)} \citep{YuanLin2006}, the {\reviseone adaptive group lasso (AdGL) \citep{WangLeng2008},} the group minimax concave penalty {\reviseone (GMCP)} \citep{BrehenyHuang2015}, and the group smoothly clipped absolute deviation {\reviseone (GSCAD)} \citep{BrehenyHuang2015}. We implemented GL, {\reviseone AdGL}, GMCP, and GSCAD using the \textsf{R} package \texttt{grpreg}, {\reviseone where the regularization parameter $\lambda > 0$ was tuned using $K$-fold CV with $K=10$. For each of these methods, the penalty was further scaled by $\sqrt{m_g}$ for each $g$th group.}  

{\reviseone AdGL adds group-specific weights $w_g$ to the groups $\bm{\beta}_g, g= 1, \ldots, G$, to counteract the well-known bias of GL. For AdGL, we chose the weights as
	\begin{align*}
		w_g = \left\{ \begin{array}{lccl} \lVert \widetilde{\bm{\beta}}_g \rVert_2^{-1}, &&& \text{if } \lVert \widetilde{\bm{\beta}}_g \rVert_2 > 0, \\ \infty, &&& \text{if } \lVert \widetilde{\bm{\beta}}_g \rVert_2 = 0,  \end{array} \right. 
	\end{align*}
	where $\widetilde{\bm{\beta}}_g$ was the GL estimator of $\bm{\beta}_g$. For GMCP and GSCAD, there is an additional parameter $\gamma$ which controls the concavity of the penalty \citep{BrehenyHuang2015}. We used the default settings of $\gamma=3$ for GMCP and $\gamma=4$ for GSCAD in \texttt{grpreg}. While it may be desirable to further tune $\gamma$, performing a two-dimensional grid search for $(\lambda, \gamma)$ with CV is very computationally expensive. In the simulation settings that we considered, we also did not find that these methods were very sensitive to the choice of $\gamma$, as long as $\gamma>1$ for GMCP, $\gamma >2$ for GSCAD, and $\gamma$ was not overwhelmingly large for either method. Thus, for GMCP and GSCAD, we opted to fix $\gamma$ and only tuned the penalty parameter $\lambda$.}

We computed the following performance metrics: mean squared error (MSE), mean squared prediction error (MSPE), {\reviseone true positive rate (TPR)}, {\reviseone true negative rate (TNR)}, {\reviseone and} precision (Prec), defined as
\begin{align*}
	& \textrm{MSE} = \frac{1}{p} \lVert \widehat{\bm{\beta}} - \bm{\beta} \rVert_2^2,~~ \textrm{MSPE} = \frac{1}{n_{\textrm{test}}} \sum_{i=1}^{n} (y_{i, \textrm{test}} - \widehat{y}_{i,\textrm{test}})^2, \\
	& \textrm{TPR} = \frac{\textrm{TP}}{\textrm{TP}+\textrm{FN}},~~ \textrm{TNR} = \frac{\textrm{TN}}{\textrm{TN}+\textrm{FP}},~~\textrm{Prec} = \frac{\textrm{TP}}{\textrm{TP}+\textrm{FP}},
\end{align*}
where TP, TN, FP, and FN are the number of true positives, true negatives, false positives, and false negatives respectively. The MSPE was computed using $n_{\textrm{test}} = 100$ out-of-sample test points $\{ (\mathbf{x}_{i,\textrm{test}}, y_{i,\textrm{test}}) \}_{i=1}^{n_{\textrm{test}}}$, and $\widehat{y}_{i,\textrm{test}} = b'(\widehat{\theta}_{i,\textrm{test}})$, where $\widehat{\theta}_i$ is the predicted natural parameter with $\mathbf{x}_{i, \textrm{test}}$ in \eqref{GLMgroups}. For the logistic regression experiments in Section \ref{logisticregression}, we also recorded the area under the receiver operating characteristic curve (AUC) on the test data.

\subsection{Grouped logistic regression} \label{logisticregression}
For logistic regression, we have $h(u) = \log \{ u / (1-u)\}$ and $b(u) = \log(1+e^{u})$ in \eqref{GLMgroups}, so that the left-hand side of \eqref{GLMgroups} is $\log(\theta_i / (1-\theta_i))$, and the responses are independently drawn from $y_i \mid \mathbf{x}_i \sim \textrm{Bernoulli}(1/(1+\exp(-\theta_i))), i = 1, \ldots, n$. We considered the following four experiments:
\begin{itemize}
	\item[] \textbf{Experiment 1 ($\boldsymbol{G<n}$)}. We set $n = 100$ and $G= 40$. We simulated the groups to have high within-group correlation. Namely, the rows of each $\mathbf{X}_g$ in \eqref{GLMgroups} were generated independently from a multivariate Gaussian with mean $\bm{0}_{m_g}$ and covariance matrix $\sigma^2 \bm{\Omega}_g$, where $\sigma^2 = 1$ and $\bm{\Omega}_g$ had all off-diagonal entries equal to 0.8 and diagonal entries equal to one. The group sizes $m_g$ were randomly chosen from $\{ 3, 4, 5 \}$, and $s_0=5$ of the vectors $\bm{\beta}_g$ were randomly chosen to be nonzero with entries randomly chosen from $\{ -2.5, -2, -1.5, 1.5, 2, 2.5 \}$. Then we modeled
	\begin{align*}
		\log \left( \frac{\theta}{1-\theta} \right) = \mathbf{x}^\top \bm{\beta}.
	\end{align*}
	\item[] \textbf{Experiment 2 ($\boldsymbol{G>n}$)}. We repeated Experiment 1 with $n = 100$, but we increased the number of groups to $G= 200$.
	\\
	\item[] \textbf{Experiment 3 (semiparametric regression)}. We set $n = 100$ and $G = 80$ and generated the entries of the $n \times G$ design matrix $\mathbf{X}$  from independent $\textrm{Uniform}(-1,1)$ random variables. Then we modeled
	\begin{align*}
		\log \left( \frac{\theta}{1-\theta} \right) = 5 \sin(3x_1) - 5 x_5 e^{0.5 x_5^2},
	\end{align*}
	i.e. only the covariates $x_1$ and $x_5$ had a non-null and nonlinear effect on the mean response, and $f_j(x_j) = 0$ for all $j \notin \{1, 5\}$. We represented each covariate as a six-term B-spline basis expansion. \\
	
	\item[] \textbf{Experiment 4 (genetic association study with $\boldsymbol{G \gg n}$)}. We set $n=100$ and $G=800$. We first generated an $n \times G$ latent matrix $\mathbf{X}$, where each $i$th row $\mathbf{x}_i$ was drawn from a multivariate Gaussian with mean $\bm{0}_{G}$ and covariance matrix $\bm{\Gamma}$, where the $(j,k)$th entry of $\bm{\Gamma}$ was $\Gamma_{jk} = 0.5^{|j-k|}$. Then each entry in $\mathbf{X}$ was trichotomized as ``0,'' ``1,'' or ``2'' according to whether it was smaller than $\Phi^{-1}(1/3)$, between $\Phi^{-1}(1/3)$ and $\Phi^{-1}(2/3)$, or greater than $\Phi^{-1}(2/3)$. Here, $\Phi^{-1}(\cdot)$ denotes the inverse cumulative distribution function (cdf) of a standard normal. Thus, the entries in our final design matrix $\mathbf{X}$ were categorical SNP variables with three levels (``0,'' ``1'', or ``2''). Letting ``2'' denote the baseline category, we then modeled
	\begin{align*}
		\log \left( \frac{\theta}{1-\theta} \right) = & ~~2.5 \mathbb{I}(x_{1} = 0) - 2.5 \mathbb{I}(x_{1} = 1) + 1.4 \mathbb{I}(x_{15}=0) + 2.2 \mathbb{I}(x_{15}=1) \\
		& - 1.6 \mathbb{I}(x_{25}=0) - 1.8 \mathbb{I}(x_{25}=1).
	\end{align*}
	i.e. only the SNPs $x_1$, $x_{15}$, and $x_{25}$ had a significant association with the phenotype.
\end{itemize}

\begin{table}[H]
	\centering
	\caption{{\reviseone Simulation results for grouped logistic regression under the SSGL, GL, AdGL, GMCP, and GSCAD models, averaged across 200 replicates. The empirical standard error is reported in parentheses below the average.}}
	\label{Table:1}
	\medskip
	\resizebox{.82\textwidth}{!}{\begin{tabularx}{\linewidth}{*{7}{p{.113\linewidth}}}
			\multicolumn{7}{c}{\textbf{Experiment 1}} \\ \toprule
			& MSE & MSPE & AUC & TPR & TNR & Prec \\ 
			\hline \hline
			SSGL & 0.350 & 0.128 & \textbf{0.901} & \textbf{0.999} & 0.580 & 0.254 \\
			& (0.044) & (0.037) & (0.066) & (0.014) & (0.017) & (0.068) \\
			\hline
			GL & \textbf{0.348} & 0.136 & 0.885 & \textbf{0.999} & 0.523 & 0.232 \\
			& (0.037) & (0.037) & (0.071) & (0.014) &(0.050) & (0.020) \\ 
			\hline
			AdGL & 0.356 & 0.136 & 0.892 & \textbf{0.999} & 0.572 & 0.251 \\ 
			& (0.039) & (0.040) & (0.071) & (0.014) &(0.037) & (0.015) \\
			\hline
			GMCP & 0.438 & 0.139 & 0.883 & 0.599 & \textbf{0.979} & \textbf{0.894} \\
			& (0.044) & (0.042) & (0.066) & (0.200) & (0.037) & (0.185) \\
			\hline
			GSCAD & 0.402 & \textbf{0.126} & 0.900 & 0.799 & 0.943 & 0.781  \\
			& (0.047) & (0.040) & (0.067) & (0.200) & (0.060) & (0.222)  \\
			\bottomrule
	\end{tabularx}}
	
	\medskip
	
	\resizebox{.82\textwidth}{!}{
		\begin{tabularx}{\linewidth}{*{7}{p{.113\linewidth}}}
			\multicolumn{7}{c}{\textbf{Experiment 2}} \\ \toprule
			& MSE & MSPE & AUC & TPR & TNR & Prec \\ 
			\hline \hline
			SSGL & \textbf{0.074} & \textbf{0.149} & \textbf{0.883} & \textbf{0.898} & 0.885 & 0.174 \\
			& (0.012) & (0.021) & (0.011) & (0.108) & (0.024) & (0.045) \\
			\hline
			GL & 0.077 & 0.158 & 0.869 & 0.706 & 0.920 & 0.223 \\
			& (0.019) & (0.027) & (0.035) & (0.300) & (0.054) & (0.064) \\
			\hline
			AdGL & 0.077 & 0.157 & 0.864 & 0.70 & 0.911 & 0.172 \\
			& (0.019) & (0.026) & (0.046) & (0.303) & (0.044) & (0.024) \\ 
			\hline
			GMCP & 0.082 & 0.156 & 0.861 & 0.596 & \textbf{0.999} & \textbf{0.989} \\
			& (0.201) & (0.008) & (0.016) & (0.201) & (0.001) & (0.074) \\
			\hline
			GSCAD & 0.075 & 0.153 & 0.863 & 0.892 & 0.964 & 0.419 \\
			& (0.010) & (0.024) & (0.047) &(0.117) & (0.015) & (0.135) \\
			\bottomrule
	\end{tabularx}}
	
	\medskip
	
	\resizebox{.82\textwidth}{!}{
		\begin{tabularx}{\linewidth}{*{6}{p{.137\linewidth}}}
			\multicolumn{6}{c}{\textbf{Experiment 3}} \\ \toprule
			& MSPE & AUC & TPR & TNR & Prec  \\ 
			\hline \hline
			SSGL & \textbf{0.109} & \textbf{0.932} & \textbf{1} & 0.829 & 0.126 \\
			& (0.015) & (0.017) & (0) & (0.034) & (0.021) \\
			\hline
			GL & 0.112 & 0.930 & \textbf{1} & 0.816 & 0.160 \\
			& (0.016) & (0.017) & (0) & (0.090) & (0.090) \\
			\hline
			AdGL & 0.110 & 0.931 & \textbf{1} & 0.816 & 0.134 \\
			& (0.015) & (0.017) & (0) & (0.033) & (0.021) \\
			\hline
			GMCP & 0.157 & 0.871 & \textbf{1} & \textbf{1} & \textbf{1} \\
			& (0.019) & (0.042) & (0) & (0) & (0) \\
			\hline
			GSCAD & 0.150 & 0.887 & \textbf{1} & \textbf{1} & \textbf{1} \\
			& (0.014) & (0.028) & (0) & (0) & (0) \\
			\bottomrule
	\end{tabularx}}
	
	\medskip
	
	\resizebox{.82\textwidth}{!}{
		\begin{tabularx}{\linewidth}{*{7}{p{.113\linewidth}}}
			\multicolumn{7}{c}{\textbf{Experiment 4}} \\ \toprule
			& MSE & MSPE & AUC & TPR & TNR & Prec \\ 
			\hline \hline
			SSGL & \textbf{0.007} & \textbf{0.180} & \textbf{0.809} & \textbf{0.840} & 0.978 & 0.138 \\
			& (0.001) & (0.022) & (0.032) & (0.167) & (0.010) & (0.032) \\
			\hline
			GL & 0.010 & 0.192 & 0.789 & 0.833 & 0.972 & 0.101 \\
			& (0.001) & (0.016) & (0.044) & (0.177) & (0.004) &(0.031) \\
			\hline
			AdGL & 0.010 & 0.191 & 0.792 & 0.825 & 0.979 & 0.134 \\
			& (0.001) & (0.016) & (0.046) & (0.189) & (0.006) & (0.065) \\
			\hline
			GMCP & 0.009 & 0.183 & 0.800 & 0.669 & \textbf{0.994} & \textbf{0.306} \\
			& (0.001) & (0.016) & (0.043) & (0.078) & (0.002) & (0.102) \\
			\hline
			GSCAD & 0.009 & 0.182 & 0.802 & 0.827 & 0.988 & 0.207 \\
			& (0.001) & (0.014) & (0.056) & (0.186) & (0.002) & (0.035) \\ 
			\bottomrule
	\end{tabularx}}
\end{table}

We repeated each of the four simulations 200 times. Table \ref{Table:1} reports our experimental results averaged across the 200 replications. Note that in Experiment 3, there is not a ``true'' $\bm{\beta}$; rather, each $j$th function $f_j(\mathbf{x}_j)$ is estimated by $\widehat{f}_j(\mathbf{x}_j) = \widetilde{\mathbf{X}}_j \widehat{\bm{\beta}}_j$, where the $(i,k)$th entry of $\widetilde{\mathbf{X}}_j$ is the $k$th B-spline basis term $g_{jk}(x_{ij})$. Thus, we do not report MSE in Experiment 3.

{\reviseone Table \ref{Table:1} shows that in} all {\reviseone four} experiments, {\reviseone SSGL had either the lowest or second lowest average MSE and MSPE. SSGL also had the highest average AUC in all experiments. These results demonstrate that SSGL performed very well for both estimation and prediction. In terms of group selection, SSGL had the highest average TPR, indicating that SSGL had the highest power to detect the truly significant groups. However, this higher TPR came at a loss of TNR and precision, where GMCP and GSCAD performed the best on average. } 

The advantages of SSGL were especially pronounced in the $G \gg n$ setting (Experiment 4), where the average MSE and MSPE were substantially lower {\reviseone for SSGL}. In Experiment 4, GL, AdGL, GMCP, and GSCAD all faced {\reviseone greater} difficulty picking up the true signals, which led to {\reviseone lower average TPR} and {\reviseone higher} estimation and prediction error. {\reviseone However,} SSGL also estimated more false positives, leading to lower precision.

\subsection{Grouped Poisson regression} \label{Poissonregression}

For Poisson regression, we have $h(u) = \log u$ and $b(u) = e^{u}$ in \eqref{GLMgroups}, so that the left-hand side of \eqref{GLMgroups} is $\log(\theta_i)$, and the response variables are independently drawn from $y_i \mid \mathbf{x}_i \sim \textrm{Poisson}(\exp(\theta_i)), i = 1, \ldots, n$. Our experiments mimicked those of Section \ref{logisticregression}, except we decreased the magnitude of the entries in the design matrix $\mathbf{X}$ and the signal sizes {\reviseone in $\bm{\beta}$ in order to ensure realistic count values}. Our four simulations were as follows:

\begin{itemize}
	\item[] \textbf{Experiment {\reviseone 5} ($\boldsymbol{G<n}$)}. With $n=100$ and $G=40$, we simulated $\mathbf{X}$ and $\bm{\beta}$ the same way as we did in Experiment 1 of Section \ref{logisticregression}, except we set $\sigma^2 = 0.3$, and the entries in the $s_0=5$ randomly chosen nonzero vectors were randomly chosen from {\reviseone $\{-1, -0.75, 0.75, 1 \}$.}
	Then we modeled
	\begin{align*}
		\log (\theta) = \mathbf{x}^\top \bm{\beta}.
	\end{align*}
	\item[] \textbf{Experiment {\reviseone 6} ($\boldsymbol{G>n}$)}. We repeated Experiment 1 with $n = 100$, but we increased the number of groups to $G= 200$.
	\item[] \textbf{Experiment {\reviseone 7} (semiparametric regression)}. We set $n = 100$ and $G = 80$ and generated the entries of the $n \times G$ design matrix $\mathbf{X}$  from independent $\textrm{Uniform}(-1,1)$ random variables. Then we modeled
	\begin{align*}
		\log \left( \theta \right) = 1.5 \sin(3x_1) - x_5 e^{0.5 x_5^2}.
	\end{align*}
	We represented each covariate as a six-term B-spline basis expansion. 
	\\
	\item[] \textbf{Experiment {\reviseone 8} (genetic association study with $\boldsymbol{G \gg n}$)}. With $n=100$ and $G=800$, we simulated the SNP categorical variables (``0'', ``1'', or ``2'') in $\mathbf{X}$ the same way that we did in Experiment 4 of Section \ref{logisticregression}. Then we modeled
	\begin{align*}
		\log \left( \theta \right) = &~~2 \mathbb{I}(x_1=0) -2 \mathbb{I}(x_1 = 1) + 1.6 \mathbb{I}(x_{15}=0) + 1.6 \mathbb{I}(x_{15}=1) \\
		& - 1.4 \mathbb{I}(x_{25}=0) -1.8 \mathbb{I}(x_{25}=1). 
	\end{align*}
\end{itemize}
Each experiment was repeated 200 times. Table \ref{Table:2} reports the results averaged across the 200 replications. As {\reviseone we explained} in Section \ref{logisticregression}, we did not report the MSE in Experiment 3.

\begin{table}[H]
	\centering
	\caption{{\reviseone Simulation results for grouped Poisson regression under the SSGL, GL, AdGL, GMCP, and GSCAD models, averaged across 200 replicates. The empirical standard error is reported in parentheses below the average.}}
	\label{Table:2}
	\medskip
	\resizebox{.82\textwidth}{!}{
		\begin{tabularx}{\linewidth}{*{6}{p{.143\linewidth}}}
			\multicolumn{6}{c}{\textbf{Experiment 5}} \\ \toprule
			& MSE & MSPE & TPR & TNR & Prec \\ 
			\hline \hline
			SSGL & \textbf{0.043} & \textbf{4.864} & \textbf{1} & 0.914 & 0.626 \\
			& (0.001) & (0.801) & (0) & (0.007) & (0.028) \\
			\hline
			GL & 0.046 & 6.343 & \textbf{1} & 0.713 & 0.333 \\
			& (0.001) & (0.452) & (0) & (0.017) & (0.009) \\ 
			\hline
			AdGL & 0.044 & 5.922 & \textbf{1} & 0.656 & 0.294 \\
			& (0.001) & (0.484) & (0) & (0.012) & (0.006) \\
			\hline
			GMCP & 0.051 & 8.216 & 0.799 & \textbf{0.971} & 0.799 \\
			& (0.001) & (0.677) & (0.032) &(0.014) & (0.033) \\
			\hline
			GSCAD & 0.051 & 8.186 & 0.799 & \textbf{0.971} & \textbf{0.800} \\
			& (0.001) & (0.254) & (0.032) & (0.008) & (0.026) \\ 
			\bottomrule
	\end{tabularx}}
	
	\medskip
	
	\resizebox{.82\textwidth}{!}{
		\begin{tabularx}{\linewidth}{*{6}{p{.143\linewidth}}}
			\multicolumn{6}{c}{\textbf{Experiment 6}} \\ \toprule
			& MSE & MSPE & TPR & TNR & Prec  \\ 
			\hline \hline
			SSGL & \textbf{0.008} & \textbf{55.32} & \textbf{0.998} & 0.850 & 0.150 \\
			& (0.001) & (27.30) & (0.020) & (0.025) & (0.041) \\
			\hline
			GL & 0.009 & 56.30 & 0.995 & 0.840 & 0.144 \\
			& (0.001) & (26.530) & (0.051) & (0.033) & (0.033) \\
			\hline
			AdGL & 0.009 & 55.90 & 0.997 & 0.852 & 0.149 \\
			& (0.001) & (27.06) & (0.032) & (0.013) & (0.016) \\
			\hline
			GMCP & 0.044 & 59.68 & 0.401 & \textbf{0.966} & \textbf{0.395} \\
			& (0.017) & (25.42) & (0.037) & (0.020) & (0.360) \\
			\hline
			GSCAD & 0.037 & 60.31 & 0.402 & 0.964 & 0.308 \\
			& (0.010) & (25.25) & (0.035) & (0.018) & (0.215) \\
			\bottomrule
	\end{tabularx}}
	
	\medskip
	
	\resizebox{.82\textwidth}{!}{
		\begin{tabularx}{\linewidth}{*{5}{p{.182\linewidth}}}
			\multicolumn{5}{c}{\textbf{Experiment 7}} \\ \toprule
			& MSPE & TPR & TNR & Prec  \\ 
			\hline \hline
			SSGL & \textbf{16.43} & \textbf{1} & 0.935 & 0.284 \\
			& (0.453) & (0) & (0.013) & (0.017) \\
			\hline
			GL & 16.58 & \textbf{1} & 0.783 & 0.106 \\
			& (0.228) & (0) & (0.005) & (0.003) \\
			\hline
			AdGL & 16.50 & \textbf{1} & 0.820 & 0.125 \\
			& (0.465) & (0) & (0.004) & (0.002) \\
			\hline
			GMCP & 18.72 & \textbf{1} & \textbf{1} & \textbf{1} \\
			& (0.166) & (0) & (0) & (0) \\
			\hline
			GSCAD & 18.72 & \textbf{1} & 0.999 & 0.998 \\
			& (0.185) & (0) & (0.001) &(0.024) \\
			\bottomrule
	\end{tabularx}}
	
	\medskip
	
	\resizebox{.82\textwidth}{!}{
		\begin{tabularx}{\linewidth}{*{6}{p{.143\linewidth}}}
			\multicolumn{6}{c}{\textbf{Experiment 8}} \\ \toprule
			& MSE & MSPE & TPR & TNR & Prec  \\ 
			\hline \hline
			SSGL & \textbf{0.001} & \textbf{7.98} & \textbf{1} & \textbf{0.999} & \textbf{0.988} \\
			& (0) & (0.639) & (0) & (0.001) & (0.088) \\
			\hline
			GL & 0.003 & 28.34 & \textbf{1} & 0.955 & 0.077 \\
			& (0.001) & (3.69) & (0) & (0.001) & (0.003) \\ 
			\hline
			AdGL & 0.003 & 26.60 & \textbf{1} & 0.955 & 0.078 \\
			& (0) & (3.89) & (0) & (0.002) & (0.006) \\
			\hline
			GMCP & 0.002 & 20.22 & 0.992 & \textbf{0.999} & \textbf{0.988} \\
			& (0.001) & (2.57) & (0.085) & (0.001) & (0.125) \\
			\hline
			GSCAD & 0.002 & 19.92 & 0.992 & \textbf{0.999} & 0.987 \\
			& (0.001) & (2.13) & (0.085) &(0.001) & (0.105) \\ 
			\bottomrule
	\end{tabularx}}
\end{table}

{\reviseone Our findings for grouped Poisson regression were largely consistent with those for grouped logistic regression. Namely, Table \ref{Table:2} shows that SSGL had the lowest average MSE and MSPE in all experiments. Thus, SSGL gave superior performance in terms of estimation and prediction. SSGL also had the highest average TPR, indicating the highest power to detect the truly nonzero groups. However, just as in grouped logistic regression, the higher average TPR came at a cost of lower average TNR and precision. }

Once again, the SSGL also demonstrated its greatest advantage over the competing methods in the $G \gg n$ setting (Experiment 4), where the {\reviseone average} MSE and MSPE were substantially lower for SSGL. {\reviseone In this $G \gg n$ setting, SSGL also had the highest average TPR, TNR, and precision.} This suggests that SSGL is especially well-suited for estimation and group selection in {\reviseone count} datasets where $G$ is much larger than $n$.

\subsection{Fully Bayesian inference} \label{MCMC:sim}

{\reviseone The simulation studies in Sections \ref{logisticregression} and \ref{Poissonregression} demonstrated that SSGL often outperforms GL, AdGL, GMCP, and GSCAD in terms of estimation, prediction, and discovery of true nonzero groups, especially when the number of predictors $p$ is larger than sample size $n$. However, these competing methods do not naturally provide uncertainty quantification of the model parameters. Thus, an additional advantage of SSGL is its ability to quantify uncertainty through its posterior distribution.
	
	In this section, we assess the performance of Gibbs sampling algorithm for SSGL introduced in Section \ref{GibbsSampler}. This Gibbs sampler is implemented in the \textsf{R} package \texttt{SSGL}. We conducted the following experiments:
	\begin{itemize}
		\item[] \textbf{Experiment 9 (grouped logistic regression).} With {\revisetwo $n=400$ and $G=250$ groups of size four (i.e. $p = 1000$)}, we simulated $\mathbf{X}$ as in Experiment 1 of Section \ref{logisticregression}. {\revisetwo We randomly chose $s_0=10$ of the vectors $\boldsymbol{\beta}_g$ to be nonzero with $\boldsymbol{\beta}_g = (-1, -0.8, 0.8, 1)^\top$, and the remaining groups were set to zero.} Then we simulated binary responses $y_i \mid \mathbf{x}_i \sim \text{Bernoulli}(1 / (1+\exp(-\theta_i))), i = 1, \ldots, n$, where $\log(\theta_i / (1 - \theta_i)) = \mathbf{x}_i^\top \bm{\beta}$. \\
		
		\item[] \textbf{Experiment 10 (grouped Poisson regression).} With $n=300$ and $G=200$ {\revisetwo groups of size four (i.e. $p=800$)}, we simulated $\mathbf{X}$ as in Experiment 5 of Section \ref{logisticregression}. {\revisetwo We randomly chose $s_0=10$ of the vectors $\boldsymbol{\beta}_g$ to be nonzero with $\boldsymbol{\beta}_g = (-0.9, -0.7, 0.7, 0.9)^\top$, and the remaining groups were set to zero.} Then we simulated count responses  $y_i \mid \mathbf{x}_i \sim \text{Poisson}(\exp(\theta_i)), i = 1, \ldots, n$, where $\log(\theta_i) = \mathbf{x}_i^\top \bm{\beta}$. 
	\end{itemize}
	After generating the data, we used the Gibbs sampling algorithm of Section \ref{GibbsSampler} to draw posterior samples. We used the same hyperparameters as those suggested in Section \ref{hyperparameters}. {\revisetwo In particular, the spike parameter $\lambda_0$ was first tuned for the SSGL MAP estimator using the EM algorithm and then fixed at this value for the MCMC algorithm.} We ran the Gibbs sampler for a total of 3000 iterations, discarding the first 1000 samples as burnin. The remaining 2000 samples were used to approximate the marginal posterior distributions for the regression coefficients. To assess the quality of the SSGL posterior approximation, we recorded the following metrics:
	\begin{enumerate}
		\item the coverage probability (CP) of the 95\% posterior credible intervals (CIs) {\revisetwo for the 40 true non-null coefficients in $\boldsymbol{\beta}$}, i.e. the proportion of CIs for the {\revisetwo nonzero regression coefficients $\{ \beta_j : \beta_j \neq 0 \}$} that contained the true $\beta_j$. These intervals were constructed using the 0.025 and 0.975 sample quantiles $[q_{0.025}(\beta_j), q_{0.975}(\beta_j)]$ of the saved 2000 MCMC samples for each $\beta_j$. 
		\item the average width of the 95\% CIs for {\revisetwo the 40 non-null regression coefficients};
		\item the average effective sample size (ESS) of the 2000 saved MCMC samples for the $\beta_j$'s, $j = 1, \ldots, p$;
		\item the maximum Monte Carlo standard error (MCSE) of $q_{0.025}(\beta_j)$ and $q_{0.975}(\beta_j)$, $j = 1, \ldots, p$. We denote these quantities by $\text{MCSE}(q_{0.025})$ and $\text{MCSE}(q_{0.975})$ respectively.
	\end{enumerate}
	We repeated Experiments 9 and 10 for 200 replications. Table \ref{Table:3} reports the CP, Width, ESS, $\text{MCSE}(q_{0.025})$ and $\text{MCSE}(q_{0.975})$ averaged across these 200 replicates. We see that {\revisetwo in both experiments}, the 95\% CIs covered the true non-null regression coefficients in $\bm{\beta}$ {\revisetwo at close to the nominal rate. {\revisetwo Table \ref{Table:3} also shows that} the average widths of the 95\% CIs for the true nonzero coefficients were not overwhelmingly conservative.} Thus, {\revisetwo this coverage was not achieved by the CIs being too wide.} Our results verify the usefulness of our SSGL Gibbs sampling algorithm for quantifying uncertainty of $\bm{\beta}$.
	
	\begin{table}[t!]
		\centering
		\caption{{\reviseone Simulation results for the SSGL Gibbs sampling algorithm,  averaged across the 200 replicates. The empirical standard error is reported in parentheses.}}
		\label{Table:3}
		\begin{tabular}{r|lcl}
			\hline
			&   \textbf{Experiment 9}  & & \textbf{Experiment 10}  \\
			\hline 
			CP & 0.960 (0.032) & & 0.937 (0.039) \\
			Width  & 2.44 (0.238) & & 1.81 (0.048) \\
			ESS &  1523.29 (13.91) & & 1255.78 (53.92) \\
			$\text{MCSE}(q_{0.025})$  & 0.034 (0.007) & & 0.033 (0.004) \\
			$\text{MCSE}(q_{0.975})$  & 0.030 (0.008) & & 0.033 (0.008) \\
			\hline
		\end{tabular} 
	\end{table}
	
	Table \ref{Table:3} shows that in Experiment 9 (grouped logistic regression), the average ESS for the 2000 saved MCMC samples was {\revisetwo 1523.29}, suggesting high efficiency of our Gibbs sampling algorithm. In Experiment 10 (grouped Poisson regression), the average ESS for the 2000 saved MCMC samples was {\revisetwo slightly lower (1255.78), and the CIs had average coverage slightly below the nominal level}. This suggests that there was some loss of efficiency from using the approximation \eqref{poisson-likelihood} for the Poisson log-likelihood in our Gibbs sampler. {\revisetwo However, the CP and ESS were still acceptable in Experiment 10. Finally, in both experiments, $\text{MCSE}(q_{0.025})$ and $\text{MCSE}(q_{0.975})$ indicated an acceptable level of precision for the 0.025 and 0.975 MCMC sample quantiles.}

	\begin{figure}[t!]
		\caption{{\reviseone Estimated marginal posterior densities for {\revisetwo four} regression coefficients from {\revisetwo one replication} of Experiment 9 (top panel) and {\revisetwo four regression coefficients from} one replication of Experiment 10 (bottom panel). The true parameter values are plotted as {\revisetwo solid} red vertical lines, {\revisetwo and the SSGL MAP estimators are plotted as dashed blue lines. The left four plots depict the results for four nonzero coefficients, while the right four plots depict the results for four null coefficients where the SSGL MAP estimate is also exactly zero.}}}
		\begin{subfigure}[h]{\textwidth}
			\centering
			\includegraphics[width=.99\textwidth]{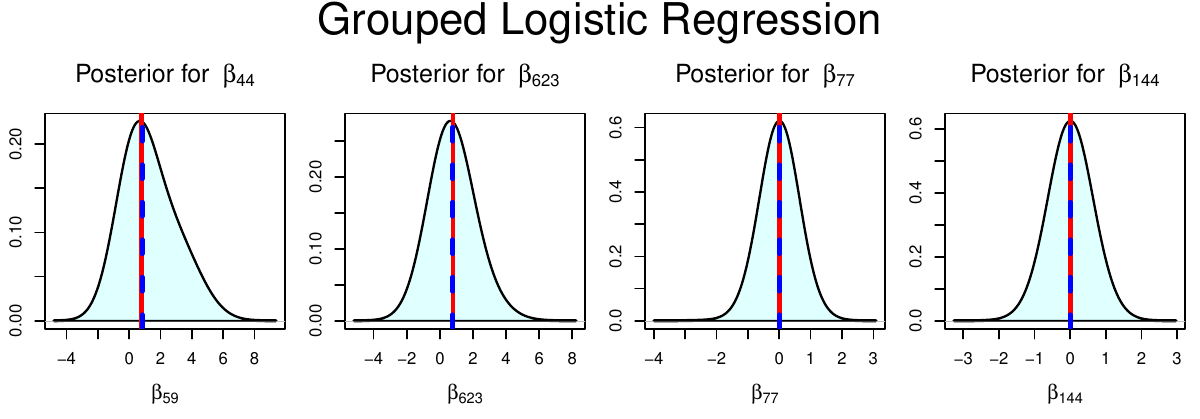}
		\end{subfigure}
		\hfill
		\begin{subfigure}[h]{\textwidth}
			\centering
			\includegraphics[width=.99\textwidth]{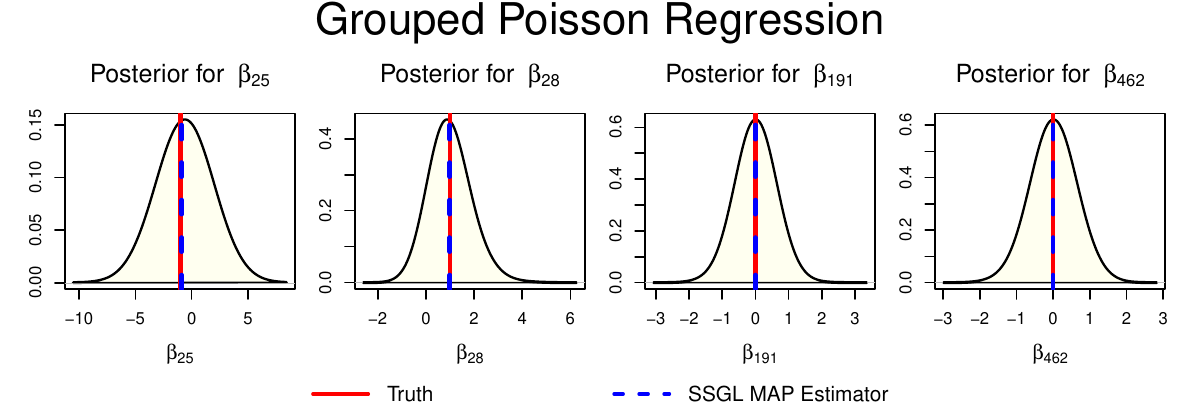}
		\end{subfigure}
		\label{fig:simulation_results}
	\end{figure}	
	
	Figure \ref{fig:simulation_results} plots the {\revisetwo kernel-smoothed} marginal posterior densities for {\revisetwo four regression coefficients} from one replication of Experiment 9 (top panel) and {\revisetwo four regression coefficients from} one replication of Experiment 10 (bottom panel). The true parameter values are plotted as {\revisetwo solid} red vertical lines, {\revisetwo and the SSGL MAP estimates are plotted as dashed vertical lines}. Figure \ref{fig:simulation_results} shows that in both grouped logistic regression and grouped Poisson regression, the SSGL posterior was able to capture the ground truth. {\revisetwo Moreover, the SSGL MAP estimates were close to their true values for the true nonzero coefficients, and they were \emph{exactly} equal to zero for the true null coefficients. This is due to the exact sparsity of the SSGL MAP estimate \eqref{MAPestimatorSSGL}.} 
	
	As discussed, the SSGL posterior mean and median are \textit{not} exactly sparse. Therefore, we generally recommend using the EM algorithm of Section \ref{EMalgorithm} to find a sparse local mode which can be used for group selection and estimation. If uncertainty quantification for $\bm{\beta}$ is also desired, then the Gibbs sampling algorithm of Section \ref{GibbsSampler} can be used for fully Bayesian inference.  
}

\section{Application to HIV drug resistance data} \label{application}

One of the challenges with drug treatments for HIV is the virus' ability to rapidly mutate and gain resistance to these drugs. The Stanford HIV Drug Resistance Database maintains isolates of HIV that were extracted from infected individuals and sequenced. In a study conducted by \citet{Rhee2006}, these isolates were used to predict resistance to 16 antiretroviral drugs used in HIV therapy. The outcome in this study was a measure of drug susceptibility, where a higher value indicated greater resistance to the drug.

For our real data application, we focus on the drug Nelfinavir, a protease inhibitor (PI), since the data from the study by \citet{Rhee2006} is publicly available.\footnote{\url{https://myweb.uiowa.edu/pbreheny/data/Rhee2006.html}. Accessed June 15, 2023.} Protease genes are made up of sequences of amino acids. A mutation occurs whenever a position in the sequence contains a different amino acid than the usual amino acid found at that position. Our dataset consists of $n=842$ isolates and $G=82$ groups, with a total of $p=361$ covariates. Each of the $G$ groups represents a specific position in the protease amino acid sequence, and within each $g$th group, the covariates are 1/0 indicator variables indicating the presence or absence of a specific amino acid mutation at the $g$th position. For example, if Valine is found at position 13 instead of the usual amino acid at position 13, then the covariate value for Valine in the group $g=13$ would be a ``1'' instead of a ``0.'' 

In \citet{Rhee2006}, a susceptibility index greater than 20 was considered to be ``highly resistant'' for PIs. Accordingly, we dichotomized the outcome into two categories according to whether the susceptibility value was greater than 20 or not. This led to 300 positive cases (``highly resistant'')  and 542 negative cases (``not highly resistant''). We then fit grouped logistic regression models to the data with the dichotomized responses.

In our study, we are mainly interested in prediction of drug resistance to Nelfinavir in HIV-infected individuals \citep{Rhee2006}. Nevertheless, group regularization can help to prevent overfitting and thus improve model generalization and classification accuracy. We examined the performance of the SSGL model on this dataset and compared it with {\reviseone GL, AdGL, GMCP, and GSCAD}. The hyperparameters, tuning parameters, {\reviseone and weights} were all chosen the same way as they were in Section \ref{illustrations}. 

To perform group selection, we fit the {\reviseone five} grouped logistic regression models to the full dataset. Next, we assessed the models' predictive power. To do so, we randomly divided the dataset into 70\% training and 30\% test data (i.e. 590 training observations and 252 test observations). We then fit the models to the training data and evaluated the MSPE and AUC on the held-out test set. We repeated this process 200 times, so that we had 200 different test sets on which to evaluate the methods.

\begin{table}[t!]  
	\centering
	\caption{Results for SSGL, GL, AdGL, GSCAD, and GMCP on the HIV drug resistance dataset. The MSPE and AUC were averaged across 200 test sets, and the empirical standard errors are shown in parentheses.}  
	\label{Table:RealData}
	\begin{tabular}{l c c c}
		\hline
		& Number of positions selected & MSPE & AUC \\ 
		\hline
		SSGL & 54 & \textbf{0.087} (0.004) & \textbf{0.951} (0.004) \\
		GL & 54 & 0.088 (0.002) & 0.945 (0.003) \\
		AdGL & 58 & 0.089 (0.003) & 0.944 (0.003) \\
		GMCP & 5 & 0.102 (0.003) & 0.925 (0.004) \\ 
		GSCAD & 11 & 0.100 (0.025) & 0.936 (0.003) \\
		\hline
	\end{tabular}
\end{table}

Our results are shown in Table \ref{Table:RealData}. SSGL selected {\reviseone 54} positions. In contrast {\reviseone to SSGL, GL, and AdGL, the methods GMCP and GSCAD selected much more parsimonious models}. However, the {\reviseone average} out-of-sample MSPE was higher and the {\reviseone average} AUC was lower {\reviseone for GMCP and GSCAD than for SSGL, GL, or AdGL}.

Despite the fact that the SSGL selected a {\reviseone less sparse model}, SSGL still {\reviseone achieved} the lowest out-of-sample MSPE and the highest out-of-sample AUC. This indicates that the SSGL model did not suffer from overfitting and possessed the best ability to correctly classify whether HIV patients were highly resistant to Nelfinavir or not. On this particular dataset, the SSGL appears to achieve the best tradeoff between {\reviseone group selection} and predictive accuracy.

\section*{Software}

An \textsf{R} package \texttt{SSGL} to implement the methodology in this paper is publicly available on the Comprehensive \textsf{R} Archive Network (CRAN).

	\section*{Acknowledgments}
	The author is grateful to the anonymous reviewers whose thoughtful feedback helped to greatly improve this paper. The author also thanks Dr. Seonghyun Jeong for helpful discussions which led to derivations of the theoretical results in this paper. 

	\bibliographystyle{apa}
\bibliography{SSGL_GLM_references}

	\begin{appendix}
		
		\setcounter{table}{0}
		
		\renewcommand{\thesection}{\Alph{section}}
		\renewcommand{\theproposition}{\thesection\arabic{proposition}}
		\renewcommand{\thetheorem}{\thesection\arabic{theorem}}
		\renewcommand{\thelemma}{\thesection\arabic{lemma}}
		\renewcommand{\theequation}{\thesection\arabic{equation}}
		\renewcommand{\thetable}{\thesection\arabic{table}}	
		
		\numberwithin{equation}{section}
		
	\section{Proofs for Section \ref{MAPcharacterization}} \label{App:A}

Before proving Theorem \ref{Th:1}, it will be necessary to define some terminology. Letting $\pi(\bm{\beta})$ denote the spike-and-slab group lasso (SSGL) prior \eqref{SSGL}, we can express the SSGL log prior as
\begin{equation*}
	\log \pi(\bm{\beta}) = \sum_{g=1}^{G} \textrm{pen}(\bm{\beta}_g),
\end{equation*}
where
\begin{equation} \label{penbetag}
	\textrm{pen}(\bm{\beta}_g) = \log \left[ (1-\theta) \bm{\Psi}(\bm{\beta}_g \mid \lambda_0) + \theta \bm{\Psi}(\bm{\beta}_g \mid \lambda_1) \right].
\end{equation}
Recall that
\begin{equation} \label{pstar}
	p_{\theta}^{\star} (\bm{\beta}_g) = \frac{\theta \bm{\Psi} (\bm{\beta}_g \mid \lambda_1)}{\theta \bm{\Psi} (\bm{\beta}_g \mid \lambda_1) + (1-\theta) \bm{\Psi} (\bm{\beta}_g \mid \lambda_0) }.
\end{equation}
The derivative of $\textrm{pen}(\bm{\beta}_g)$ with respect to $\lVert \bm{\beta}_g \rVert_2$ is 
\begin{equation} \label{derviativebetag}
	\frac{\partial \textrm{pen}(\bm{\beta}_g)}{\partial \lVert \bm{\beta}_g \rVert_2} = - \lambda_{\theta}^{\star}(\bm{\beta}_g),
\end{equation}
where
\begin{equation} \label{lambdastar}
	\lambda_{\theta}^{\star} (\bm{\beta}_g) = \lambda_1 p_{\theta}^{\star}(\bm{\beta}_g) + \lambda_0 \left[ 1 - p_{\theta}^{\star} (\bm{\beta}_g) \right].
\end{equation}
Under the objective function \eqref{logposterior-main}, it can then be seen from \eqref{llgradient} and \eqref{derviativebetag} that any local maximizer {\picotwo $\widehat{\boldsymbol{\beta}}$} of \eqref{logposterior-main} must satisfy
\begin{equation} \label{necessary}
	\mathbf{X}_g^\top  \textrm{diag} \{ \xi'(\mathbf{X}{\picotwo \widehat{\bm{\beta}}}) \}  \left( \mathbf{Y} - {\pico b'( \xi(\mathbf{X} {\picotwo \widehat{\bm{\beta}}}))} \right) - \lambda_{\theta}^{\star}({\picotwo \widehat{\bm{\beta}}_g}) \partial \lVert \bm{\beta}_g \rVert_2 = \bm{0}_{m_g} \textrm{for all } g \in \{ 1, \ldots, G \},
\end{equation}
where $\partial f$ denotes the subdifferential of $f$. From \eqref{necessary}, we can obtain the necessary first-order Karush-Kuhn-Tucker (KKT) conditions for $\widehat{\bm{\beta}}$ to be a local maximizer of \eqref{logposterior-main}, 
\begin{equation} \label{KKTll}
	\begin{array}{lcl}
		\mathbf{X}_g^\top \textrm{diag} \{ \xi'(\mathbf{X}\widehat{\bm{\beta}}) \} ( \mathbf{Y} - {\pico b'( \xi (\mathbf{X}\widehat{\bm{\beta}}))} ) - \lambda_{\theta}^{\star} (\widehat{\bm{\beta}}_g) \frac{\widehat{\bm{\beta}}_g}{\lVert \widehat{\bm{\beta}}_g \rVert_2} = \bm{0}_{m_g}, & & \textrm{for } \widehat{\bm{\beta}}_g \neq \bm{0}_{m_g}, \\
		\big\lVert \mathbf{X}_g^\top \textrm{diag}\{ \xi'(\mathbf{X}\widehat{\bm{\beta}})\} ( \mathbf{Y} - {\pico b'(\xi(\mathbf{X}\widehat{\bm{\beta}})))} \big\rVert_2 \leq \lambda_{\theta}^{\star} (\bm{0}_{m_g}) & & \textrm{for } \widehat{\bm{\beta}}_g = \bm{0}_{m_g}.
	\end{array}
\end{equation}

\paragraph{Proof of Theorem \ref{Th:1}}	
{\pico Recall that $\bm{\beta}_0 = ( \bm{\beta}_{0S_0}^\top, \bm{0}^\top )^\top$.} To prove Theorem \ref{Th:1}, we do the following two steps. First, we consider the subspace of vectors $\mathcal{T} = \{ \bm{\beta} \in \mathbb{R}^{p}: \bm{\beta}_{S_0^c} = \bm{0} \}$. We show that on $\mathcal{T}$, there exists a local maximizer {\pico $\widehat{\bm{\beta}} = (\widehat{\bm{\beta}}_{S_0}^\top, \bm{0}^\top)^\top$} of \eqref{logposterior-main} under the SSGL prior \eqref{SSGL} such that $\lVert \widehat{\bm{\beta}} - \bm{\beta}_0 \rVert_2 = O_p(\epsilon_n)$. Next, we show that $\widehat{\bm{\beta}}$ is a strict local maximizer of \eqref{logposterior-main} under the SSGL prior \eqref{SSGL} on the \emph{entire} parameter space $\mathbb{R}^{p}$. \\

\noindent \emph{Step 1 (local consistency)}. First we solve the constrained penalized likelihood estimation problem on the subspace $\mathcal{T}= \{ \bm{\beta} \in \mathbb{R}^{p} : \bm{\beta}_{S_0^c} = \bm{0} \}$. Then on $\mathcal{T}$, any estimator of $\bm{\beta}_0$ must be of the form $\widehat{\bm{\beta}} = (\widehat{\bm{\beta}}_{S_0}^\top, \bm{0}^\top)^\top$. We first show the existence of such a $\widehat{\bm{\beta}}$ to the optimization problem,
\begin{equation} \label{constrainedopt}
	\widehat{\bm{\beta}} = \argmax_{\bm{\beta}} Q_1(\bm{\beta}) = \argmax_{\bm{\beta}} \left\{ \ell_n(\bm{\beta}) + \sum_{g=1}^{s_0} \textrm{pen} (\bm{\beta}_g ), ~~ \bm{\beta}_{S_0^c} = \bm{0} \right\},
\end{equation}
where $\textrm{pen}(\bm{\beta}_g)$ is defined as in \eqref{penbetag} and $\widehat{\bm{\beta}}$ satisfies 
\begin{equation} \label{MAPdistance}
	\lVert \widehat{\bm{\beta}} - \bm{\beta}_0 \rVert_2 = O_p(\epsilon_n),
\end{equation}
with $\epsilon_n = (s_0 \log G / n)^{1/2}$ and $\bm{\beta}_0 = ( \bm{\beta}_{0S_0}^\top, \bm{0}^\top )^\top$. Let $\bm{\Delta} = ( \bm{\Delta}_{S_0}^\top, \bm{0}^\top )^\top$, where $\lVert \bm{\Delta}_{S_0} \rVert_2 = C$. In order to establish \eqref{MAPdistance}, it suffices to show that, for some small $\varepsilon > 0$ and sufficiently large $C > 0$,
\begin{equation} \label{highprobevent}
	P \left\{ \sup_{\lVert \bm{\Delta}_{S_0} \rVert_2 = C} Q_1(\bm{\beta}_0 + \epsilon_n \bm{\Delta} ) < Q_1 (\bm{\beta}_0) \right\} \geq 1 - \varepsilon, 
\end{equation}
{\pico as $n \rightarrow \infty$.} If \eqref{highprobevent} holds, this will imply that with probability at least $1-\varepsilon$, there exists a local maximum inside the ball $\{  \bm{\beta}_0 + \epsilon_n \bm{\Delta} : \lVert \bm{\Delta}_{S_0} \rVert_2 \leq C \}$ {\pico as $n \rightarrow \infty$}. In order to establish \eqref{highprobevent}, we must show that with probability tending to one and sufficiently large $C >0$,
\begin{align*} \label{difflessthanzero}
	Q_1(\bm{\beta}_0 + \epsilon_n \bm{\Delta}) - Q_1 (\bm{\beta}_0) &  = \left[ \ell_n(\bm{\beta}_0 + \epsilon_n \bm{\Delta}) - \ell_n(\bm{\beta}_0) \right] + \sum_{g=1}^{s_0} \left[ \textrm{pen}(\bm{\beta}_{0g} + \epsilon_n \bm{\Delta}_g) - \textrm{pen}( \bm{\beta}_{0g}) \right] \\
	& \overset{\Delta}{=} (\textrm{I}) + (\textrm{II}) < 0, \numbereqn
\end{align*}
{\pico as $n \rightarrow \infty$.} We bound the terms in \eqref{difflessthanzero} separately. We first focus on bounding $(\textrm{I}) := \ell_n (\bm{\beta}_0 + \epsilon_n \bm{\Delta}) - \ell_n(\bm{\beta}_0)$. Let $\nabla_{S_0}$ and $\nabla_{S_0}^2$ denote the partial derivatives with respect to $\bm{\beta}_{0S_0}$. By a Taylor expansion and the fact that $\bm{\beta}_0 = ( \bm{\beta}_{0S_0}^\top, \bm{0}^\top )^\top$, we have
\begin{align*} \label{J1J2}
	(\textrm{I}) & = \epsilon_n \nabla_{S_0} \ell_n (\bm{\beta}_0) ^\top \bm{\Delta}_{S_0} + \frac{\epsilon_n^2}{2} \bm{\Delta}_{S_0}^\top \nabla^2_{S_0} {\pico \ell_n} (\widetilde{\bm{\beta}}) \bm{\Delta}_{S_0} \\
	& \overset{\Delta}{=} J_1 + J_2, \numbereqn
\end{align*}
where $\widetilde{\bm{\beta}}$ lies on the line segment between $\bm{\beta}_0$ and $\bm{\beta}_0 + \epsilon_n \bm{\Delta}$. Next, we bound the terms $J_1$ and $J_2$ in \eqref{J1J2} separately. We have
\begin{align*}
	\nabla_{S_0}\ell_n (\bm{\beta}_0) = \mathbf{X}_{S_0}^\top \textrm{diag}\{ \xi'(\mathbf{X}\bm{\beta}_0)\} ( \mathbf{Y} - {\pico b'(\xi(\mathbf{X}\bm{\beta}_0))}).
\end{align*} 
Let $d_2$ denote the largest diagonal entry in $\textrm{diag}( \xi'(\mathbf{X} \bm{\beta}_0) )$, which we know is finite and positive due to Assumption {\pico (A4)(ii)}. Then we have, for some $A > 0$,
\begin{align*} \label{Opepsilon}
	 P\left( \lVert \nabla_{S_0} \ell_n(\bm{\beta}_0) \rVert_2 > A n \epsilon_n \right) & \leq P \left( \big\lVert \frac{1}{n} \mathbf{X}_{S_0}^\top (\mathbf{Y} - {\pico b'( \xi (\mathbf{X}\bm{\beta}_0)))} \big\rVert_2^2 > A^2 \epsilon_n^2 / d_2^2 \right) \\
	&  \leq \frac{d_2^2 n}{A^2 s_0 \log G} \mathbb{E} \left\{ \sum_{g=1}^{s_0} \sum_{k=1}^{m_g} \left[ \frac{1}{n} \sum_{i=1}^{n} x_{igk} \left( y_i - {\pico b'( \xi (\mathbf{x}_i^\top \bm{\beta}_0))} \right) \right]^2 \right\} \\
	&  = \frac{d_2^2}{A^2 s_0 \log G} \textrm{tr} \left( \frac{1}{n} \mathbf{X}_{S_0}^\top {\pico \bm{\Omega}(\bm{\beta}_0)}  \mathbf{X}_{S_0} \right) \\
	& \leq \frac{d_2^2}{A^2 s_0 \log G} \left( \sum_{g=1}^{s_0} m_g \right) \lambda_{\max} \left( \frac{1}{n} \mathbf{X}_{S_0}^\top {\pico \bm{\Omega}(\bm{\beta}_0)} \mathbf{X}_{S_0} \right) \\
	& \leq \frac{d_2^2 s_0 m_{\max}}{A^2 s_0 \log G} \lambda_{\max} \left( \frac{1}{n} \mathbf{X}_{S_0}^\top {\pico \bm{\Omega}(\bm{\beta}_0)} \mathbf{X}_{S_0} \right) \\
	& {\pico \lesssim} \frac{d_2^2 \log n}{A^2 \log G} \lambda_{\max} \left( \frac{1}{n} \mathbf{X}_{S_0}^\top {\pico \bm{\Omega}(\bm{\beta}_0)} \mathbf{X}_{S_0} \right) \rightarrow 0~~{\pico \textrm{as}~~n \rightarrow \infty}, \numbereqn
\end{align*}
where the last inequality of the display uses Assumption (A2) that $m_{\max} = O(\log n \wedge (\log G / \log n))$ and Assumption {\pico (A3)(ii)} that $\lambda_{\max} \left( n^{-1} \mathbf{X}_{S_0}^\top {\pico \bm{\Omega}(\bm{\beta}_0)} \mathbf{X}_{S_0} \right)$ ${\pico = o(\log G / \log n)}$. Therefore, from \eqref{Opepsilon}, we have that $\lVert \nabla_{S_0}\ell_n (\bm{\beta}_0) \rVert_2 = O_p(n \epsilon_n)$, and so an upper bound for $J_1$ in \eqref{J1J2} is
\begin{align*} 
	J_1 \leq | J_1 | \leq \epsilon_n \lVert \nabla_{S_0} \ell_n (\bm{\beta}_0) \rVert_2 \lVert \bm{\Delta}_{S_0} \rVert_2 = A n \epsilon_n^2 \lVert \bm{\Delta}_{S_0} \rVert_2.
\end{align*}
for some $A > 0$. We also have by Assumption {\pico (A3)(ii)} that
\begin{align*} 
	J_2 & = - \frac{\epsilon_n^2}{2} \bm{\Delta}_{S_0}^\top \left\{ \mathbf{X}_{S_0}^\top \bm{\Sigma}(\widetilde{\bm{\beta}}) \mathbf{X}_{S_0} \right\} \bm{\Delta}_{S_0} \\
	& \leq -\frac{\epsilon_n^2}{2} \lambda_{\min} \left( \mathbf{X}_{S_0}^\top \bm{\Sigma} ( \widetilde{\bm{\beta}} ) \mathbf{X}_{S_0} \right) \lVert \bm{\Delta}_{S_0} \rVert_2^2   \\
	& {\pico \lesssim} - \frac{{\pico n \epsilon_n^2}}{2} \lVert \bm{\Delta}_{S_0} \rVert_2^2. 
\end{align*}
Therefore, an upper bound for $(\textrm{I})$ in \eqref{difflessthanzero} is
\begin{equation} \label{upperboundI}
	(\textrm{I})~{\pico \lesssim}~A n \epsilon_n^2 \lVert \bm{\Delta}_{S_0} \rVert_2 - \frac{n \epsilon_n^2}{2} \lVert \bm{\Delta}_{S_0} \rVert_2^2
\end{equation}
Now we turn our attention to upper-bounding $(\textrm{II})$ in \eqref{difflessthanzero}. We first focus on bounding each of the summands in (II) from above. We first rewrite $\textrm{pen}(\bm{\beta}_g)$ in \eqref{penbetag} as
\begin{align*}
	\textrm{pen}(\bm{\beta}_g ) = -\lambda_1 \lVert \bm{\beta}_g \rVert_2 - \log \left[ p_{\theta}^{\star} (\bm{\beta}_g) \right],
\end{align*} 
where $p_{\theta}^{\star}(\bm{\beta}_g)$ is defined as in \eqref{pstar}. Therefore, for any one group $g$,
{\small
\begin{align*} \label{upperboundIIpt1}
\textrm{pen}(\bm{\beta}_{0g} + \epsilon_n \bm{\Delta}_g ) - \textrm{pen}(\bm{\beta}_{0g} ) & = \lambda_1 \left\{ \lVert \bm{\beta}_{0g} \rVert_2 - \lVert \bm{\beta}_{0g} + \epsilon_n \bm{\Delta}_g \rVert_2 \right\} + \left[ \log\left( p_{\theta}^{\star}(\bm{\beta}_{0g}) \right) - \log \left( p_{\theta}^{\star} (\bm{\beta}_{0g} + \epsilon_n \bm{\Delta}_g) \right) \right] \\
	& \leq \lambda_1 \epsilon_n \lVert \bm{\Delta}_g \rVert_2 - \left[ \log ( p_{\theta}^{\star} (\bm{\beta}_{0g} + \epsilon_n \bm{\Delta}_g )) - \log (p_{\theta}^{\star}(\bm{\beta}_{0g})) \right] \\
	& \leq \lambda_1 \epsilon_n \lVert \bm{\Delta}_g \rVert_2 - \epsilon_n \left( \frac{\partial p_{\theta}^{\star}(\widetilde{\bm{\beta}}_g)}{\partial \widetilde{\bm{\beta}}_g} \right)^\top \bm{\Delta}_g, \\ & \qquad \qquad \textrm{where } \widetilde{\bm{\beta}}_g \textrm{ lies on line segment between } \bm{\beta}_{0g} \textrm{ and } \bm{\beta}_{0g}+\epsilon_n \bm{\Delta}_g, \\
	& \leq \epsilon_n \lVert \bm{\Delta}_g \rVert_2 \left\{ \lambda_1 + \bigg\lVert \frac{\partial p_{\theta}^{\star}(\widetilde{\bm{\beta}}_g)}{\partial \widetilde{\bm{\beta}}_g} \bigg\rVert_2  \right\}, \numbereqn
\end{align*}}
We examine the second term in brackets in \eqref{upperboundIIpt1}. Noting that 
\begin{align} \label{pthetastar}
	p_{\theta}^{\star}({\picotwo \bm{\beta}_g}) = \frac{1}{1 + \left( \frac{1-\theta}{\theta} \right) \left( \frac{\lambda_0}{\lambda_1} \right)^{m_g} \exp \left[ - (\lambda_0-\lambda_1) \lVert \bm{\beta}_g \rVert_2 \right] } := \frac{1}{1+c(\bm{\beta}_g)},
\end{align}
we can see that when $\bm{\beta}_g \neq \bm{0}_{m_g}$,  
\begin{align*}
	\frac{\partial p_{\theta}^{\star}(\bm{\beta}_g)}{\partial \bm{\beta}_g} = \frac{(\lambda_0-\lambda_1)c(\bm{\beta}_g)}{(1+c(\bm{\beta}_g))^2} \times \frac{\bm{\beta}_g}{\lVert \bm{\beta}_g \rVert_2}.
\end{align*}
Note that 
\begin{align*}
	\frac{(\lambda_0 - \lambda_1)c(\bm{\beta}_g)}{(1+c(\bm{\beta}_g))^2} < (\lambda_0 - \lambda_1) c(\bm{\beta}_g) \prec 1, 
\end{align*}
since the exponent term $\exp[-(\lambda_0-\lambda_1) \lVert \bm{\beta}_g \rVert_2]$ in $c(\bm{\beta}_g)$ decays faster than $(\lambda_0-\lambda_1)$ multiplied by the other terms in $c(\bm{\beta}_g)$ as $n \rightarrow \infty$. Since it must be that $\widetilde{\bm{\beta}}_g$ in \eqref{upperboundIIpt1} is a nonzero vector, we have that for large $n$,
\begin{align*}
	\bigg\lVert	\frac{\partial p_{\theta}^{\star}(\widetilde{\bm{\beta}}_g)}{\partial \widetilde{\bm{\beta}}_g} \bigg\rVert_2  < \bigg\lVert \frac{\widetilde{\bm{\beta}}_g}{\lVert \widetilde{\bm{\beta}}_g \rVert_2}  \bigg\rVert_2 = 1,  
\end{align*}
and so we can further bound \eqref{upperboundIIpt1} from above by $\epsilon_n ( 1 + \lambda_1 ) \lVert \bm{\Delta}_g \rVert_2$. Therefore, an upper bound for $(\textrm{II})$ in \eqref{difflessthanzero} is 
\begin{align} \label{upperboundIIpt2}
	(\textrm{II})~{\pico \lesssim}~\epsilon_n  \sum_{g=1}^{s_0} (1+\lambda_1) \lVert \bm{\Delta}_g \rVert_2 & \leq s_0 \epsilon_n (1+\lambda_1) \max_{1 \leq g \leq s_0} \lVert \bm{\Delta}_g \rVert_2  \leq s_0 \epsilon_n (1+\lambda_1) \lVert \bm{\Delta}_{S_0} \rVert_2. \numbereqn
\end{align}
Combining the upper bounds in \eqref{upperboundI} and \eqref{upperboundIIpt2}, it is clear that 
\begin{align*}
	(\textrm{I}) + (\textrm{II})~{\pico \lesssim}~A n \epsilon_n^2 \lVert \bm{\Delta}_{S_0} \rVert_2 - \frac{n \epsilon_n^2}{2} \lVert \bm{\Delta}_{S_0} \rVert_2^2 + s_0 \epsilon_n (1+\lambda_1) \lVert \bm{\Delta}_{S_0} \rVert_2,
\end{align*}
and the right-hand side can be made negative by choosing $C := \lVert \bm{\Delta}_{S_0} \rVert_2$ large enough. For large $C > 0$, the negative term above is the dominating term. This proves \eqref{difflessthanzero}, and thus, $\widehat{\bm{\beta}} = (\widehat{\bm{\beta}}_{S_0}^\top, \bm{0}^\top )^\top$ is a local maximum over the subspace $\mathcal{T} = \{ \bm{\beta} \in \mathbb{R}^{p} : \bm{\beta}_{S_0^c} = \bm{0} \}$ {\pico as $n \rightarrow \infty$}. \\

\noindent \emph{Step 2 (sparsity)}. Since $\log \pi(\bm{\beta})$ is a nonconvex function, the KKT conditions \eqref{KKTll} only give necessary conditions for $\widehat{\bm{\beta}}$ from Step 1 to be a local maximum on $\mathcal{T}$. However, by suitably modifying Theorem 1 of \citet{FanLv2011} to the grouped GLM setting with the log SSGL prior as the penalty function, we can also obtain \emph{sufficient} conditions for $\widehat{\bm{\beta}} \in \mathcal{T}$ from Step 1 to be a strict local maximum over \emph{all} of $\mathbb{R}^{p}$. This will be the case if, {\pico for sufficiently large $n$,} $\widehat{\bm{\beta}}$ from Step 1 satisfies
\begin{equation} \label{sufficientcondition1}
	\lambda_{\max} \left\{ {\picotwo \nabla_g^2 \ell_n(\widehat{\bm{\beta}})} + \nabla^2 \textrm{pen}(\widehat{\bm{\beta}}_g ) \right\} < 0~~ \textrm{for all } {\picotwo \widehat{\boldsymbol{\beta}}_g \neq \bm{0}_{m_g}}, g \in S_0,
\end{equation}
{\picotwo where $\nabla_g^2$ denotes the second partial derivative with respect to $\widehat{\boldsymbol{\beta}}_g$,} $\textrm{pen}(\bm{\beta}_g)$ is {\picotwo the penalty function} in \eqref{penbetag}, and
\begin{equation} \label{sufficientcondition2}
	\lVert \mathbf{X}_g^\top \textrm{diag} ( \xi'(\mathbf{X} \widehat{\bm{\beta}}) ) [ \mathbf{Y} - {\pico b'( \xi({\picotwo \mathbf{X}\widehat{\bm{\beta}}}))]} \rVert_2 < \lambda_{\theta}^{\star}(\bm{0}_{m_g})~~ \textrm{for all } g \in S_0^c.
\end{equation}
That is, we require the KKT second order sufficiency condition to hold {\pico as $n \rightarrow \infty$}, and the second inequality in \eqref{KKTll} has to hold with a \textit{strict} inequality. 

We first prove \eqref{sufficientcondition1}. {\picotwo For $\widehat{\boldsymbol{\beta}}_g \neq \bm{0}_{m_g}$,} we have {\picotwo by \eqref{llhessian} and the chain rule for second derivatives (with $f(\mathbf{u}) = \text{pen}(\mathbf{u})$ and $\mathbf{u} = g(\mathbf{v}) = \lVert \mathbf{v} \rVert_2$) that}
\begin{align*} \label{second-order}
	& {\picotwo \nabla_g^2} \ell_n (\widehat{\bm{\beta}}) + \nabla^2 \textrm{pen} (\widehat{\bm{\beta}}_g ) \\
	& \qquad = -\mathbf{X}_g^\top \bm{\Sigma}(\widehat{\bm{\beta}}) \mathbf{X}_g {\picotwo~+~p_{\theta}^{\star}(\widehat{\bm{\beta}}_g) [1 - p_{\theta}^{\star}(\widehat{\bm{\beta}}_g)] (\lambda_0 - \lambda_1)^2 \frac{\widehat{\boldsymbol{\beta}}_g \widehat{\boldsymbol{\beta}}_g^\top}{\lVert \widehat{\boldsymbol{\beta}}_g \rVert_2^2}} {\picotwo - ~ \lambda_{\theta}^{\star}(\widehat{\boldsymbol{\beta}}_g) \left[ \frac{1}{\lVert \widehat{\bm{\beta}}_g \rVert_2} \mathbf{I}_{m_g} - \frac{\widehat{\bm{\beta}}_g \widehat{\bm{\beta}}_g^\top}{\lVert \widehat{\bm{\beta}}_g \rVert_2^3} \right].} \numbereqn
\end{align*}
{\picotwo Note that $\lambda_{\min}( \lVert \widehat{\boldsymbol{\beta}}_g \rVert_2^{-1} \mathbf{I}_{m_g} - \lVert \boldsymbol{\widehat{\beta}}_g \rVert_2^{-3} \boldsymbol{\widehat{\beta}}_g \boldsymbol{\widehat{\beta}}_g^\top) = 0$, since $\mathbf{I}_{m_g} - \rVert \widehat{\boldsymbol{\beta}}_g \rVert_2^{-2} \widehat{\boldsymbol{\beta}}_g \widehat{\boldsymbol{\beta}}_g^\top$ is the annihilator matrix.
	Using the fact that $\lambda_{\max}(-\mathbf{A}) = -\lambda_{\min}(\mathbf{A})$ for a square matrix $\mathbf{A}$, it follows from \eqref{second-order} that  
	\begin{align*}
		 \lambda_{\max} \left\{ {\picotwo \nabla_g^2} \ell_n(\widehat{\bm{\beta}}) + \nabla^2 \textrm{pen} (\widehat{\bm{\beta}}_g ) \right\} & \leq -\lambda_{\min} \left( \mathbf{X}_g^\top \bm{\Sigma}(\widehat{\bm{\beta}}) \mathbf{X}_g \right) + ~p_{\theta}^{\star}(\widehat{\bm{\beta}}_g) [1 - p_{\theta}^{\star}(\widehat{\bm{\beta}}_g)] (\lambda_0 - \lambda_1)^2  \\
		&  \lesssim -n + \frac{c(\widehat{\bm{\beta}}_g) (\lambda_0-\lambda_1)^2}{(1+c(\widehat{\bm{\beta}}_g))^2}, ~~ \text{where } c(\bm{\beta}_g) \text{ is as in } (\text{A}.15) \\
		& \leq - n + c(\widehat{\bm{\beta}}_g) (\lambda_0-\lambda_1)^2 \\
		& \lesssim -n + 1 < 0,
	\end{align*}
	for sufficiently large $n$, where we used Assumption (A3)(ii) in the third line and the fact that given our choices of hyperparameters for $(\theta, \lambda_0, \lambda_1)$ in Theorem \ref{Th:1}, $c(\boldsymbol{\beta}_g) (\lambda_0 - \lambda_1)^2 \prec 1$ for any $\boldsymbol{\beta}_g \neq \bm{0}_{m_g}$ in the last line.}
Thus, we have shown {\pico that} \eqref{sufficientcondition1} {\pico holds for large $n$}. 

Next, we prove \eqref{sufficientcondition2}. It suffices to show that as $n \rightarrow \infty$,
\begin{equation*}
	P \left( \exists~g \in S_0^c,~~\bigg\lVert \frac{\partial \ell_n (\widehat{\bm{\beta}})}{\partial \bm{\beta}_g} \bigg\rVert_2 \geq \lambda_{\theta}^{\star}(\bm{0}_{m_g}) \right) \rightarrow 0.
\end{equation*}
To ease the notation, let $h_g(\widehat{\bm{\beta}}) = \partial \ell_n (\widehat{\bm{\beta}}) / \partial \bm{\beta}_g$. By a Taylor expansion around $\bm{\beta}_{0}$ and the fact that $\widehat{\bm{\beta}}_{0S_0^c} = \bm{0}$, we have
\begin{equation} \label{hg}
	h_g(\widehat{\bm{\beta}}) = h_g (\bm{\beta}_0) + \nabla_{S_0}h_g(\widetilde{\bm{\beta}})^\top (\widehat{\bm{\beta}}_{S_0} - \bm{\beta}_{0S_0}),
\end{equation}
where $\widetilde{\bm{\beta}}$ lies on the line segment between $\bm{\beta}_0$ and $\widehat{\bm{\beta}}$. By \eqref{hg}, we have that
{\small
\begin{align*} \label{K1K2}
	& P \left( \exists~g \in S_0^c,~~ \lVert h_g(\widehat{\bm{\beta}}) \rVert_2 \geq \lambda_{\theta}^{\star}(\bm{0}_{m_g}) \right) \\
	& \qquad \leq P \left( \exists~g \in S_0^c,~~ \lVert h_g (\bm{\beta}_0) \rVert_2 \geq \frac{\lambda_{\theta}^{\star}(\bm{0}_{m_g})}{2} \right) + P \left( \exists~g \in S_0^c,~~\lVert \nabla_{S_0} h_g(\widetilde{\bm{\beta}})^\top (\widehat{\bm{\beta}}_{S_0} - \bm{\beta}_{0S_0}) \rVert_2 \geq \frac{\lambda_{\theta}^{\star}(\bm{0}_{m_g})}{2} \right) \\
	& \qquad \leq \sum_{g=s_0+1}^{G} P \left( \lVert h_g(\bm{\beta}_0) \rVert_2 \geq \frac{\lambda_{\theta}^{\star} (\bm{0}_{m_g})}{2} \right) + \sum_{g=s_0+1}^{G} P \left( \lVert \nabla_{S_0} h_g(\widetilde{\bm{\beta}})^\top ( \widehat{\bm{\beta}}_{S_0} - \bm{\beta}_{0S_0} ) \rVert_2 \geq \frac{\lambda_{\theta}^{\star}(\bm{0}_{m_g})}{2}  \right)  \\
	& \qquad \overset{\Delta}{=} T_1 + T_2. \numbereqn
\end{align*}}
We will bound each of the terms in \eqref{K1K2} from above separately. First note that since $\lambda_0 = (1-\theta)/\theta \asymp G^{c}, c > 2$, and $\lambda_1 \asymp n^{-1}$, we have that for some constant $M>0$,
\begin{align*}
	p_{\theta}^{\star}(\bm{0}_{m_g}) = \frac{1}{1 + \left( \frac{1-\theta}{\theta}\right) \left( \frac{\lambda_0}{\lambda_1} \right)^{m_g} } = \frac{1}{1 + M n\lambda_0^{m_g+1}},
\end{align*}
and thus,
\begin{align*} \label{lowerboundlambdastar}
	\lambda_{\theta}^{\star}(\bm{0}_{m_g}) & = \lambda_1 p_{\theta}^{\star}(\bm{0}_{m_g}) + \lambda_0 \left[ 1 - p_{\theta}^{\star} (\bm{0}_{m_g}) \right] \\
	& = \frac{\lambda_1 + \lambda_0 (M n \lambda_0^{m_g+1})}{1+M n \lambda_0^{m_g+1}} > \lambda_0 \left( \frac{Mn\lambda_0^{m_g+1}}{1+Mn \lambda_0^{m_g+1}} \right) > \frac{\lambda_0}{2}, \numbereqn
\end{align*}
for large $n$.

We first examine each of the summands in $T_1$ in \eqref{K1K2}. Since all of the entries of $\mathbf{X}$ satisfy {\pico $| x_{ij} | = O(\log G)$} (by Assumption {\pico (A3)(i)}), we have that for {\pico some constant $D>0$} and any $g \in \{ 1, \ldots, G \}$, $\lVert \mathbf{X}_g \rVert_{2,\infty} \leq {\pico D \log G} \sqrt{m_g} \leq {\pico D \log G} \sqrt{m_{\max}}$. Thus, in light of \eqref{lowerboundlambdastar}, we have that for sufficiently large $n$,
\begin{align*} \label{K1upperboundptI}
	P \left( \lVert h_g(\bm{\beta}_0) \rVert_2 \geq \frac{\lambda_{\theta}^{\star}(\bm{0}_{m_g})}{2} \right) & < P \left( \lVert \mathbf{X}_g \rVert_{2, \infty} \lVert \mathbf{Y} - {\pico b'( \xi (\mathbf{X}\bm{\beta}_0))} \rVert_2 \geq \frac{\lambda_0}{2} \right) \\
	& \leq P \left( \lVert \mathbf{Y} - {\pico b'( \xi (\mathbf{X}\bm{\beta}_0))} \rVert_2 \geq \frac{\lambda_0}{2 D {\pico \log G} \sqrt{m_{\max}}} \right) \\
	& \leq 2 \exp \left( -\frac{\lambda_0^2 / ( 4 D^2 (\log G)^2 m_{\max})}{2 [ 2nG + L \lambda_0 / (D {\pico \log G} \sqrt{m_{\max}})]} \right) \\
	& = 2 \exp \left( - \frac{\lambda_0^2}{16 D^2 (\log G)^2 m_{\max} n G + 8 L \lambda_0 D \log G \sqrt{m_{\max}}} \right) \numbereqn
\end{align*} 
where we used Assumption {\pico (A4)(i)} and the Bernstein inequality (specifically, Lemma 2.2.11 of \citet{vanderVaartWellner2006}). Thus, from \eqref{K1upperboundptI}, we have as an upper bound for $T_1$ in \eqref{K1K2},
\begin{equation} \label{K1upperboundptII}
	T_1 < 2 \exp \left( \log G - \frac{\lambda_0^2}{16 D^2 (\log G)^2 m_{\max} n G + 8 L \lambda_0 D \log G \sqrt{m_{\max}}} \right) \rightarrow 0,
\end{equation}
where we used the fact that $\lambda_0 \asymp G^{c}, c > 2$, and Assumptions (A1) and (A2) that $G \gg n$ and $m_{\max} = O(\log n \wedge (\log G/ \log n))$. Thus, the second term in the exponent of \eqref{K1upperboundptII}  dominates the first term as $n \rightarrow \infty$. 

We next examine each of the summands in $T_2$ in \eqref{K1K2}. In light of \eqref{lowerboundlambdastar} that $\lambda_{\theta}^{\star}(\bm{0}_{m_g}) > \lambda_0 / 2$, we have that for each $g \in S_0^c$ and sufficiently large $n$,
\begin{align*} \label{K2upperboundptI}
	P \left( \lVert \nabla_{S_0} h_g(\widetilde{\bm{\beta}})^\top (\widehat{\bm{\beta}}_{S_0} - \bm{\beta}_{0S_0}) \lVert_2 \geq \frac{\lambda_{\theta}^{\star}(\bm{0}_{m_g})}{2} \right) & <	P \left( \lVert \nabla_{S_0} h_g(\widetilde{\bm{\beta}})^\top (\widehat{\bm{\beta}}_{S_0} - \bm{\beta}_{0S_0}) \lVert_2 \geq \frac{\lambda_0}{4} \right) \\
	&  \leq P\left( \lVert \nabla_{S_0} h_g(\widetilde{\bm{\beta}}) \rVert_{2, \infty} \lVert \widehat{\bm{\beta}}_{S_0} - \bm{\beta}_{0S_0} \rVert_2 \geq \frac{\lambda_0}{4} \right) \\
	&  \leq P \left( \lVert \mathbf{X}_{S_0}^\top \bm{\Sigma}(\widetilde{\bm{\beta}}) \mathbf{X}_g \rVert_{2, \infty} \lVert \widehat{\bm{\beta}}_{S_0} - \bm{\beta}_{0S_0} \rVert_2 \geq \frac{\lambda_0}{4} \right) \\
	& \lesssim \frac{4 n \epsilon_n}{\lambda_0}, \numbereqn
\end{align*}
To arrive at \eqref{K2upperboundptI}, we used the fact that $\nabla_{S_0} h_g(\widetilde{\bm{\beta}}) = \mathbf{X}_{S_0}^\top \bm{\Sigma}(\widetilde{\bm{\beta}}) \mathbf{X}_g$ in the second line. In the last line, we used Assumption {\pico (A3)(iii)} that for all $g \in \mathcal{S}_0^{c}$, $\lVert \mathbf{X}_{S_0}^\top \bm{\Sigma}(\widetilde{\bm{\beta}}) \mathbf{X}_g \rVert_{2, \infty} = O(n)$, the fact that $\lVert \widehat{\boldsymbol{\beta}}_{S_0} - \boldsymbol{\beta}_{0S_0} \rVert_2 = O_p(\epsilon_n)$, and Markov's inequality. From \eqref{K2upperboundptI} and the fact that $\lambda_0 \asymp G^{c}, c > 2$, and $G \gg n$, we have as an upper bound for $T_2$ in \eqref{K1K2},
\begin{align} \label{K2upperboundptII}
	T_2 \prec \frac{4 Gn \epsilon_n}{\lambda_0} \prec \frac{G n}{G^{c}} = \frac{n}{G^{c-1}} \rightarrow 0 \textrm{ as } n \rightarrow \infty.
\end{align}
Combining \eqref{K1K2}, \eqref{K1upperboundptII}, and \eqref{K2upperboundptII} allows us to conclude that \eqref{sufficientcondition2} holds {\pico for sufficiently large $n$}, and the proof is finished.  \hfill $\square$

\section{Proofs for Section \ref{fullposteriorcharacterization}} \label{App:B}

To prove Theorems \ref{Th:2} and \ref{Th:3}, we follow the proof strategy of \citet{JeongGhosal2021}. However, a crucial difference is that we study an \emph{absolutely continuous} spike-and-slab
prior in this paper, whereas \citet{JeongGhosal2021} prove their results for a \emph{point-mass} spike-and-slab prior. The continuous nature of the SSGL requires it to be handled quite differently from the point mass prior of \citet{JeongGhosal2021}. In particular, since SSGL \eqref{SSGL} puts zero probability on exactly sparse configurations of $\bm{\beta}$, we need to instead rely on notions of ``approximate'' sparsity to appropriately bound the approximation error.

We first define some necessary notation. Let $f_i$ be the density of $y_i$ belonging to the exponential family \eqref{exponentialfamily}, where the natural parameter $\theta_i$ is related to the grouped covariates $\mathbf{x}_i = (\mathbf{x}_{i1}^\top, \ldots, \mathbf{x}_{iG}^\top)^\top$ through \eqref{GLMgroups}. Denote $f_{0i}$ analogously with true natural parameter $\theta_{0i}$. Then the joint densities are $f = \prod_{i=1}^{n} f_i$ and $f_0 = \prod_{i=1}^{n} f_{0i}$. We define the average squared Hellinger metric between $f$ and $f_0$ as $H_n^2(f, f_0) = n^{-1} \sum_{i=1}^{n} H^2(f_i, f_{0i})$, where $H(f_i, f_{0i}) = (\int (\sqrt{f_i} - \sqrt{f_{0i}})^2)^{1/2}$ is the Hellinger distance between $f_i$ and $f_{0i}$. The Kullback-Leibler (KL) divergence and KL variation between $f_i$ and $f_{0i}$ are given by $K(f_{0i}, f_i) = \int f_{0i} \log (f_{0i}/f_i)$ and $V(f_{0i}, f_i) = \int f_{0i} (\log (f_{0i}/f_i) - K(f_{0i}, f_i))^2$ respectively. For a set $\Omega$ with semimetric $d$, the $\varepsilon$-covering number $N(\varepsilon, \Omega, d)$ is the minimal number of $d$-balls of radius $\varepsilon$ needed to cover $\Omega$. Meanwhile, the $\varepsilon$-packing number for $\Omega$, denoted by $D(\varepsilon, \Omega, d)$, is the maximal number of $d$ balls of radius $\varepsilon$ needed to cover $\Omega$.

\subsection{Proof of Theorem \ref{Th:2}}

We first prove a lemma that is needed to prove Theorem \ref{Th:2}.

\begin{lemma}[evidence lower bound] \label{Lemma:B1}
	Assume the same setup as Theorem \ref{Th:2}. Then $\sup \mathbb{P}_0(\mathcal{E}_n^c) \rightarrow 0$, where the set $\mathcal{E}_n$ is defined as
	\begin{equation} \label{Enevent}
		\mathcal{E}_n = \left\{ \int \prod_{i=1}^{n} \frac{f_i(y_i)}{f_{0i}(y_i)} d \Pi(\bm{\beta}) \geq e^{-C_1 n \epsilon_n^2} \right\},
	\end{equation}
	for some constant $C_1 > 0$ and $\epsilon_n^2 = s_0 \log G / n$.
\end{lemma}

\begin{proof}
	By Lemma 10 of \citet{GhosalVanDerVaart2007}, this statement will be proven if we can show that
	\begin{equation} \label{KLcondition}
		\Pi \left( \mathcal{B}_n \right) \gtrsim e^{-C_1 n \epsilon_n^2},
	\end{equation}
	where the set $\mathcal{B}_n$ is defined as the event,
	\begin{equation}
		\mathcal{B}_n = \left\{ \frac{1}{n} \sum_{i=1}^{n} K(f_{0i}, f_i) \leq \epsilon_n^2, \frac{1}{n} \sum_{i=1}^{n} V(f_{0i}, f_i) \leq \epsilon_n^2 \right\}. 
	\end{equation}
	As established in Lemma 1 of \citet{JeongGhosal2021}, the KL divergence and KL variation for the exponential family \eqref{exponentialfamily} is given by
	\begin{align*}
		K(f_{0i}, f_i) & = (\theta_{0i} - \theta_i) b'(\theta_{0i}) - b(\theta_{0i}) + b(\theta_i), \\
		V(f_{0i}, f_i) & = b''(\theta_{0i}) (\theta_i - \theta_{0i})^2.
	\end{align*}
	Recalling that $\theta_i = \xi (\mathbf{x}_i^\top \bm{\beta})$ and $\theta_{0i} = \xi(\mathbf{x}_i^\top \bm{\beta}_0)$ where $\xi = (h \circ b')^{-1}$, we have by Taylor expansion in $\mathbf{x}_i^\top \bm{\beta}$ at $\mathbf{x}_i^\top \bm{\beta}_0$ that
	\begin{align*}
		\max \left\{ K(f_{0i}, f_i), V(f_{0i}, f_i) \right\} & \leq b''(\theta_{0i}) (\xi'(\mathbf{x}_i^\top \bm{\beta}_0))^2 (\mathbf{x}_i^\top \bm{\beta} - \mathbf{x}_i^\top \bm{\beta}_0)^2 + o((\mathbf{x}_i^\top \bm{\beta} - \mathbf{x}_i^\top \bm{\beta}_0)^2) \\
		&  = (h^{-1})'(\mathbf{x}_i^\top \bm{\beta}_0) \xi'(\mathbf{x}_i^\top \bm{\beta}_0) (\mathbf{x}_i^\top \bm{\beta} - \mathbf{x}_i^\top \bm{\beta}_0 )^2 + o((\mathbf{x}_i^\top \bm{\beta} - \mathbf{x}_i^\top \bm{\beta}_0)^2).	
	\end{align*} 
	Note that $h^{-1}$ and $\xi$ are strictly increasing. Thus, if $\max_{1 \leq i \leq n} \{ (h^{-1})' (\mathbf{x}_i^\top \bm{\beta}_0) \xi'(\mathbf{x}_i^\top \bm{\beta}_0)  (\mathbf{x}_i^\top \bm{\beta} - \mathbf{x}_i^\top \bm{\beta}_0)^2 \} \leq \epsilon_n^2 \rightarrow 0$, then $| \mathbf{x}_i^\top \bm{\beta} - \mathbf{x}_i^\top \bm{\beta}_0 | \rightarrow 0$ for all $1 \leq i \leq n$. Recall {\pico $\bm{\Omega}(\bm{\beta}_0) = \textrm{diag} \{ (h^{-1})'(\mathbf{X}\bm{\beta}_0) \} \textrm{diag} \{ \xi'(\mathbf{X}\bm{\beta}_0) \}$}. Then both $n^{-1} \sum_{i=1}^{n} K(f_{0i}, f_i)$ and $n^{-1} \sum_{i=1}^{n} V(f_{0i}, f_i)$ can be bounded above by a constant multiple of $n^{-1} \lVert {\pico \bm{\Omega}^{1/2}(\bm{\beta}_0)} \mathbf{X}(\bm{\beta} - \bm{\beta}_0) \rVert_2^2$ on $\mathcal{B}_n$.
	
	Thus, for sufficiently large $n$, we have for some constants $\widetilde{C}_2, C_2 > 0$,
	\begin{align*} \label{PiBnlowerbound1}
		\Pi(\mathcal{B}_n) & \geq \Pi \left( \frac{1}{n} \lVert {\pico \bm{\Omega}^{1/2} (\bm{\beta}_0)} \mathbf{X}(\bm{\beta} - \bm{\beta}_0) \rVert_2^2 \leq \widetilde{C}_2^2  \epsilon_n^2 \right) \\
		& \geq \Pi \left( \lambda_{\max} \left( \frac{1}{n} \mathbf{X}_g^\top {\pico \bm{\Omega}(\bm{\beta}_0)} \mathbf{X}_g \right) \left( \sum_{g=1}^{G} \lVert \bm{\beta}_g - \bm{\beta}_{0g} \rVert_2 \right)^2 \leq \widetilde{C}_2^2 \epsilon_n^2 \right) \\
		& \geq \Pi \left( \sum_{g=1}^{G} \lVert \bm{\beta}_g - \bm{\beta}_{0g} \rVert_2 \leq \frac{C_2 \sqrt{s_0}}{\sqrt{n}} \right) \\
		& \geq \Pi \left( \sum_{g \in S_0} \lVert \bm{\beta}_g - \bm{\beta}_{0g} \rVert_2 \leq \frac{C_2 \sqrt{s_0}}{2 \sqrt{n}} \right) \Pi \left( \sum_{g \in S_0^c} \lVert \bm{\beta}_g - \bm{\beta}_{0g} \rVert_2 \leq \frac{C_2 \sqrt{s_0}}{2 \sqrt{n}} \right) \\
		& \geq \prod_{g \in S_0} \Pi \left(\lVert \bm{\beta}_g - \bm{\beta}_{0g} \rVert_2^2 \leq \frac{C_2^2}{4 n} \right) \prod_{g \in S_0^c} \Pi \left( \lVert \bm{\beta}_{S_0^c} \rVert_2^2 \leq \frac{C_2^2 s_0}{4n (G-s_0)} \right), \numbereqn
	\end{align*} 
	where we used Assumption {\pico (B3)(ii)} in the third inequality and the Cauchy-Schwarz inequality in the fifth inequality. Arguing similarly as in (D.17)-(D.20) in the proof of Theorem 2 of \citet{BaiMoranAntonelliChenBoland2022}, we have that
	\begin{equation*}
		\prod_{g \in S_0} \Pi \left( \lVert \bm{\beta}_g - \bm{\beta}_{0g} \rVert_2^2 \leq \frac{C_2^2}{4n} \right) \geq \theta^{s_0} e^{-\lambda_1 \lVert \bm{\beta}_{S_0} \rVert_2} e^{-\lambda_1 C_2 / 2 \sqrt{n}} \prod_{g \in S_0} \frac{C_g}{m_g!} \left( \frac{\lambda_1 C_2}{2s_0 \sqrt{n}} \right)^{m_g},
	\end{equation*}
	where $C_g = 2^{-m_g} \pi^{-(m_g-1)/2} [ \Gamma((m_g+1)/2)]^{-1}$ and
	\begin{equation*}
		\prod_{g \in S_0^c} \Pi \left( \lVert \bm{\beta}_g \rVert_2^2 \leq \frac{C_2^2 s_0}{4 n (G-s_0)} \right) \geq (1-\theta)^{G-s_0} \left[ 1 - \frac{4 n G m_{\max}(m_{\max}+1)}{\lambda_0^2 C_2^2 s_0} \right]^{G-s_0}.
	\end{equation*}
	Thus, a lower bound on \eqref{PiBnlowerbound1} is
	{\small 
	\begin{align} \label{PiBnlowerbound2}
		\Pi(\mathcal{B}_n)\geq &~\theta^{s_0} (1-\theta)^{G-s_0} e^{-\lambda_1 \lVert \bm{\beta}_{S_0} \rVert_2} e^{-\lambda_1 C_2 / 2 \sqrt{n}} \left[ 1 - \frac{4nG m_{\max}(m_{\max}+1)}{\lambda_0^2 C_2^2 s_0} \right]^{G-s_0} \times \prod_{g \in S_0} \frac{C_g}{m_g!} \left( \frac{\lambda_1 C_2}{2 s_0 \sqrt{n}} \right)^{m_g}. 
	\end{align}}
	Now, since $\lambda_0 = (1-\theta) / \theta \asymp G^c, c>2$, and {\pico $s_0 = o(n^{1/2}/\log G)$} and $m_{\max} = O(\log n \wedge (\log G / \log n))$ by Assumption (A2), we have that
	\begin{align*}
		\left[ 1 - \frac{4nG m_{\max}(m_{\max}+1)}{\lambda_0^2 C_2^2 s_0} \right]^{G-s_0} \gtrsim \left[ 1 - \frac{1}{G-s_0} \right]^{G-s_0} \rightarrow e^{-1} \textrm{ as } n \rightarrow \infty.
	\end{align*} 
	Furthermore, $\theta \asymp \frac{1}{G^c+1}$, and thus,
	\begin{align*}
		(1-\theta)^{G-s_0} \gtrsim \left( 1 - \frac{1}{G-s_0} \right)^{G-s_0} \rightarrow e^{-1}.
	\end{align*}
	Thus, using Assumption (B4) that $\lVert \bm{\beta}_0 \rVert_{\infty} = O(\log G)$ and $\lVert \bm{\beta}_{S_0} \rVert_2 \leq \sqrt{s_0} \lVert \bm{\beta}_0 \rVert_{\infty}$, we can further lower bound \eqref{PiBnlowerbound2} as 
	\begin{align*}
		\Pi(\mathcal{B}_n) \gtrsim (G^c+1)^{-s_0} e^{-\lambda_1 \left( \sqrt{s_0} \log G + C_1 / 2 \sqrt{n} \right)} \prod_{g \in S_0} \frac{C_g}{m_g!} \left( \frac{\lambda_1 C_2}{2 s_0 \sqrt{n}} \right)^{m_g},
	\end{align*}
	and since $\lambda_1 \asymp 1/n$, we have
	\begin{align} \label{PiBnlowerbound3}
		& -\log \Pi(\mathcal{B}_n) \lesssim 2 c s_0 \log G + \sqrt{s_0} \log G + \sum_{g \in S_0} \left[ \log (m_g!) - \log (C_g) + m_g \log \left( \frac{2 s_0 \sqrt{n}}{\lambda_1 C_2} \right) \right]. 
	\end{align}
	Similar to (D.24) in the proof of Theorem 2 of \citet{BaiMoranAntonelliChenBoland2022}, the first term of \eqref{PiBnlowerbound3} can be shown to be the dominating term, and thus, $-\log \Pi (\mathcal{B}_n) \lesssim n \epsilon_n^2$. This establishes \eqref{KLcondition} and the proof is finished. \hfill $\square$
\end{proof}

\paragraph{Proof of Theorem 2}
We follow the proof strategy of Theorem 2 in \citet{JeongGhosal2021} but work with the generalized dimensionality $| \nu (\bm{\beta}) |$ \eqref{generalizeddimensionality}. Let $\mathcal{E}_n$ be the event defined in \eqref{Enevent}, where $\epsilon_n = (s_0 \log G /n)^{1/2}$. Define the set $\mathcal{A}_n = \{ \bm{\beta} : | \nu(\bm{\beta}) | \leq K_1 s_0 \}$, where $K_1 \geq C_1 \vee 1$, and $C_1$ is the constant in Lemma \ref{Lemma:B1}. Then we have
\begin{align*}
	\mathbb{E}_0 \Pi (\mathcal{A}_n^c \mid \mathbf{Y}) \leq \mathbb{E}_0 \Pi \left( \mathcal{A}_n^c \mid \mathbf{Y} \right) \mathbbm{1}_{\mathcal{E}_n} + \mathbb{P}_0 (\mathcal{E}_n^c).
\end{align*}
By Lemma \ref{Lemma:B1}, $\mathbb{P}_0(\mathcal{E}_n^c) \rightarrow 0$ as $n \rightarrow \infty$, so it suffices to show that $\mathbb{E}_0 \Pi ( \mathcal{A}_n^c \mid \mathbf{Y}) \mathbbm{1}_{\mathcal{E}_n} \rightarrow 0$. Note that
\begin{equation} \label{posterioreventAnc}
	\Pi(\mathcal{A}_n^c \mid \mathbf{Y}) = \frac{\int_{\mathcal{A}_n^c} \prod_{i=1}^{n} \frac{f_i(y_i)}{f_{0i}(y_i)} d \Pi (\bm{\beta})}{\int \prod_{i=1}^{n} \frac{f_i(y_i)}{f_{0i}(y_i)} d \Pi(\bm{\beta}) }.
\end{equation}
On the set $\mathcal{E}_n$, the denominator in \eqref{posterioreventAnc} is bounded below by $e^{-C_1 n \epsilon_n^2}$ by Lemma \ref{Lemma:B1}. On the other hand, an upper bound for the expected value of the numerator is
\begin{equation} \label{numeratorupperboundI}
	\mathbb{E}_0 \left(\int_{\mathcal{A}_n^c} \prod_{i=1}^{n} \frac{f_i (y_i)}{f_{0i}(y_i)} d \Pi(\bm{\beta}) \right) \leq \int_{\mathcal{A}_n^c} d \Pi (\bm{\beta}) = \Pi ( | \bm{\beta} | > K_1 s_0).
\end{equation}
Now, since $\pi(\bm{\beta}_g \mid \theta) < 2 \theta C_g \lambda_1^{m_g} e^{-\lambda_1 \lVert \bm{\beta}_g \rVert_2}$ for all $\lVert \bm{\beta}_g \rVert_2 > \omega_g$, we have (arguing as in (D.30) of \citet{BaiMoranAntonelliChenBoland2022}) that
{\small
\begin{align*}
	\Pi(|\nu(\bm{\beta})| > K_1 s_0 ) & \leq \sum_{S: |S| > C_3 s_0} 2^{|S|} \theta^{|S|} \left\{ \prod_{g \in S} \int_{\lVert \bm{\beta}_g \rVert_2 > \omega_g} C_g \lambda_1^{m_g} e^{-\lambda_1 \lVert \bm{\beta}_g \rVert_2} d \bm{\beta}_g \right\}  \times \left\{ \prod_{g \in S^c} \int_{\lVert \bm{\beta}_g \rVert_2 \leq w_g} \pi(\bm{\beta}_g) d \bm{\beta}_g \right\} \\
	& \lesssim \sum_{S: |S| > K_1 s_0} \theta^{|S|}, 
\end{align*}}
where we bounded the first bracketed term from above as $\prod_{g \in S} (1/n)^{m_g} \leq n^{-|S|}$ and the second bracketed term from above by one. Now, since $\theta < 1/G^2$, we have
\begin{align*} \label{numeratorupperboundII}
	\sum_{S:|S| > K_1 s_0} \theta^{|S|} \leq \sum_{k= \lfloor C_1 s_0 \rfloor + 1}^{G} {G \choose k} \left( \frac{1}{G^2} \right)^{k}  \leq \sum_{k = \lfloor K_1 s_0 \rfloor + 1}^{G} \left( \frac{e}{kG} \right)^{k}  & < \sum_{k=\lfloor K_1 s_0 \rfloor +1}^{G} \left( \frac{e}{G \lfloor K_1 s_0 \rfloor + 1} \right)^{k} \\
	& \leq G^{-( \lfloor K_1 s_0 \rfloor +1)} \\
	& \lesssim e^{-K_1 n \epsilon_n^2}. \numbereqn
\end{align*}
Combining \eqref{posterioreventAnc}-\eqref{numeratorupperboundII} and Lemma \ref{Lemma:B1} gives $\mathbb{E}_0 \Pi(\mathcal{A}_n^c \mid \mathbf{Y}) \mathbbm{1}_{\mathcal{E}_n} \prec e^{-(K_1-C_1) n \epsilon_n^2}$ $\rightarrow 0$, since $K_1 > C_1$. This completes the proof. \hfill $\square$

\subsection{Proof of Theorem \ref{Th:3}}

We first prove a lemma which verifies that the prior puts sufficient mass on the event that the generalized dimensionality $\lvert \bm{\nu} (\bm{\beta}) \rvert$ \eqref{generalizeddimensionality} equals $s_0$, where $s_0$ is the true number of nonzero groups in $\bm{\beta}_0$.

\begin{lemma} \label{Lemma:B2}
	Assume the same setup as Theorem \ref{Th:2}. Then for some constant $C_3 > 0$,
	\begin{align*}
		\Pi ( \lvert \bm{\nu} (\bm{\beta}) \rvert = s_0 ) \geq e^{-C_3 n \epsilon_n^2}
	\end{align*}
\end{lemma} 
\begin{proof}
	Note that $\pi(\bm{\beta}_g) \geq \Psi (\bm{\beta}_g \mid \lambda_0)$ for all $\bm{\beta}_g$. Further, for $\bm{\beta}_g \sim \Psi (\bm{\beta}_g \mid \lambda_0)$, we have that $\lVert \bm{\beta}_g \rVert_2$ follows a gamma distribution with shape parameter $m_g$ and scale parameter $\lambda_0$. Thus,
	\begin{align*} \label{Pgreaterthanomegag}
		\Pi(\lVert \bm{\beta}_g \rVert_2 > \omega_g) & \geq \int_{\omega_g}^{\infty} \frac{1}{\Gamma(m_g) \lambda_0^{m_g}}  x^{m_g-1} e^{-x / \lambda_0} dx \\
		& \geq \frac{1}{\Gamma(m_g) \lambda_0^{m_g}} \int_{\lambda_0}^{\infty} x^{m_g-1} e^{-x/\lambda_0} dx \\
		& \geq \frac{1}{\Gamma(m_g) \lambda_0} \int_{\lambda_0}^{\infty} e^{-x / \lambda_0} dx \\
		& = \frac{e^{-1}}{\Gamma(m_g)} \geq \frac{e^{-1}}{\Gamma(m_{\max})} > \frac{e^{-1}}{m_{\max}!}.  \numbereqn
	\end{align*} 
	Meanwhile,
	\begin{align*} \label{Plessthanomegag}
		\Pi (\lVert \bm{\beta}_g \rVert_2 \leq \omega_g) & \geq \int_{0}^{\omega_g} \frac{1}{\Gamma(m_g) \lambda_0^{m_g}}x^{m_g-1} e^{-x/\lambda_0} dx \\
		& = 1 - \int_{\omega_g}^{\infty} \frac{1}{\Gamma(m_g) \lambda_0^{m_g}} x^{m_g-1} e^{-x/\lambda_0} dx  \\
		& \geq 1 - \exp(-\lambda_0 \omega_g + m_{\max}) \\
		& \geq 1 - \frac{1}{G-s_0}, \numbereqn	
	\end{align*}
	where we used the tail bound for the gamma density on p. 29 of \citet{BoucheronLugosiMassart2013} and the inequality $1+x-\sqrt{1+2x} \geq (x-1)/2$ for $x > 0$ (similarly as in \citet{NingJeongGhosal2020}) to bound the integral in the second line from above by $\exp(-\lambda_0 \omega_g + m_{\max})$.
	
	From \eqref{Pgreaterthanomegag}-\eqref{Plessthanomegag}, we can bound $\Pi(|\bm{\nu}(\bm{\beta})| = s_0)$ from below as
	\begin{align*}
		\Pi(|\bm{\nu}(\bm{\beta})| = s_0) & \geq  \left( \frac{e^{-1}}{m_{\max}!} \right)^{s_0} \left( 1 - \frac{1}{G-s_0} \right)^{G-s_0} \\
		& \gtrsim \left( \frac{e^{-1}}{m_{\max}^{m_{\max}}} \right)^{s_0},
	\end{align*}
	where we used the facts that $(1-1/(G-s_0))^{G-s_0} \rightarrow e^{-1}$ as $n \rightarrow \infty$, and $x! \leq x^{x}$ in the second line of the display. Therefore, we see that for sufficiently large $n$,
	\begin{align}
		-\log \Pi(\lvert \bm{\nu}(\bm{\beta}) \rvert = s_0) \leq s_0 m_{\max} \log m_{\max} + s_0 \lesssim s_0 \log G,
	\end{align}
	by Assumptions (A1) and (A2) that $G \gg n$ and $m_{\max} = O(\log n \wedge (\log G / \log n))$. This proves the lemma. \hfill $\square$
\end{proof}

\paragraph{Proof of Theorem \ref{Th:3}}
The proof of Theorem \ref{Th:3} is broken up into two parts and follows the proof strategy of Theorems 2 and 3 of \citet{JeongGhosal2021}. In the first part, we establish the posterior contraction rate in terms of the average squared Hellinger metric. In the second part, we convert the Hellinger contraction rate result to a contraction rate for the regression coefficients themselves.
\vspace{.5cm}

\noindent \emph{Step 1: Hellinger contraction rate}. We first show that there exists $C_4 > 0$ such that
\begin{equation} \label{Hellingercontraction}
	\mathbb{E}_0 \Pi \left( \bm{\beta} \in \mathbb{R}^{p}: H_n(\bm{\beta}, \bm{\beta}_0) > C_4 \epsilon_n \mid \mathbf{Y} \right) \rightarrow 0 \textrm{ as } n \rightarrow \infty.
\end{equation} 
Define the set $\mathcal{A}_n = \{ \bm{\beta} \in \mathbb{R}^{p} : | \bm{\nu}(\bm{\beta}) | \leq K_1 s_0 \}$, where $K_1$ is the constant from Theorem \ref{Th:2}. Then for every $\epsilon> 0$,
\begin{align*}
	& \mathbb{E}_0 \Pi \left( \bm{\beta} \in \mathbb{R}^{p} : H_n(\bm{\beta}, \bm{\beta}_0) > \epsilon \mid \mathbf{Y} \right) \\
	& \qquad \leq \mathbb{E}_0 \Pi \left( \bm{\beta} \in \mathcal{A}_n : H_n(\bm{\beta}, \bm{\beta}_0) > \epsilon \mid \mathbf{Y} \right) \mathbbm{1}_{\mathcal{E}_n} + \mathbb{E}_0 \Pi (\mathcal{A}_n^c \mid \mathbf{Y}) + \mathbb{P}_0 \mathcal{E}_n^c,
\end{align*}
where $\mathcal{E}_n$ is in the event in \eqref{Enevent}. By Lemma \ref{Lemma:B1} and Theorem \ref{Th:2}, the second and third terms on the right-hand side above tend to zero uniformly over $\bm{\beta}_0$. Hence, in order to prove \eqref{Hellingercontraction}, we only focus on the first term. That is, it suffices to show that for $\epsilon = C \epsilon_n$, where $C >0$ and $\epsilon_n = (s_0 \log G /n)^{1/2}$,
\begin{equation} \label{Hellingercontraction2}
	\mathbb{E}_0 \Pi \left( \bm{\beta} \in \mathcal{A}_n : H_n(\bm{\beta}, \bm{\beta}_0) > \epsilon \mid \mathbf{Y} \right) \textrm{ as } n \rightarrow \infty,
\end{equation}
uniformly over $\bm{\beta}_0$. Define the sieve,
\begin{equation} \label{sieve}
	\mathcal{A}_n^{\star} = \left\{ \bm{\beta} \in \mathcal{A}_n : \lVert \bm{\beta} - \bm{\beta}_0 \rVert_2 {\pico \leq G^{M}} \right\},
\end{equation}
{\pico where $M>0$ is defined as in Assumption {\pico (B3)(ii)}.} By Lemma 2 of \citet{GhosalVanDerVaart2007}, there exists a test function $\varphi_n$ such that for any $\bm{\beta}_1$ with $H_n(\bm{\beta}_0, \bm{\beta}_1) > \epsilon$,
\begin{align*}
	\mathbb{E}_0 \varphi_n \leq \exp(-n \epsilon^2/2), \sup_{\bm{\beta} : H_n(\bm{\beta}, \bm{\beta}_1) \leq \epsilon / 18} \mathbb{E}_{\bm{\beta}} (1-\varphi_n) \leq \exp(-n \epsilon^2 /2).
\end{align*}
We want to use Lemma 9 of \citet{GhosalVanDerVaart2007}. In order to do so, we need to show that $\log N (\epsilon_n / 36, \mathcal{A}_n^{\star}, H_n) \lesssim n \epsilon_n^2$ for $\epsilon_n = (s_0 \log G / n)^{1/2}$. Since the squared Hellinger distance is bounded by one half of the KL divergence, we have by a Taylor expansion that for any $\bm{\beta}_1, \bm{\beta}_2 \in \mathcal{A}_n^{\star}$,
\begin{align*}
	 H^2 (f_{\bm{\beta}_1,i}, f_{\bm{\beta}_2,i} ) &  \leq \frac{K(f_{\bm{\beta}_1, i}, f_{\bm{\beta}_2,i})}{2} = \frac{(h^{-1})'(\mathbf{x}_i^\top \bm{\beta}_{1}) \xi'(\mathbf{x}_i^\top \bm{\beta}_1)}{2} (\mathbf{x}_i^\top \bm{\beta}_1 - \mathbf{x}_i^\top \bm{\beta}_2)^2 + o((\mathbf{x}_{i}^\top \bm{\beta}_1 - \mathbf{x}_i^\top \bm{\beta}_2)^2).
\end{align*}
It then follows that
\begin{align*}
	H_n^2(f_{\bm{\beta}_1}, f_{\bm{\beta}_2}) & = \frac{1}{2n} \lVert {\pico \bm{\Omega}^{1/2} (\bm{\beta}_1)} \mathbf{X} (\bm{\beta}_1 - \bm{\beta}_2) \rVert_2^2 + o \left( \lVert \mathbf{X}(\bm{\beta}_1 - \bm{\beta}_2) \rVert_2^2 \right) \\
	& \leq \frac{1}{2} \max_{1 \leq g \leq G} \left( \lambda_{\max} \left( \frac{1}{n} \mathbf{X}_g^\top {\pico \bm{\Omega}(\bm{\beta}_1)} \mathbf{X}_g \right)  \right) \left( \sum_{g=1}^{G} \lVert \bm{\beta}_{1g} - \bm{\beta}_{2g} \rVert_2 \right)^2 + o \left( \lVert \mathbf{X}(\bm{\beta}_1 - \bm{\beta}_2) \rVert_2^2 \right) \\
	& \lesssim G \log G \lVert \bm{\beta}_1 - \bm{\beta}_2 \rVert_2^2 + \lVert \mathbf{X} \rVert_{2, \infty}^2 \lVert \bm{\beta}_1 - \bm{\beta}_2 \rVert_2^2 \\
	& \lesssim G^2 \lVert \bm{\beta}_1 - \bm{\beta}_2 \rVert_2^2,
\end{align*}
where we used Assumption {\pico (B3)(ii)} that $\lambda_{\max}(n^{-1} \mathbf{X}_g^\top {\pico \bm{\Omega}(\bm{\beta}_1)} \mathbf{X}_g) \lesssim \log G$ and the fact that all of the entries of $\mathbf{X}$ are uniformly bounded, and thus, $\lVert \mathbf{X} \rVert_{2, \infty} \lesssim G^{1/2}$. Thus, for some $C_5 > 0$, we can bound $N(\epsilon_n /36, \mathcal{A}_n^{\star}, H_n)$ from above by
\begin{align} \label{entropyupperbound1}
	N(\epsilon_n /36, \mathcal{A}_n^{\star}, H_n) & \leq N \left( \frac{C_5 \epsilon_n}{36 G}, \mathcal{A}_n^{\star}, \lVert \cdot \rVert_2 \right)  = N \left( \frac{C_5 \epsilon_n}{36G}, \left\{ \bm{\beta} \in \mathcal{A}_n, \lVert \bm{\beta} - \bm{\beta}_0 \rVert_2 \leq G^{{\pico M}} \right\}, \lVert \cdot \rVert_2 \right).
\end{align}
Denote by $\widetilde{\omega}$, 
\begin{align*} 
	\widetilde{\omega} = \frac{1}{\lambda_0 - \lambda_1} \log \left[ \frac{1-\theta}{\theta} \frac{\lambda_0^{m_{\max}}}{\lambda_1^{m_{\max}}} \right],
\end{align*}
It is clear that given our choice of hyperparameters for $(\theta, \lambda_0, \lambda_1)$, the threshold $\omega_g$ in \eqref{omegag} is an increasing function of $m_g$. Therefore,
\begin{align*}
	 \{ \bm{\beta} \in \mathcal{A}_n, \lVert \bm{\beta} - \bm{\beta}_0 \rVert_2 \leq G^{{\pico M}} \} & \subset \left\{ \bm{\beta} \in \mathbb{R}^{p} : \lVert \bm{\beta}_{S_0} - \bm{\beta}_{0S_0} \rVert_2 \leq G^{{\pico M}}, \textrm{ and } \lVert \bm{\beta}_g \rVert_2 \leq \widetilde{\omega} \textrm{ for all } g \in S_0^c \right\} \\
	& \subset \left\{  \lVert \bm{\beta}_{S_0} - \bm{\beta}_{0S_0} \rVert_2 \leq G^{{\pico M}} \right\} \times [-\widetilde{\omega}, \widetilde{\omega} ]^{G-K_1 s_0}.
\end{align*} 
By Lemma S4 of \citet{ShenSolisLemusDeshpande2022}, for any set of the form $E = A \times [-\delta, \delta]^{Q-s} \subset \mathbb{R}^{Q}$ where $A \subset \mathbb{R}^{s}$ and $s < Q$, 
\begin{align*}
	D (\epsilon, E, \lVert \cdot \rVert_2) \leq D( (1-T^{-1})^{1/2} \epsilon, A, \lVert \cdot \rVert_2 ),
\end{align*}
if $\delta < \epsilon / (2[T(Q-s)]^{1/2})$, for some $T>1$, where $D$ denotes the packing number. In order to use Lemma A4 of \citet{ShenSolisLemusDeshpande2022}, we can check that for sufficiently large $n$ and $T=2$,
\begin{align*}
	\widetilde{\omega} = \frac{1}{\lambda_0 - \lambda_1} \log \left[ \frac{1-\theta}{\theta} \frac{\lambda_0^{m_{\max}}}{\lambda_1^{m_{\max}}}  \right] \lesssim \frac{ m_{\max} \log G}{G^2} < \frac{C_5 \epsilon_n / 36 G}{2 [ 2 T(G-s_0)]^{1/2}},
\end{align*} 
where we used Assumptions (A1)-(A2) and the fact that $\lambda_0 = (1-\theta) / \theta \asymp p^{c}, c >2$, and $\lambda_1 \asymp 1/n$. Hence, by Lemma S4 of \citet{ShenSolisLemusDeshpande2022} and the fact that we can upper bound the covering number by the packing number, we can further upper bound \eqref{entropyupperbound1} by 
\begin{align*} \label{entropyupperbound2}
	& {G \choose K_1 s_0} D \left( \frac{C_5 \epsilon_n}{36G \sqrt{2}}, \left\{ \lVert \bm{\beta}_S - \bm{\beta}_{0S} \rVert_2 \leq G^{{\pico M}} \right\}, \lVert \cdot \rVert_2 \right) \leq {G \choose K_1 s_0} \left( \frac{108 \sqrt{2} G^{{\pico M}+1}}{C_5 \epsilon_n} \right)^{K_1 s_0}, \numbereqn
\end{align*} 
noting that on the set $\mathcal{A}_n$, $|S| \leq K_1 s_0$. From \eqref{entropyupperbound1}-\eqref{entropyupperbound2}, we have that
\begin{align} \label{metricentropybound}
	\log N( \epsilon_n / 36, \mathcal{A}_n^{\star}, H_n ) & \leq K_1 s_0 \log G + K_1 s_0 \left[ ({\pico M} + 1) \log G + \log (108 \sqrt{2n} ) \right]  \lesssim n \epsilon_n^2, 
\end{align}  
where we used the fact that ${G \choose K_1 s_0} \leq G^{K_1 s_0}$. Having established \eqref{metricentropybound}, we can now use Lemma 9 of \citet{GhosalVanDerVaart2007}, which implies that for every $\epsilon > \epsilon_n$, there exists a test $\bar{\varphi}_n$ such that
\begin{align*}
	\mathbb{E}_0 \bar{\varphi}_n \leq \frac{1}{2} \exp \left( C_6 n \epsilon_n^2 - n \epsilon / 2 \right), \sup_{\bm{\beta} \in \mathcal{A}_n^{\star} : H_n (\bm{\beta}, \bm{\beta}_0) > \epsilon} \mathbb{E}_{\bm{\beta}} (1- \bar{\varphi}_n) \leq \exp(- n \epsilon^2/2).
\end{align*}
for some $C_6 > 0$. By Lemma \ref{Lemma:B2}, $\Pi(| \bm{\nu}(\bm{\beta}) | = s_0) \geq e^{-C_3 n \epsilon_n^2}$ for some $C_3 > 0$, and thus, 
\begin{align*}
	& \mathbb{E}_0 \Pi \left( \bm{\beta} \in \mathcal{A}_n : H_n (\bm{\beta}, \bm{\beta}_0) > \epsilon \mid \mathbf{Y} \right) \\
	& \qquad \leq \mathbb{E}_0 \Pi \left( \bm{\beta} \in \mathcal{A}_n : H_n (\bm{\beta}, \bm{\beta}_0) > \epsilon \mid \mathbf{Y} \right) \mathbbm{1}_{\mathcal{E}_n} (1- \bar{\varphi}_n) + \mathbb{E}_0 \bar{\varphi}_n \\
	& \qquad \leq \left\{ \sup_{\bm{\beta} \in \mathcal{A}_n^{\star}: H_n (\bm{\beta}, \bm{\beta}_0) > \epsilon} \mathbb{E}_{\bm{\beta}} (1-\bar{\varphi}_n) + \Pi(\mathcal{A}_n \setminus \mathcal{A}_n^{\star}) \right\} e^{C_3 n \epsilon_n^2} + \mathbb{E}_0 \bar{\varphi}_n.
\end{align*}
All of the terms in the display except $\Pi ( \mathcal{A}_n \setminus \mathcal{A}_n^{\star} ) e^{C_3 n \epsilon_n^2}$ go to zero by choosing $\epsilon = C_4 \epsilon_n$ for a sufficiently large $C_4 > C_3$. To complete the proof then, we must show that
\begin{equation} \label{completeHellingerproof}
	\Pi(\mathcal{A}_n \setminus \mathcal{A}_n^{\star}) e^{C_3 n \epsilon_n^2} \rightarrow 0 \textrm{ as } n \rightarrow \infty.
\end{equation}
To establish \eqref{completeHellingerproof}, note that
\begin{align*} \label{completeHellingerproofPtII}
	\Pi(\mathcal{A}_n \setminus \mathcal{A}_n^{\star}) & = \Pi ( \bm{\beta} \in \mathbb{R}^{p} : | \bm{\nu}(\bm{\beta}) | \leq K_1 s_0, \lVert \bm{\beta} - \bm{\beta}_0 \rVert_2 > G^{{\pico M}} ) \\
	& \leq \Pi ( \lVert \bm{\beta} - \bm{\beta}_0 \rVert_2 > G^{{\pico M}} ) \\
	& \leq \sum_{g \in S_0} \Pi \left( \lVert \bm{\beta}_g \rVert_2 > G^{{\pico M}} - \lVert \bm{\beta}_{0g} \rVert_2 \right) + \sum_{g \in S_0^c} \Pi \left( \lVert \bm{\beta}_g \rVert_2 > G^{{\pico M}} \right). \numbereqn
\end{align*}
We examine each of the terms in \eqref{completeHellingerproofPtII} separately. Notice that $\pi(\bm{\beta}_g ) \leq \Psi (\bm{\beta}_g \mid \lambda_1)$ for all $\bm{\beta}_g$. Moreover, for $\Psi (\bm{\beta}_g \mid \lambda_1)$, $\lVert \bm{\beta}_g \rVert_2$ follows a gamma distribution with shape parameter $m_g$ and scale parameter $\lambda_1$ Letting $X_g$ denote a gamma random variable with shape and scale parameters $m_g$ and $\lambda_1$, and recalling that ${\pico M} \geq 1$, we have
\begin{align*} \label{completeHellingerproofPtIII}
	\sum_{g \in S_0} \Pi \left( \lVert \bm{\beta}_g \rVert_2 > G^{{\pico M}} - \lVert \bm{\beta}_{0g} \rVert_2 \right) & \leq \sum_{g \in \mathcal{S}_0} P(X_g >  G^{{\pico M}} - \lVert \bm{\beta}_{0g} \rVert_2) \\
	& \leq \sum_{g \in \mathcal{S}_0} P(X_g >  G^{{\pico M}} - {\pico C_7} \sqrt{G} \log G ) \\
	&  \leq \sum_{g \in S_0} \exp \left[ - \lambda_1 (G^{{\pico M}} - {\pico C_7} \sqrt{G} \log G) + m_{\max} \right] \\
	&  = \exp \left[ - \lambda_1 G^{{\pico M}} + \lambda_1 {\pico C_7} \sqrt{G} \log G + m_{\max} + \log s_0 \right], \numbereqn
\end{align*} 
where in the second line, we used the fact that by Assumption (B4), $\lVert \bm{\beta}_{0g} \rVert_2 \leq \sqrt{G} \lVert \bm{\beta}_0 \rVert_{\infty} \leq {\pico C_7} \sqrt{G} \log G$, for some constant ${\pico C_7} > 0$. In the third line of the display, we used same gamma density tail bound that we used to prove \eqref{Plessthanomegag}. Using a similar argument as \eqref{completeHellingerproofPtIII}, we also have
\begin{align*} \label{completeHellingerproofPtIV}
	\sum_{g \in S_0^c} \Pi \left( \lVert \bm{\beta}_g \rVert_2 > G^{{\pico M}} \right) & \leq \sum_{g \in S_0^c} P(X_g > G^{{\pico M}}) \\
	& \leq \sum_{g \in S_0^c} \exp \left[ -\lambda_1 G^{{\pico M}} + m_{\max} \right] \\
	& = \exp \left[ -\lambda_1 G^{{\pico M}} + m_{\max} + \log (G- s_0) \right]. \numbereqn
\end{align*} 
Combining \eqref{completeHellingerproofPtII}-\eqref{completeHellingerproofPtIV}, we have that for sufficiently large $n$,
\begin{align} \label{completeHellingerproofPtV}
	\Pi (\mathcal{A}_n \setminus \mathcal{A}_n^{\star}) & \leq 2 \exp \left[ -\lambda_1 G^{{\pico M}} + \lambda_1 {\pico C_7} \sqrt{G} \log G + \log G + m_{\max} \right]  \lesssim \exp(-C_4 s_0 \log G), 
\end{align}
where we can choose $C_4 > C_3$ so that $-\lambda_1 G^{{\pico M}}$ is the dominating term in the first line of the display, and further, $G^{{\pico M}} \gg C_4 s_0 n \log G$ when $n$ is large. Thus, from \eqref{completeHellingerproofPtV}, we can see that \eqref{completeHellingerproof} holds, and we have established the posterior contraction rate with respect to the average squared Hellinger metric \eqref{Hellingercontraction}.
\vspace{.5cm}

\noindent \emph{Step 2: Contraction rate for the regression coefficients}. We first define the uniform compatibility number $\phi_1$ as
\begin{align*}
	\phi_1 (s; {\pico \bm{\Omega}}) = \inf_{\bm{\beta} : 1 \leq | \bm{\nu}(\bm{\beta}) | \leq s} \frac{\lVert {\pico \bm{\Omega}^{1/2}} \mathbf{X}\bm{\beta} \rVert_2 | \bm{\nu} (\bm{\beta}) |^{1/2}}{n^{1/2} \lVert \bm{\beta} \rVert_1},
\end{align*}
and the smallest scaled singular value $\phi_2$ as
\begin{align*}
	\phi_2(s; {\pico \bm{\Omega}}) = \inf_{\bm{\beta}: 1 \leq | \bm{\nu} (\bm{\beta}) | \leq s} \frac{\Vert {\pico  \bm{\Omega}^{1/2}} \mathbf{X} \bm{\beta} \rVert_2}{n^{1/2} \lVert \bm{\beta} \rVert_2}.
\end{align*}
According to Theorem 3 of \citet{JeongGhosal2021}, as long as
\begin{equation} \label{sufficientconditionHellingertoL2}
	\frac{s_0^2 \log G \lVert \mathbf{X} \rVert_{\max}}{\phi_1^2 \left( (K_2+1)s_0; {\pico \bm{\Omega}(\bm{\beta}_0)} \right)} = o(n),
\end{equation}
then for some $C_8 > 0$,
\begin{equation} \label{HellingertoL2}
	\left\{ \bm{\beta} : H_n(\bm{\beta}, \bm{\beta}_0) \leq C_4 \epsilon_n \right\} \subset \left\{ \bm{\beta} : \lVert \bm{\beta} - \bm{\beta}_0 \rVert_2 \leq \frac{{\pico C_8} \epsilon_n}{\phi_2 \left( (K_2+1) s_0 ; {\pico \bm{\Omega}(\bm{\beta}_0)} \right) } \right\},
\end{equation}
where $\epsilon_n = (s_0 \log G /n)^{1/2}$. Notice that for any set $S$ such that $s := |S| \leq (K_2+1) s_0$, we have
\begin{align*} \label{phillowerbound}
	\phi_2 ( s;  {\pico \bm{\Omega}(\bm{\beta}_0)}) & = \inf_{\bm{\beta}: 1 \leq | \bm{\nu}(\bm{\beta}) | \leq s} \frac{\lVert {\pico \bm{\Omega}^{1/2} (\bm{\beta}_0)} \mathbf{X} \bm{\beta} \rVert_2}{\lVert \bm{\beta} \rVert_2 n^{1/2}} \\
	& \geq \frac{[\lambda_{\min} \left( \mathbf{X}_S^\top {\pico \bm{\Omega}(\bm{\beta}_0)} \mathbf{X}_S \right) ]^{1/2} \lVert \bm{\beta} \rVert_2}{\lVert \bm{\beta} \rVert_2 n^{1/2}} \\
	& = \left[ \lambda_{\min} \left( \frac{1}{n} \mathbf{X}_S^\top {\pico \bm{\Omega}(\bm{\beta}_0)} \mathbf{X}_S \right) \right]^{1/2} {\pico \gtrsim}~ 1, \numbereqn
\end{align*}
by Assumption {\pico (B3)(ii)}. We also have {\pico $\lVert \mathbf{X} \rVert_{\max} \lesssim \log G$} by Assumption {\pico (B3)(i)}, and $\phi_1(s; {\pico \bm{\Omega}}) \geq \phi_2 (s; {\pico \bm{\Omega}})$. {\pico It follows from \eqref{phillowerbound} that}
\begin{align*}
	\frac{s_0^2 \log G \lVert \mathbf{X} \rVert_{\max}}{\phi_1^2 \left( (K_2+1)s_0; {\pico \bm{\Omega}(\bm{\beta}_0)} \right)} \lesssim \frac{s_0^2 {\pico (\log G)^2}}{\phi_2^2 \left( (K_2 +1) s_0 ; {\pico \bm{\Omega}(\bm{\beta}_0)} \right) } \lesssim s_0^2 {\pico (\log G)^2}.
\end{align*}
Thus, by Assumption (A2) that $s_0 = {\pico o(n^{1/2} / \log G)}$, it must be that \eqref{sufficientconditionHellingertoL2} holds. Thus, for some $K_3 > 0$, we have by \eqref{HellingertoL2} and Theorem \ref{Th:2} that
\begin{align*}
	\mathbb{E}_0 \Pi \left( \bm{\beta} : \lVert \bm{\beta} - \bm{\beta}_0 \rVert_2 \leq K_3 \epsilon_n \mid \mathbf{Y} \right) \geq \Pi ( \bm{\beta} : H_n (\bm{\beta}, \bm{\beta}_0 ) \leq C_4 \epsilon_n ) \rightarrow 1 \textrm{ as } n \rightarrow \infty,
\end{align*}
where the smallest scaled singular value $\phi_2 ( (K_2+1) s_0 ; {\pico \bm{\Omega}(\bm{\beta}_0)})$ is a {\pico positive constant (by \eqref{phillowerbound})} that is absorbed into the constant $K_3$. We are done.	\hfill $\square$

\subsection{Proof of Proposition \ref{Prop:1}}

{\picotwo 
	Throughout, we use $P_0$ and $\mathbb{E}_0$ to respectively denote the probability and expectation operators under the law induced by the model $\mathbf{Y} \sim \mathcal{N}_n(\boldsymbol{\beta}_0, \mathbf{I}_n)$.
	
	The posterior contraction proof in Proposition \ref{Prop:1} follows a similar strategy as the proof of Theorem 7 in \citet{CastilloSchmidtHieberVanDerVaart2015}. However, it should be noted that \citet{CastilloSchmidtHieberVanDerVaart2015} deal exclusively with the case where $\boldsymbol{\beta}_0 = (\beta_{01}, \ldots,\beta_{0n})^\top$ is the zero vector, i.e. $\boldsymbol{\beta}_0 = \boldsymbol{0}_n$, and the \emph{individual} regression coefficients $\beta_{0i}, i = 1, \ldots, n$, are endowed with \emph{univariate} $\text{Laplace}(0, \lambda^{-1})$ priors. In contrast, we consider the \textit{grouped} scenario where $\boldsymbol{\beta}_0 = ( \boldsymbol{\beta}_{01}^\top, \ldots, \boldsymbol{\beta}_{0G}^\top )^\top$ with $\boldsymbol{\beta}_{01} \neq \boldsymbol{0}_{m_1}$, and the groups $\boldsymbol{\beta}_{0g}, g = 1, \ldots, G$, are endowed with independent \emph{multivariate} Laplace priors \eqref{multivariateLaplace}.
	
	\paragraph{Proof of Proposition \ref{Prop:1}}
	
	We first prove \eqref{group-lasso-MAP-risk}. Let $\mathbf{Y} = (\mathbf{Y}_1^\top, \ldots, \mathbf{Y}_G^\top)^\top$, where the indices of $\mathbf{Y}_g$ correspond to those of $\boldsymbol{\beta}_{0g}$ in $\boldsymbol{\beta}_0 = (\boldsymbol{\beta}_{01}^\top, \ldots, \boldsymbol{\beta}_{0G}^\top)^\top$ for all $g = 1, \ldots, G$. When $\mathbf{Y} \sim \mathcal{N}_n ( \boldsymbol{\beta}_0, \mathbf{I}_n)$, the group lasso estimator has a closed form,
	\begin{equation*}
		\widehat{\boldsymbol{\beta}}_g = \left( 1 - \frac{\lambda}{\lVert \mathbf{Y}_g \rVert_2} \right)_{+} \mathbf{Y}_g, ~ \text{ for all } ~ g = 1, \ldots, G,
	\end{equation*} 
	where $x_{+} = \min \{ x, 0 \}$. Therefore, we must have $\lVert \widehat{\boldsymbol{\beta}}_g \rVert_2 < \lVert \mathbf{Y}_g \rVert_2$ for all $g \in \{ 1, \ldots, G \}$. We can decompose the expected squared $\ell_2$ risk as
	\begin{align} \label{totalriskdecomposition}
		\mathbb{E}_0 \lVert \widehat{\boldsymbol{\beta}} - \boldsymbol{\beta}_0 \rVert_2^2 = \mathbb{E}_0 \lVert \widehat{\boldsymbol{\beta}}_1 - \boldsymbol{\beta}_{01} \rVert_2^2 + \sum_{g = 2}^{G} \mathbb{E}_0 \lVert \widehat{\boldsymbol{\beta}}_g \rVert_2^2,
	\end{align}
	We first upper-bound the first term in \eqref{totalriskdecomposition}. Since $\mathbf{Y}_1 \sim \mathcal{N}_m (\boldsymbol{\beta}_{01}, \mathbf{I}_m)$, $\lVert \widehat{\boldsymbol{\beta}}_1 \rVert_2 < \lVert \mathbf{Y}_1 \rVert_2$, and $\widehat{\boldsymbol{\beta}}_1 = \boldsymbol{0}_m$ when $\lVert \mathbf{Y}_1 \rVert_2 \leq \lambda$, we have
	\begin{align*} \label{upperboundrisk1}
		\mathbb{E}_0 \lVert \widehat{\boldsymbol{\beta}}_1 - \boldsymbol{\beta}_{01} \rVert_2^2 & \leq 2 \mathbb{E}_0 \lVert \mathbf{Y}_1 - \boldsymbol{\beta}_{01} \rVert_2^2 +  2 \mathbb{E}_{0} \lVert \widehat{\boldsymbol{\beta}}_1 - \mathbf{Y}_1 \rVert_2^2 \\
		& \leq 2m + 2 \mathbb{E}_0 \{ \lVert \mathbf{Y}_1 \rVert_2^2 \mathbb{I}( \lVert \mathbf{Y}_1 \rVert_2 \leq \lambda) \} + 2 \mathbb{E}_0 \{ \lambda^2 \mathbb{I}(\lVert \mathbf{Y}_1 \rVert_2 > \lambda) \} \\
		& \leq 2m +  2 \lambda^2 + 2 \lambda^2 P_0 (  \lVert \mathbf{Y}_1 \rVert_2 > \lambda ) \\
		& \leq 2 m + 4 \lambda^2  \asymp 2 \log n, \numbereqn 
	\end{align*} 
	since $m$ is constant and $\lambda \asymp \sqrt{2 \log n}$. Next, we upper-bound the second term in \eqref{totalriskdecomposition}. Again, using the fact that $\lVert \widehat{\boldsymbol{\beta}}_g \rVert_2 < \lVert \mathbf{Y}_g \rVert_2$ and $\widehat{\boldsymbol{\beta}}_g = \boldsymbol{0}_m$ when $\lVert \mathbf{Y}_g \rVert_2 \le \lambda$, each of the summands can be upper-bounded as
	\begin{align*} 
		\mathbb{E}_0 \lVert \widehat{\boldsymbol{\beta}}_g \rVert_2^2 & \leq \mathbb{E}_0 \{ \lVert \mathbf{Y}_g \rVert_2^2 \mathbb{I}(\lVert \mathbf{Y}_g \rVert_2^2 > \lambda^2 ) \}.
	\end{align*}
	Since $\mathbf{Y}_g \sim \mathcal{N}_m(\boldsymbol{0}_m, \mathbf{I}_m)$ for all $g \neq 1$, $\lVert \mathbf{Y}_g \rVert_2^2 \sim \chi_m^2$. Letting $\Gamma(a,x) = \int_{x}^{\infty} t^{a-1} e^{-t} dt$ denote the incomplete gamma function, we proceed to upper-bound the individual summands as 
	\begin{align*} \label{upperboundrisk2}
		\mathbb{E}_0 \lVert \widehat{\boldsymbol{\beta}}_g \rVert_2^2 & \leq \int_{\lambda^2}^{\infty} \left[ \Gamma \left( \frac{m}{2} \right) \right]^{-1} \left( \frac{u}{2} \right)^{m/2} e^{-u/2} du \\
		& = 2 \left[ \Gamma(\frac{m}{2}) \right]^{-1} \Gamma \left( \frac{m}{2} +1, \lambda^2 \right) \\
		& \leq 2 \left[ \Gamma \left(\frac{m}{2} \right) \right]^{-1} \left( \frac{m}{2} +1 \right) e^{-\lambda^2} (\lambda^2)^{m/2} ~ \text{ for large } n \\ 
		& \asymp n^{-2}(\log n)^{m/2}, \numbereqn
	\end{align*}
	where we used the bound $\Gamma(a, x) \leq  a e^{-x} x^{a-1}$ when $a \geq 1$ and $ x > a$ \citep{Pinelis2020} in the third line, and the fact that $m$ is fixed and $\lambda \asymp \sqrt{2 \log n}$ in the last line. Combining \eqref{totalriskdecomposition}-\eqref{upperboundrisk2} gives
	\begin{align*}
		\mathbb{E}_0 \lVert \widehat{\boldsymbol{\beta}} - \boldsymbol{\beta}_0 \rVert_2^2  \lesssim 2 \log n + (G-1) n^{-2} (\log n)^{m/2} = 2 \log n + o(1) ~ \text{ as } n \rightarrow \infty, 
	\end{align*}
	where we used the fact that $G = n/m$ and $m$ is a constant. This completes the proof of \eqref{group-lasso-MAP-risk}.
	
	Next, we prove \eqref{group-lasso-posterior-contraction}. We follow the proof strategy of Theorem 7 in \citet{CastilloSchmidtHieberVanDerVaart2015}. According to Lemma 7.1 of \cite{CastilloVanDerVaart2012}, $\mathbb{E}_0 \Pi ( \boldsymbol{\beta} : \lVert \boldsymbol{\beta} - \boldsymbol{\beta}_0 \rVert_2 < s_n \mid \mathbf{Y} ) \rightarrow 0$ for any $s_n$ for which there exists a sequence $r_n$ such that
	\begin{equation} \label{CV12}
		\frac{\Pi ( \boldsymbol{\beta}:  \lVert \boldsymbol{\beta} - \boldsymbol{\beta}_0 \rVert_2 < s_n)}{\Pi ( \boldsymbol{\beta}: \lVert \boldsymbol{\beta} - \boldsymbol{\beta}_0 \rVert_2 < r_n )} e^{ 2 r_n^2} = o(1).
	\end{equation} 
	Since the prior is $\boldsymbol{\beta}_g \overset{iid}{\sim} \Psi(\boldsymbol{\beta}_g \mid \lambda)$ for all $g = 1, \ldots, G$, $\lVert \boldsymbol{\beta}_1 \rVert_2 \sim \text{Gamma}(m, \lambda)$ and $\sum_{g=1}^{G} \lVert \boldsymbol{\beta}_g \rVert_2 \sim \text{Gamma}(n, \lambda)$ where $n = mG$. 
	
	We first upper-bound the numerator in \eqref{CV12}. Since $\lVert \boldsymbol{\beta}_0 \rVert_2 = \lVert \boldsymbol{\beta}_{01} \rVert_2 = L$ and $\lVert \boldsymbol{\beta} - \boldsymbol{\beta}_0 \rVert_2 \geq \lVert \boldsymbol{\beta}_1 - \boldsymbol{\beta}_{01} \rVert_2$, we have that for sufficiently large $n$ such that $s_n > L$,
	\begin{align*} \label{CV12numbound}
		\Pi ( \boldsymbol{\beta} : \lVert \boldsymbol{\beta} - \boldsymbol{\beta}_0 \rVert_2 < s_n ) & \leq \Pi ( \boldsymbol{\beta}_1 :  \lVert \boldsymbol{\beta}_1 - \boldsymbol{\beta}_0 \rVert_2 < s_n ) \\
		& \leq \Pi \left( \boldsymbol{\beta}_1 : \big| \lVert \boldsymbol{\beta}_1 \rVert_2 - \lVert \boldsymbol{\beta}_{01} \rVert_2 \big| < s_n \right) \\
		& = \Pi ( \boldsymbol{\beta}_1 : \lVert \boldsymbol{\beta}_1 \rVert_2 < s_n + L ) \\
		& = \int_{0}^{s_n+L} \frac{\lambda^m}{\Gamma(m)} u^{m-1} e^{-\lambda u} du \\
		& \leq \frac{[\lambda(s_n + L)]^{m}}{\Gamma(m+1)}. \numbereqn
	\end{align*}
	Next, we lower-bound the denominator in \eqref{CV12}. For sufficiently large $n$ so that $r_n > L$, we have
	\begin{align*} \label{CV12denbound}
		\Pi ( \boldsymbol{\beta} : \lVert \boldsymbol{\beta} - \boldsymbol{\beta}_0 \rVert_2 < r_n ) & \geq \Pi ( \boldsymbol{\beta}: \lVert \boldsymbol{\beta} \rVert_2 + \lVert \boldsymbol{\beta}_0 \rVert_2 < r_n ) \\
		& = \Pi ( \boldsymbol{\beta} : \lVert \boldsymbol{\beta} \rVert_2 < r_n - L ) \\
		& \geq \Pi \left(  \boldsymbol{\beta} : \sum_{g=1}^{G} \lVert \boldsymbol{\beta}_g \rVert_2 < r_n - L \right) \\
		& = \int_{0}^{r_n-L} \frac{\lambda^n}{\Gamma(n)} u^{n-1} e^{-\lambda u} du \\
		& \geq \frac{\lambda^n}{n \Gamma(n)} e^{-\lambda (r_n - L)} (r_n-L)^{n}. \numbereqn
	\end{align*} 
	Combining \eqref{CV12}-\eqref{CV12denbound} gives
	\begin{align*} \label{CV12pt2}
		\frac{\Pi ( \boldsymbol{\beta} : \lVert \boldsymbol{\beta} - \boldsymbol{\beta}_0 \rVert_2 < s_n )}{\Pi ( \boldsymbol{\beta} : \lVert \boldsymbol{\beta} - \boldsymbol{\beta}_0 \rVert_2 < r_n)} & \leq n \lambda^{m-n} \frac{\Gamma(n)}{\Gamma(m+1)} \frac{(s_n+L)^{m}}{(r_n-L)^{n}} e^{\lambda (r_n - L)} \\
		& \lesssim e^{\lambda \sqrt{n} r_n} \left( \frac{s_n}{r_n} \right)^{n} \\
		& = \exp \left[ \lambda \sqrt{n} r_n - n \log \left( \frac{r_n}{s_n} \right) \right]. \numbereqn 
	\end{align*}
	Setting $r_n = \sqrt{n} (\lambda^{-1} \wedge 1)$ and $s_n = \delta r_n$ for $\delta$ small enough in \eqref{CV12pt2} satisfies \eqref{CV12}, and thus, by Lemma 7.1 of \citet{CastilloVanDerVaart2012},
	\begin{align*}
		\mathbb{E}_0 \Pi \left( \boldsymbol{\beta} : \lVert \boldsymbol{\beta} - \boldsymbol{\beta}_0 \rVert_2 < \delta \sqrt{n} \left( \frac{1}{\lambda} \wedge 1 \right) ~\bigg|~ \mathbf{Y} \right) \rightarrow 0 ~ \textrm{ as } ~ n \rightarrow \infty.
	\end{align*}
	By setting $\lambda \asymp \sqrt{2 \log n}$ in the above display, we obtain the posterior contraction rate of $\sqrt{n / \log n}$ in \eqref{group-lasso-posterior-contraction}. \hfill $\square$
}	

\section{Additional details for the EM algorithm in Section \ref{EMalgorithm}} \label{App:C}

\subsection{Derivation of the EM algorithm}

Recall that the log-likelihood function $\ell_n(\bm{\beta})$ is given in \eqref{ll}. With the reparameterization of the hierarchical SSGL prior in \eqref{SSGLrepar} and the $\mathcal{B}(a,b)$ prior \eqref{thetaprior} on $\theta$, we can write the log-posterior $\log \pi(\bm{\beta}, \theta \mid \mathbf{Y})$ in \eqref{logposteriorII} as
\begin{align*} \label{logposterior}
	\log \pi ( \bm{\beta}, \theta \mid \mathbf{Y} ) = &~ \ell_n (  \bm{\beta} )  + \displaystyle \sum_{g=1}^{G} \log \left( (1-\gamma_g ) \lambda_0^{m_g} e^{-\lambda_0 \lVert \bm{\beta}_g \rVert_2} + \gamma_g \lambda_1^{m_g} e^{-\lambda_1 \lVert \bm{\beta}_g \rVert_2} \right) \\
	&~ + \left(a-1+\sum_{g=1}^{G} \gamma_g \right) \log \theta + \left(b-1 + p -\sum_{g=1}^{G} \gamma_g \right) \log (1-\theta). \numbereqn
\end{align*}
From \eqref{logposterior}, it is straightforward to verify that $\mathbb{E}[\gamma_g \mid \mathbf{Y}, \bm{\beta}, \theta ] = p_g^{\star} (\bm{\beta}_g, \theta)$, where $p_g^{\star} (\bm{\beta}_g, \theta)$ is as in \eqref{Estep}.

In the E-step of our EM algorithm, we treat $\bm{\gamma}$ as missing data and take expectation with respect to $\gamma_g$ for all $g = 1, \ldots, G$, holding the parameters $(\bm{\beta}, \theta)$ fixed at their previous values. That is, we compute $p_g^{\star (t-1)} := p^{\star} ( \bm{\beta}_g^{(t-1)}, \theta^{(t-1)} ) = \mathbb{E} [ \gamma_g \mid \mathbf{Y}, \bm{\beta}^{(t-1)}, \theta^{(t-1)} ]$ for all $g = 1, \ldots, G$, given the previous estimates $(\bm{\beta}^{(t-1)}, \theta^{(t-1)})$. 

For the M-step, we then maximize the following function:
\begin{align*} \label{Mstep}
	& \mathbb{E} [ \log \pi ( \bm{\beta}, \theta \mid \mathbf{Y} ) \mid \bm{\beta}^{(t-1)}, \theta^{(t-1)} ] =  \ell_n ( \bm{\beta}) - \sum_{g=1}^{G} \lambda_g^{\star (t-1)} \lVert \bm{\beta}_g \rVert_2  \\
	& \qquad + \left(a-1 + \sum_{g=1}^{G} p_g^{\star (t-1)} \right) \log \theta + \left(b-1+ G -\sum_{g=1}^{G} p_g^{\star (t-1)} \right) \log (1- \theta),  \numbereqn
\end{align*}
where $\lambda_g^{\star (t-1)} = \lambda_1 p_g^{\star (t-1)} + \lambda_0 (1- p_g^{\star (t-1)})$. From \eqref{Mstep}, it is easy to derive the update for $\theta$ in \eqref{thetaupdate} by taking the derivative of \eqref{Mstep} with respect to $\theta$. The update for $\bm{\beta}$ in \eqref{betaupdate} is obtained by isolating the terms on the right-hand side of \eqref{Mstep} that only depend on $\bm{\beta}$. We discuss the M-step for updating $\bm{\beta}$ in detail in the next section.

\subsection{M-step for updating the regression coefficients}

In the M-step of our EM algorithm, we need to solve the optimization problem \eqref{betaupdate}, or equivalently,
\begin{equation} \label{coefficientsMstep} 
	\bm{\beta}^{(t)} = \displaystyle \argmin_{ \beta} \left\{ - \ell_n ( \bm{\beta} ) + \sum_{g=1}^{G} \lambda_g^{\star (t-1)} \lVert \bm{\beta}_g \rVert_2  \right\},
\end{equation}
where $-\ell_n(\bm{\beta})$ is the negative of the log-likelihood in \eqref{ll}. While \eqref{coefficientsMstep} may appear to be intractable, we can apply the standard iteratively reweighted least squares (IRLS) algorithm \citep{McCullaghNelder1989} for GLMs to efficiently solve \eqref{coefficientsMstep}.

The IRLS algorithm in GLMs is based on a quadratic approximation of the negative log-likelihood. Denote the mean response as $\mu_i = b'(\theta_i)$, and let $V(\mu_i) = b'' ((b')^{-1} (\mu_i))$ be the variance function, where $b$ is the cumulant function in \eqref{exponentialfamily}. With the link function $h$ in \eqref{GLMgroups}, we define the ``working response'' vector $\mathbf{Z}$ at the $k$th iteration of the optimization problem \eqref{coefficientsMstep} as $\mathbf{Z} = \mathbf{X} \bm{\beta}^{(k)} + \bm{\zeta}^{(k)}$, where $\zeta_i^{(k)} = h'(\mu_i^{(k)})(y_i - \mu_i^{(k)}),$ $i = 1, \ldots, n$. Similarly, we define the weights matrix $\mathbf{W} = \textrm{diag}(w_1^{(k)}, \ldots, w_n^{(k)})$, where the weights are $w_i^{(k)} = [ V(\mu_i^{(k)})(h'(\mu_i^{(k)}))^2]^{-1}$. As shown in \citet{McCullaghNelder1989}, the negative log-likelihood for any GLM in the exponential dispersion family (1.1) can then be approximated as
\begin{equation} \label{IRLS}
	- \ell_n (\bm{\beta}) \approx \frac{1}{2} ( \mathbf{Z} - \mathbf{X} \bm{\beta} )^\top \mathbf{W} ( \mathbf{Z} - \mathbf{X} \bm{\beta} ).
\end{equation}
Thus, substituting the approximation \eqref{IRLS} for $-\ell_n ( \bm{\beta})$ in \eqref{coefficientsMstep}, we see that the M-step \eqref{coefficientsMstep} for $\bm{\beta}$ can alternatively be written as
\begin{equation} \label{regularizedIRLS}
	\bm{\beta}^{(t)} = \argmin_{\bm{\beta}} \left\{ \frac{1}{2} ( \mathbf{Z} - \mathbf{X} \bm{\beta} )^\top \mathbf{W} ( \mathbf{Z} - \mathbf{X} \bm{\beta} ) + \sum_{g=1}^{G} \lambda_g^{\star (t-1)} \lVert \bm{\beta}_g \rVert_2 \right\}.
\end{equation}
With the quadratic approximation to the negative log-likelihood,  \eqref{regularizedIRLS} is now a standard linear regression model with a group lasso penalty \citep{YuanLin2006} and group-specific weights $\lambda_g^{\star (t-1)}$ for each $g$th group. Solving \eqref{regularizedIRLS} can be done with any standard block coordinate descent algorithm for the regular group lasso \citep{YuanLin2006} in linear regression. In particular, if the canonical link function $h = (b')^{-1}$ is used, then the Fisher information matrix and the negative Hessian matrix are equal, and we use the majorization-minimization (MM) algorithm in \citet{BrehenyHuang2015} to solve \eqref{regularizedIRLS}. On the other hand, if a non-canonical link function is used, then we use the least squares approximation (LSA) approach of \citet{WangLeng2007} with a group lasso penalty.

{\picotwo The SSGL MAP algorithm is on average slower than the numerical algorithms used for group lasso, group SCAD, or group MCP. This is because \emph{each} iteration of the EM algorithm requires numerically solve the optimization problem \eqref{coefficientsMstep} in the M-step. Therefore, if the EM algorithm takes $B$ total iterations to converge, we expect the average runtime of the SSGL method to be on the order of $B$ times slower than that for group lasso, group SCAD, and group MCP. However, we observed that our EM algorithm tended to converge within a small number of iterations (typically fewer than 10), at least in all of the simulation settings we considered. Thus, the computation time for SSGL for fixed $\lambda_0$ was still reasonably quick. For example, in Simulations 4 and 8 of Section \ref{illustrations} where $p = 1600$, the SSGL runtime for any fixed $\lambda_0$ was under 30 seconds in all replications.
	
	To further accelerate the computational efficiency for SSGL, we could employ alternative optimization strategies that obviate the need to introduce latent indicator variables. For example, we could potentially approximate the SSGL prior with a neuronized prior \citep{ShinLiu2022}, which is a product of a Gaussian weight variable and a scale variable transformed from Gaussian via an activation function. This approximation would allow for a fast coordinate ascent algorithm which avoids the need to marginalize out latent indicator variables \citep{ShinLiu2022}. We leave the investigation of more computationally efficient algorithms for SSGL for future work.}

\section{Complete Gibbs sampling algorithms for Section 5.3} \label{App:D}

{\pico 
	Recall that $\bm{\beta} = (\bm{\beta}_1^\top, \ldots, \bm{\beta}_G^\top)^\top$, where $\bm{\beta}_g \in \mathbb{R}^{m_g}, g = 1, \ldots, G$, and $\bm{\tau} = (\tau_1, \ldots, \tau_G)^\top$ is defined in (5.14). Let $\mathcal{PG}(a, b)$ denote the P\'{o}lya-gamma distribution \citep{PolsonScottWindle2013}, and let $\mathcal{GIG}(a,b,c)$ denote the generalized inverse Gaussian distribution with density function $f(x; a, b, c) \propto {\picotwo x^{a-1} e^{-(b/x+cx)/2}}$. We define $\boldsymbol{\kappa} = (\kappa_1, \ldots, \kappa_n)^\top$, where for $i = 1, \ldots, n$,
	\begin{equation} \label{kappavectors}
		\kappa_i = \left\{ \begin{array}{ll} y_i - 0.5 & ~~ \text{for logistic regression,} \\
			0.5(y_i - M) & ~~ \text{for Poisson regression.} \end{array}  \right.
	\end{equation}
	We also define $\bm{\omega} = (\omega_1, \ldots, \omega_n)^\top$, where for $i = 1, \ldots, n$,
	\begin{equation} \label{omegavectors}
		\left\{ \begin{array}{ll} \omega_i \sim \mathcal{PG}(1, \mathbf{x}_i^\top \bm{\beta}) & ~~ \text{for logistic regression,}  \\ \omega_i \sim \mathcal{PG}(M, \mathbf{x}_i^\top \bm{\beta} - \log M) & ~~ \text{for Poisson regression.} \end{array} \right.  
	\end{equation}
	Finally, define the vector $\mathbf{Z} = (z_1, \ldots, z_n)^\top$, where for $i = 1, \ldots, n$,  
	\begin{equation} \label{zvectors}
		z_i = \left\{ \begin{array}{ll} \kappa_i / \omega_i & ~~ \text{for logistic regression,} \\
			\kappa_i / \omega_i + \log M & ~~ \text{for Poisson regression,} \end{array} \right.
	\end{equation}
	and $\bm{\kappa}$ is defined as in \eqref{kappavectors}. 
	
	The Gibbs sampler for grouped logistic or Poisson regression with the SSGL prior (5.4) is given in Algorithm \ref{algorithm1}.} {\pico When $p = \sum_{g=1}^{G} m_g$ is larger than $n$, we can use the fast sampling algorithm of \citet{bhattacharya2016fast} to sample $\bm{\beta}^{(t)}$ in Step 4 of Algorithm \ref{algorithm1} in $\mathcal{O}(n^2p)$ time instead of $\mathcal{O}(p^3)$ time. This sampling algorithm is given in Algorithm \ref{algorithm2}.}

\begin{algorithm}[t!]
	\begin{flushleft}
		{\pico \textbf{Input:} \hspace{.15cm} Initial values $\bm{\beta}^{(0)}$, $\theta^{(0)}$, $\bm{\tau}^{(0)}$, $T$ (number of MCMC samples to run),  \\
			\hspace{1.17cm} $B$ (number of samples to discard as burn-in)} \\
		\vspace{.1cm}
		
		{\pico \textbf{Output:} MCMC samples for regression coefficients $\bm{\beta}$}
		\vspace{.3cm}
		
		{\pico \textbf{for} $t = 1, \ldots, T$ \textbf{do}}
		\begin{enumerate}
			\item {\pico \textbf{for} $i = 1, \ldots, n$ \textbf{do}} \\
			
			\hspace{.5cm} {\pico i. \hspace{.01cm} Sample $\bm{\omega}^{(t)}$ as in \eqref{omegavectors} and set $\bm{\Omega} = \text{diag}\{ \bm{\omega}^{(t)} \}$}
			\vspace{.1cm}
			
			\hspace{.5cm} {\pico ii. Update $\mathbf{Z}^{(t)}$ as in \eqref{zvectors}}
			\vspace{.1cm}
			
			\item {\pico \textbf{for} $g = 1, \ldots, G$ \textbf{do}} \\
			\hspace{.5cm} {\pico i. \hspace{.01cm} Sample $\gamma_g^{(t)} \sim \text{Bernoulli} \left( \frac{\pi_1}{\pi_1 + \pi_0} \right)$, where $\pi_1 = \theta^{(t-1)} \lambda_1^{m_g+1} \exp \left( - \lambda_1^2 \tau_g^{(t-1)} / 2 \right)$} 
			
			\hspace{.85cm} {\pico and $\pi_0 = \left(1-\theta^{(t-1)}\right) \lambda_0^{m_g+1} \exp \left( - \lambda_0^2 \tau_g^{(t-1)} / 2 \right)$}
			
			\hspace{.5cm} ii. Sample $\tau_g^{(t)} \sim \mathcal{GIG} \left( \frac{1}{2}, \lVert \bm{\beta}_g^{(t-1)} \rVert_2^2, (\lambda_g^{\star})^2 \right)$, where {\picotwo $\lambda_g^{\star} = \gamma_g^{(t)} \lambda_1 + (1-\gamma_g^{(t)}) \lambda_0$}
			
			\item {\pico Sample $\theta^{(t)} \sim \text{Beta} \left( a+ \sum_{g=1}^{G} \gamma_g^{(t)},~b + G - \sum_{g=1}^{G} \tau_g^{(t)} \right)$}
			\item {\pico Sample $\bm{\beta}^{(t)} \sim \mathcal{N} \left( \bm{\mu}_{\bm{\beta}}, \bm{\Sigma}_{\bm{\beta}} \right)$, where $\bm{\mu}_{\bm{\beta}} = \bm{\Sigma}_{\bm{\beta}} \mathbf{X}^\top \bm{\Omega} \mathbf{Z}$, $\bm{\Sigma}_{\bm{\beta}} = ( \mathbf{X}^\top \bm{\Omega} \mathbf{X} + \mathbf{D}_{\bm{\tau}}^{-1} )^{-1}$, and $\mathbf{D}_{\bm{\tau}} = \text{Bdiag}~( \tau_1^{(t)} \mathbf{I}_{m_1}, \ldots, \tau_G^{(t)} \mathbf{I}_{m_G} )$}
		\end{enumerate}
		\item 
		{\pico \textbf{return} $\bm{\beta}^{(B+1)}, \ldots, \bm{\beta}^{(T)}$}
	\end{flushleft}
	\caption{ {\pico MCMC algorithm for grouped logistic and grouped Poisson regression with the SSGL prior}} \label{algorithm1}
\end{algorithm}

\begin{algorithm}[H]
	\begin{flushleft}
		{\pico \textbf{Input:} Most recent updates of $\bm{\Omega}$, $\mathbf{D}_{\bm{\tau}}$, and $\mathbf{Z}$} \\
		
		{\pico \textbf{Output:} An exact sample of $\bm{\beta}$ in Step 4 of Algorithm \ref{algorithm1}}
		
		\begin{enumerate}
			\item {\pico Set $\widetilde{\mathbf{X}} = \bm{\Omega}^{1/2} \mathbf{X}$}
			\item {\pico Sample $\mathbf{m} \sim \mathcal{N}(\bm{0}, \bm{D}_{\bm{\tau}})$ and $\bm{\delta} \sim \mathcal{N}(\bm{0}, \mathbf{I}_n)$ independently}
			\item {\pico Set $\mathbf{v} = \widetilde{\mathbf{X}} \mathbf{m}+ \bm{\delta}$ and $\mathbf{v}^{\star} = \bm{\Omega}^{1/2} \mathbf{Z} - \mathbf{v}$}   
			\item {\pico Set $\mathbf{A} = \widetilde{\mathbf{X}} \mathbf{D}_{\bm{\tau}} \widetilde{\mathbf{X}}^\top + \mathbf{I}_n$} 
			\item {\pico Set $\mathbf{w} = \mathbf{A}^{-1} \bm{v}^{\star}$}
			\item {\pico Set $\bm{\beta} = \mathbf{m} + \bm{D}_{\bm{\tau}} \widetilde{\mathbf{X}}^\top \mathbf{w}$}
		\end{enumerate}
		
		{\pico \textbf{return} $\bm{\beta}$}
	\end{flushleft}
	\caption{{\pico Algorithm for sampling $\bm{\beta}$ in Step 4 of Algorithm \ref{algorithm1} when $p > n$}} \label{algorithm2}
\end{algorithm}

\section{Simulation studies for negative binomial regression with a log link} \label{App:E}

Since the mean and variance are equal for the Poisson distribution, Poisson regression may be inappropriate for modeling count data that exhibit overdispersion. In this case, it may be more appropriate to use negative binomial regression. That is, we assume that $y_i \sim NB(\alpha, \mu_i), i = 1, \ldots, n$, where $\alpha > 0$ is a size parameter, and the probability density function (pdf) for each $y_i$ is
\begin{align*}
	f (y_i \mid \alpha, \mu_i ) = \frac{\Gamma(y_i + \alpha)}{y_i! \Gamma(\alpha)} \left( \frac{\mu_i}{\mu_i + \alpha} \right)^{y_i} \left( \frac{\alpha}{\mu_i + \alpha} \right)^{\alpha}, \hspace{.3cm} y_i = 0, 1, 2, \ldots
\end{align*}
Since $\textrm{Var}(y_i) = \mu_i + \mu_i^2/ \alpha$, the negative binomial distribution is better equipped to handle overdispersed count data than the Poisson distribution. 

For negative binomial distributed responses, the natural parameter is $\theta_i = \log \left( \mu_i / (\mu_i + \alpha) \right)$. With the link function $h(u) = \log(u)$, we have $b(u) = - \alpha \log(1-e^{u})$ in \eqref{exponentialfamily}. It is readily seen that $\mathbb{E}(y_i) = b'(\theta_i) = \mu_i$. For the function $\xi = (h \circ b')^{-1}$, we have $\xi(u) = -\log(\alpha e^{-u}+1)$. Since $\xi(u) \neq u$, negative binomial regression with the log link is an example of a GLM with a \textit{non}-canonical link function. 

We now present simulation results in grouped negative binomial regression for the SSGL model \eqref{SSGL}. We compared our results to group lasso ({\pico GL}), {\pico adaptive group lasso (AdGL)}, group SCAD ({\pico GSCAD}), and group MCP ({\pico gMCP}). {\pico We used the same methods for choosing the hyperparameters, regularization parameters, and weights as those described in Section \ref{setup}.} We fixed $\alpha = 1$ and generated the responses independently as $y_i \mid \mathbf{x}_i \sim NB (\alpha, \exp(\theta_i)), i = 1, \ldots, n$. We conducted the following simulation studies:
\begin{itemize}
	\item[] \textbf{Experiment 1}. We set $n = 500$ and $G = 30$. The design matrix $\mathbf{X}$ and the regression coefficients $\bm{\beta}$ were generated the same way as in Experiment 1 in Section \ref{Poissonregression}. Then we modeled
	\begin{align*}
		\log(\theta) = \mathbf{x}^\top \bm{\beta}.
	\end{align*}
	\item[] \textbf{Experiment 2}. We set $n = 500$ and $G = 30$. We generated the entries of the $n \times G$ design matrix $\mathbf{X}$ from independent $\textrm{Uniform}(-1,1)$ random variables. Then we modeled
	\begin{align*}
		\log \left( \theta \right) = 1.5 \sin(3x_1) - x_5 e^{0.5 x_5^2}.
	\end{align*}
	We represented each covariate as a six-term B-spline basis expansion. 
\end{itemize}
We repeated each experiment 200 times. Table \ref{Table:3} reports our results averaged across the 200 replications. In Experiment 1, {\pico SSGL had the lowest MSE and the second lowest MSPE. Meanwhile, all five methods had a TPR of 1, indicating that all methods had equally high power to detect the significant groups in this particular setting. However, SSGL had the highest average TNR and precision, indicating superior overall group selection performance.}

In Experiment 2, {\pico SSGL also had the lowest MSPE, indicating superior average prediction on out-of-sample data. In this experiment, all five methods once again had a TPR of 1. However, SSGL had the highest average TNR and precision, indicating that SSGL was the best suited for function selection in this particular generalized additive model.}

\begin{table}[t!]
	\centering
	\caption{Simulation results for grouped negative binomial regression under the SSGL, GL, {\pico AdGL,} GMCP, and GSCAD models, averaged across 200 replicates. The empirical standard error is reported in parentheses below the average.}
	\label{Table:E}
	\medskip
	\begin{tabularx}{\linewidth}{*{6}{p{.143\linewidth}}}
		\multicolumn{6}{c}{\textbf{Experiment 1}} \\ \toprule
		& MSE & MSPE & TPR & TNR & Prec \\ 
		\hline \hline
		SSGL & \textbf{0.008} & 30.04 & \textbf{1} & \textbf{0.999} & \textbf{0.999} \\
		& (0.006) & (10.56) & (0) & (0.007) & (0.029) \\
		\hline
		GL & 0.029 & \textbf{29.43} & \textbf{1} & 0.387 & 0.255 \\
		& (0.008) & (9.52) & (0) & (0.146) & (0.051)  \\ 
		\hline
		AdGL & 0.018 & 31.09 & \textbf{1} & 0.809 & 0.562 \\
		& (0.008) & (12.32) & (0) & (0.119) & (0.180) \\
		\hline
		GMCP & 0.012 & 31.15 & \textbf{1} & 0.892 & 0.705 \\
		& (0.005) & (15.98) & (0) & (0.094) & (0.192) \\
		\hline
		GSCAD & 0.013 & 31.31 & \textbf{1} & 0.702 & 0.434 \\
		&  (0.005) & (14.59) & (0) & (0.133) & (0.130) \\
		\bottomrule
	\end{tabularx}
	
	\medskip
	
	\begin{tabularx}{\linewidth}{*{5}{p{.18\linewidth}}}
		\multicolumn{5}{c}{\textbf{Experiment 2}} \\ \toprule
		& MSPE & TPR & TNR & Prec \\ 
		\hline \hline
		SSGL & \textbf{17.43} & \textbf{1} & \textbf{1} & \textbf{1}  \\
		& (5.09) & (0) & (0) & (0) \\
		\hline
		GL & 18.35 & 1 & 0.383 & 0.114 \\
		& (5.85) & (0) & (0.184) & (0.044) \\
		\hline
		AdGL & 18.16 & \textbf{1} & 0.759 & 0.296 \\
		& (5.47) & (0) & (0.140) & (0.186) \\		
		\hline
		GMCP & 17.52 & \textbf{1} & 0.921 & 0.695 \\
		& (5.10) & (0) & (0.123) & (0.317) \\
		\hline
		GSCAD & 17.49 & 1 & 0.790 & 0.363 \\
		& (5.33) & (0) & (0.160) & (0.236) \\
		\bottomrule
	\end{tabularx}
\end{table}
	
\end{appendix}
	
\end{document}